\documentclass[11pt,english,table]{article}
\usepackage{lmodern}

\usepackage[T1]{fontenc}
\usepackage[latin9]{inputenc}
\usepackage{geometry}
\usepackage{verbatim}
\usepackage{bm}
\usepackage{amsmath}
\usepackage{amsthm}
\usepackage{amssymb}
\usepackage{graphicx}
\usepackage{rotfloat}
\usepackage{setspace}
\usepackage[authoryear]{natbib}
\makeatletter
\theoremstyle{plain}
\newtheorem{prop}{\protect\propositionname}
\theoremstyle{plain}
\newtheorem{thm}{\protect\theoremname}
\theoremstyle{plain}
\newtheorem{lem}{\protect\lemmaname}

\setcitestyle{round}
\usepackage{lscape}
\usepackage{longtable}
\usepackage{xcolor}
\usepackage{rotating}

\usepackage{array}
\usepackage{tabularx}\usepackage{multirow}\usepackage{booktabs}
\usepackage{ragged2e}\newcolumntype{C}[1]{>{\centering\arraybackslash}p{#1}}
\usepackage{rotfloat}
\newcolumntype{J}[1]{>{\justify\arraybackslash}p{#1}}
\newcolumntype{R}[1]{>{\RaggedLeft\arraybackslash}p{#1}}
\newcolumntype{Q}[1]{>{\columncolor{Gray}\RaggedLeft\arraybackslash}p{#1}}
\newcolumntype{L}[1]{>{\RaggedRight\arraybackslash}p{#1}}
\newcolumntype{G}{@{\extracolsep{0.5cm}}l@{\extracolsep{0pt}}}%
\newcolumntype{P}[1]{>{\centering\arraybackslash}p{#1}}
\newcolumntype{Y}{>{\centering\arraybackslash}X}
\newcommand{\nhphantom}[1]{\sbox0{#1}\hspace{-\the\wd0}} 

\AtBeginDocument{

}

\usepackage[colorlinks,
            linkcolor=blue,
            anchorcolor=blue,
            citecolor=blue]{hyperref}

\usepackage{babel}

\makeatother

\usepackage{babel}
\providecommand{\lemmaname}{Lemma}
\providecommand{\propositionname}{Proposition}
\providecommand{\theoremname}{Theorem}

\begin{document}
\title{Cluster GARCH\emph{\normalsize{}\bigskip{}
}}
\author{\textbf{Chen Tong}$^{\dagger}$\textbf{ }and\textbf{ Peter Reinhard
Hansen}$^{\ddagger}$ and\textbf{ Ilya Archakov}{\normalsize{}$^{\mathsection}$}\thanks{We are grateful for many valuable comments made by participants at
the 2023 conference ``Robust Econometric Methods in Financial Econometrics''.
Chen Tong acknowledges financial support from the Youth Fund of the
National Natural Science Foundation of China (72301227), and the Ministry
of Education of China, Humanities and Social Sciences Youth Fund (22YJC790117).
Corresponding author: Peter Reinhard Hansen. Email: hansen@unc.edu.}\bigskip{}
 \\
\\
 {\normalsize{}$^{\dagger}$}\emph{\normalsize{}School of Economics,
Xiamen University}{\normalsize{} }\\
 {\normalsize{}$^{\ddagger}$}\emph{\normalsize{}University of North
Carolina, Chapel Hill}{\normalsize{} }\\
 {\normalsize{}$^{\mathsection}$}\emph{\normalsize{}York University,
Toronto, Canada}{\normalsize{} }\emph{\normalsize{}\medskip{}
 }}
\date{\emph{\normalsize{}\today}}
\maketitle
\begin{abstract}
We introduce a novel multivariate GARCH model with flexible convolution-$t$
distributions that is applicable in high-dimensional systems. The
model is called \emph{Cluster GARCH} because it can accommodate cluster
structures in the conditional correlation matrix and in the tail dependencies.
The expressions for the log-likelihood function and its derivatives
are tractable, and the latter facilitate a score-drive model for the
dynamic correlation structure. We apply the Cluster GARCH model to
daily returns for 100 assets and find it outperforms existing models,
both in-sample and out-of-sample. Moreover, the convolution-$t$ distribution
provides a better empirical performance than the conventional multivariate
$t$-distribution.

\bigskip{}
\end{abstract}
\textit{\small{}{\noindent}Keywords:}{\small{} Multivariate GARCH,
Score-Driven Model, Cluster Structure, Block Correlation Matrix, Heavy
Tailed Distributions.}{\small\par}

\noindent \textit{\small{}{\noindent}JEL Classification:}{\small{}
G11, G17, C32, C58 }{\small\par}

\clearpage{}

\section{Introduction}

Univariate GARCH models have enjoyed considerable empirical success
since they were introduced in \citet{Engle:1982} and refined in \citet{bollerslev:86}.
In contrast, the success of multivariate GARCH models has been more
moderate due to a number of challenges, see e.g. \citet{BauwensLaurentRombouts:2006}.
A common approach to modeling covariance matrices is to model variances
and correlations separately, as is the case in the Constant Conditional
Correlation (CCC) model by \citet{Bollerslev1990} and the Dynamic
Conditional Correlation (DCC) model by \citet{Engle2002}. See also
\citet{EngleSheppard:2001}, \citet{TseTsui:2002}, \citet{Aielli:2013},
\citet{EngleLedoitWolf:2019}, and \citet{PakelShephardSheppardEngle:2021}.
While univariate conditional variances can be effectively modeled
using standard GARCH models, the modeling of dynamic conditional correlation
matrices necessitates less intuitive choices to be made. One challenge
is that the number of correlations increases with the square of the
number of variables, a second challenge is that the conditional correlation
matrix must be positive semidefinite, and a third challenge is to
determine how correlations should be updated in response to sample
information.

In this paper, we develop a novel dynamic model of the conditional
correlation matrix, the \emph{Cluster GARCH} model, which has three
main features. First, use convolution-$t$ distributions, which is
a flexible class of multivariate heavy-tailed distributions with tractable
likelihood expressions. The multivariate $t$-distributions are nested
in this framework, but a convolution-$t$ distribution can have heterogeneous
marginal distributions and cluster-based dependencies. For instance,
convolution-$t$ distributions can generate the type of sector-specific
price jumps reported in \citet{AndersenDingTodorov:2024}. Second,
the dynamic model is based on the score-driven framework by \citet{CrealKoopmanLucas:2013},
which leads to closed-form expressions for all key quantities. Third,
the model can be combined with a block correlation structure that
makes the model applicable to high-dimensional systems. This partitioning,
defining the block structure, can also be interpreted as a second
type of cluster structure.

Heavy-tailed distributions are common in financial returns, and many
empirical studies adopt the multivariate $t$-distribution to model
vectors of financial series, e.g., \citet{Kotz2004}, \citet{Harvey2013},
and \citet{IbragimovIbragimovWalden:2015}. An implication of the
multivariate $t$-distribution is that all standardized returns have
identical and time-invariant marginal distributions. This is a restrictive
assumption, especially in high dimensions. The convolution-$t$ distributions
by \citet{HansenTong:2024} relax these assumptions, and one of the
main contributions of this paper is to incorporate this class of distributions
into a tractable multivariate GARCH model. A convolution-$t$ distribution
is a convolution of multivariate $t$-distributions. In the Cluster
GARCH model, standardized returns are time-varying linear combinations
of independent $t$-distributions, which can have different degrees
of freedom. This leads to dynamic and heterogeneous marginal distributions
for standardized returns, albeit the conventional multivariate $t$-distribution
is nested in this framework as a special case. We focus on three particular
types of convolution-$t$ distributions, labelled Canonical-Block-$t$,
Cluster-$t$, and Hetero-$t$. These all have relatively simple log-likelihood
functions, such that we can obtain closed-form expressions for the
first two derivatives, score and information matrix, of the conditional
log-likelihood functions. These are used in our score-driven model
for the time-varying correlation structure, which is a key component
of the Cluster GARCH model. 

High-dimensional correlation matrices can be modeled using a parsimonious
block structure for the conditional correlation matrix. The DCC model
is parsimonious but greatly limits the way the conditional covariance
matrix can be updated. Without additional structure, the number of
latent variables increases with $n^{2}$, where $n$ is the number
of assets. This number becomes unmanageable once $n$ is more than
a single digit, and maintaining a positive definite correlation matrix
can be challenging too. The correlation structure in the Block DECO
model by \citet{EngleKelly:2012} is an effective way to reduce the
dimension of the estimated parameters. However, the estimation strategy
in \citet{EngleKelly:2012} was based on an ad-hoc averaging of within-block
correlations for an auxiliary DCC model, and they did not fully utilize
the simplifications offered by the block structure.\footnote{They derived likelihood expressions for the case with $K=2$ blocks.
For more two blocks, $K>2$, they resort to a composite likelihood
evaluation.} The model proposed in this paper draws on recent advances in correlation
matrix analysis by \citet[2024]{ArchakovHansen:Correlation}\nocite{ArchakovHansen:CanonicalBlockMatrix}.
We will, in some specifications, adopt the block parameterization
of the conditional correlation matrix, used in \citet{ArchakovHansenLundeMRG},
which has (at most) $K\left(K+1\right)/2$ free parameters where $K$
is the number of blocks. This approach guarantees a positive definite
correlation matrix and the likelihood evaluation is greatly simplified.
Overall, the Cluster GARCH offers a good balance between flexibility
and computational feasibility in high dimensions.

We adopt the convenient parametrization of the conditional correlation
matrix, $\gamma(C)$, which is defined by taking the matrix logarithm
of the correlation matrix, $C$, and stacking the off-diagonal elements
of $\log C$ into the vector, $\gamma\in\mathbb{R}^{d}$, where $d=n(n-1)/2$.
This parametrization was introduced in \citet{ArchakovHansen:Correlation}
and the mapping $C\mapsto\gamma(C)$ is one-to-one between the set
of non-singular correlation matrices $\mathcal{C}_{n\times n}$ and
$\mathbb{R}^{d}$. So, the inverse mapping, $C(\gamma)$, will always
yield a positive definite correlation matrix and any non-singular
correlation matrix can be generated in this way. The parametrization
can be viewed as a generalization of Fisher\textquoteright s Z-transformation
to the multivariate case. It has attractive finite sample properties,
which makes it suitable for an autoregressive model structure, see
\citet{ArchakovHansen:Correlation}.%

A block correlation structure arises when variables can be partitioned
into clusters, $K$ say, and the correlation between two variables
is determined by their cluster assignments. When $C$ has a block
structure, then $\log C$ also has a block structure. This leads to
a new parametrization of block correlation matrices, which defines
a one-to-one mapping $C\mapsto\eta(C)$ between the set of non-singular
block correlation matrices $\mathcal{C}_{n\times n}$ and $\mathbb{R}^{d}$
with $d=K\left(K+1\right)/2$. We adopt the canonical representation
by \citet{ArchakovHansen:CanonicalBlockMatrix}, which is a quasi-spectral
decomposition of block matrices that diagonalizes the matrix with
the exception of a small $K\times K$ submatrix. This decomposition
makes the model parsimonious and greatly simplifies the evaluations
of the log-likelihood function. This parameterization of block correlation
matrices is more general than the factor-based approach to parametrizing
block correlation matrices.\footnote{The factor-induced block structure, see \citet{CrealTsay2015}, \citet{OpschoorLucasBarraVanDick:2021},
and \citet{OhPatton2023}, entails superfluous restrictions on $C$,
see \citet{TongHansen:2023}. Both approach simplifies the computation
of $\det C$ and $C^{-1}$, but only the parametrization based on
the canonical representation simplifies the evaluation of the likelihood
function for the convolution-$t$ distributions.}

Our paper contributes to the literature on score-driven model for
dynamics of covariance matrices. Using the multivariate $t$-distribution,
\citet{CrealKoopmanLucasJBES:2012} and \citet{HafnerWang:2023} proposed
score-driven model for time-varying covariance and correlation matrix,
respectively.\footnote{The model by \citet{HafnerWang:2023} update parameters using the
unscaled score, \textit{i.e.,} they did not use the information matrix.} \citet{OhPatton2023} proposed a score-driven dynamic factor copula
models with skew-$t$ copula function, however, the analytical information
matrices in these copula models are not available. Using realized
measures of the covariance matrix, \citet{GorgiHansenJanusKoopman:2019}
proposed the Realized Wishart-GARCH, which relies on a Wishart distribution
for realized covariance matrices and on a Gaussian distribution for
returns. \citet{OpschoorJanusLucasVanDick:2017} constructed a multivariate
HEAVY model based on Heavy-tailed distributions for both returns and
the realized covariances. An aspect, which sets the Cluster GARCH
apart from the existing literature, is that the model is based on
the convolution-$t$ distributions, which includes the Gaussian distribution
and the multivariate $t$-distributions as special cases. The block
structures we impose on the correlation matrix in some specifications,
was previously used in \citet{ArchakovHansenLundeMRG}. Their model
used the Realized GARCH framework with a Gaussian specification, whereas
we adopt the score-driven framework for convolution-$t$ distributions,
and do require realized volatility measures in the modeling.

We conduct an extensive empirical investigation on the performance
of our dynamic model for correlation matrices. The sample period spans
the period from January 3, 2005 to December 31, 2021. The new model
is applicable to high dimensions, and we consider a ``small universe''
with $n=9$ assets and a ``large universe'' with $n=100$ assets.
The small universe allows us compare the new models with a range of
existing models, as most of these are not applicable to the large
universe. We also undertake an more detailed specification analysis
with the small universe. The nine stocks are from three sectors, three
from each sector, which motivates certain block and cluster structures.
First, we find that the convolution-$t$ distribution offers a better
fit than the conventional $t$-distribution. Overall, the Cluster-$t$
distribution has the largest log-likelihood value. Second, we find
that score-driven models successfully captures the dynamic variation
in the conditional correlation matrix. The new score-driven models
outperform traditional DCC models when based on the same distributional
assumptions, and the proposed score-driven model with a sector motivated
block correlation matrix has the smallest BIC.

The large universe with $n=100$ stocks poses no obstacles for the
Cluster GARCH model. We used the sector classification of the stocks
to define the block structure in the correlation matrix. We also used
the sector classification to explore possible cluster structures in
the tail-dependencies, which are related to parameters in the convolution-$t$
distribution. With $K=10$ sectors this reduces the number of free
parameters in the correlation matrix from 4950 to 55, and the model
estimation is very fast and stable, in part because the required computations
only involve $K\times K$ matrices (instead of $n\times n$ matrices).
For the large universe, the empirical results favor the Hetero-$t$
specification, which entails a convolutions of a large number of univariate
$t$-distributions. We also find that \emph{correlation targeting},
which is analogous to variance targeting in GARCH models, is beneficial.

The rest of this paper is organized as follows: In Section \ref{sec:NewPaBlockCorr}
we introduce a new parametrization of block correlation matrices,
based on \citet{ArchakovHansen:Correlation} and \citet{ArchakovHansen:CanonicalBlockMatrix}.
In Section 3, we introduce the convolution-$t$ distributions. We
derive the score-driven models in Section 4, and we obtain analytical
expressions for the score and information matrix for the convolution-$t$
distributions, including the special case where $C$ has a block structure.
Some details about practical implementation are given in Section 5.
The empirical analysis is presented in Section 6 and includes in-sample
and out-of-sample evaluations and comparisons. All proofs are given
in the Appendix.

\section{The Theoretical Model\label{sec:NewPaBlockCorr}}

Consider an $n$-dimensional time-series, $R_{t}$, $t=1,2,\ldots,T$,
and let $\{\mathcal{F}_{t}\}$ be a filtration to which $R_{t}$ is
adapted, i.e. $R_{t}\in\mathcal{F}_{t}$. We denote the conditional
mean by $\mu_{t}=\mathbb{E}(R_{t}|\mathcal{F}_{t-1})$ and the conditional
covariance matrix by $\Sigma_{t}=\mathrm{var}(R_{t}|\mathcal{F}_{t-1})$.
With $\Lambda_{\sigma_{t}}\equiv\mathrm{diag}(\sigma_{1t},\ldots,\sigma_{nt})$,
where $\sigma_{it}^{2}=\mathrm{var}(R_{it}|\mathcal{F}_{t-1})$, $i=1,\ldots,n$,
it follows that the conditional correlation matrix is given by
\[
C_{t}=\Lambda_{\sigma_{t}}^{-1}\Sigma_{t}\Lambda_{\sigma_{t}}^{-1}.
\]
Initially, we take $\mu_{t}$ and $\Lambda_{\sigma_{t}}$ as given
and focus on the dynamic modeling of $C_{t}$. We are particularly
interested in the case where $n$ is large. We define the following
standardized variables with a dynamic correlation matrix $C_{t}$,
\[
Z_{t}=\Lambda_{\sigma_{t}}^{-1}(X_{t}-\mu_{t}).
\]

To simplify the notation, we omit subscript-$t$ in most of Sections
\ref{sec:NewPaBlockCorr} and \ref{sec:Distributions} and reintroduce
it again in Section \ref{sec:Score-Driven-Models} where the dynamic
model is presented.

\subsection{Block Correlation Matrix}

If $n$ is relatively small, we can model the dynamic correlation
matrix using $d=n(n-1)/2$ latent variables. Additional structure
on $C$ is required when $n$ is larger, because the number of latent
variables becomes unmanageable. Additional structure can be imposed
using a block structures on $C$, as in \citet{EngleKelly:2012}. 

A block correlation matrix is characterized by a partitioning of the
variables into clusters, such that the correlation between two variables
is solely determined by their cluster assignments. Let $K$ be the
number of clusters, and let $n_{k}$ be the number of variables in
the $k$-th cluster, $k=1,\ldots,K$, such that $n=\sum_{k=1}^{K}n_{k}$.
We let $\bm{n}=\left(n_{1},n_{2},\ldots,n_{K}\right)^{\prime}$ be
the vector with cluster sizes and sort the variables such that the
first $n_{1}$ variables are those in the first cluster , the next
$n_{2}$ variables are those in the second cluster, and so forth.
Then $C=\mathrm{corr}(Z)$ will have the following block structure
\begin{equation}
\ensuremath{C=\left[\begin{array}{cccc}
C_{[1,1]} & C_{[1,2]} & \cdots & C_{[1,K]}\\
C_{[2,1]} & C_{[2,2]}\\
\vdots &  & \ddots\\
C_{[K,1]} &  &  & C_{[K,K]}
\end{array}\right]},\label{eq:BlockC}
\end{equation}
where $C_{[k,l]}$ is an $n_{k}\times n_{l}$ matrix given by
\[
C_{[k,l]}=\left[\begin{array}{ccc}
\rho_{kl} & \cdots & \rho_{kl}\\
\vdots & \ddots & \vdots\\
\rho_{kl} & \cdots & \rho_{kl}
\end{array}\right],\text{ for }k\neq l\qquad\text{and }C_{[k,k]}=\left[\begin{array}{cccc}
1 & \rho_{kk} & \cdots & \rho_{kk}\\
\rho_{kk} & 1 & \ddots\\
\vdots & \ddots & \ddots\\
\rho_{kk} &  &  & 1
\end{array}\right].
\]
Each block, $C_{[k,l]}$, has just one correlation coefficient, such
that the block structure reduces the number of unique correlations
from $n\left(n-1\right)/2$ to at most $K\left(K+1\right)/2$.\footnote{This is based on the general case that the number of assets in each
group is at least two. When there are $\tilde{K}\leq K$ groups with
only one asset, this number become $K\left(K+1\right)/2-\tilde{K}$.
The reason for the distinction between these two cases is that an
$1\times1$ diagonal block has no correlation coefficients.} This number does not increase with $n$, and this makes it possible
to scale the model to accommodate high-dimensional correlation matrices. 

Below we derive score-driven models for unrestricted correlation matrices
and for the case where $C$ has a block structure. time{]}.\footnote{It is unproblematic to extend the model to allow for some missing
observations and occasional changes in the cluster assignments. }

\subsection{Parametrizing the Correlation Matrix\label{subsec:PaBlock}}

We parameterize the correlation matrix with the vector
\begin{equation}
\gamma(C)\equiv{\rm vecl}\left(\log C\right)\in\mathbb{R}^{d},\qquad d=n(n-1)/2,\label{eq:LogCtrans}
\end{equation}
where ${\rm vecl}(\cdot)$ extracts and vectorizes the elements below
the diagonal and $\log C$ is the matrix logarithm of the correlation
matrix.\footnote{For a nonsingular correlation matrix, we have $\log C=Q\log\Lambda Q^{\prime}$,
where $C=Q\Lambda Q^{\prime}$ is the spectral decomposition of $C$,
so that $\Lambda$ is a diagonal matrix with the eigenvalues of $C$.} The following example illustrates this parametrization for an $3\times3$
correlation matrix:{\small{}
\[
{\rm vecl}\left[\log\left(\begin{array}{ccc}
1.0 & \bullet & \bullet\\
0.5 & 1.0 & \bullet\\
0.3 & 0.7 & 1.0
\end{array}\right)\right]={\rm vecl}\left[\left(\begin{array}{ccc}
-0.15 & \phantom{-}\bullet & \phantom{-}\bullet\\
\phantom{-}0.53 & -0.47 & \phantom{-}\bullet\\
\phantom{-}0.13 & \phantom{-}0.85 & -0.34
\end{array}\right)\right]=\left(\begin{array}{c}
0.53\\
0.13\\
0.85
\end{array}\right)=:\gamma.
\]
}This parametrization is convenient because it guarantees a unique
positive definiteness correlation matrix, $C(\gamma)$ for any vector
$\gamma,$ without imposing superfluous restrictions on the correlation
matrix, see \citet{ArchakovHansen:Correlation}.

For a block correlation matrix the logarithmic transformation preserves
the block structure as illustrated in the following example:{\small{}
\[
\ensuremath{\underbrace{\left[\begin{array}{ccccccc}
\cellcolor{black!10}1.0 & \cellcolor{black!10}0.8 & 0.4 & 0.4 & \cellcolor{black!05}0.2 & \cellcolor{black!05}0.2 & \cellcolor{black!05}0.2\\
\cellcolor{black!10}0.8 & \cellcolor{black!10}1.0 & 0.4 & 0.4 & \cellcolor{black!05}0.2 & \cellcolor{black!05}0.2 & \cellcolor{black!05}0.2\\
0.4 & 0.4 & \cellcolor{black!10}1.0 & \cellcolor{black!10}0.6 & 0.1 & 0.1 & 0.1\\
0.4 & 0.4 & \cellcolor{black!10}0.6 & \cellcolor{black!10}1.0 & 0.1 & 0.1 & 0.1\\
\cellcolor{black!05}0.2 & \cellcolor{black!05}0.2 & 0.1 & 0.1 & \cellcolor{black!10}1.0 & \cellcolor{black!10}0.3 & \cellcolor{black!10}0.3\\
\cellcolor{black!05}0.2 & \cellcolor{black!05}0.2 & 0.1 & 0.1 & \cellcolor{black!10}0.3 & \cellcolor{black!10}1.0 & \cellcolor{black!10}0.3\\
\cellcolor{black!05}0.2 & \cellcolor{black!05}0.2 & 0.1 & 0.1 & \cellcolor{black!10}0.3 & \cellcolor{black!10}0.3 & \cellcolor{black!10}1.0
\end{array}\right]}_{=C}\quad\underbrace{\left[\begin{array}{ccccccc}
\cellcolor{black!10}-.59 & \cellcolor{black!10}1.02 & .251 & .251 & \cellcolor{black!05}.115 & \cellcolor{black!05}.115 & \cellcolor{black!05}.115\\
\cellcolor{black!10}1.02 & \cellcolor{black!10}-.59 & .251 & .251 & \cellcolor{black!05}.115 & \cellcolor{black!05}.115 & \cellcolor{black!05}.115\\
.251 & .251 & \cellcolor{black!10}-.29 & \cellcolor{black!10}.626 & .036 & .036 & .036\\
.251 & .251 & \cellcolor{black!10}.626 & \cellcolor{black!10}-.29 & .036 & .036 & .036\\
\cellcolor{black!05}.115 & \cellcolor{black!05}.115 & .036 & .036 & \cellcolor{black!10}-.09 & \cellcolor{black!10}.259 & \cellcolor{black!10}.259\\
\cellcolor{black!05}.115 & \cellcolor{black!05}.115 & .036 & .036 & \cellcolor{black!10}.259 & \cellcolor{black!10}-.09 & \cellcolor{black!10}.259\\
\cellcolor{black!05}.115 & \cellcolor{black!05}.115 & .036 & .036 & \cellcolor{black!10}.259 & \cellcolor{black!10}.259 & \cellcolor{black!10}-.09
\end{array}\right]}_{=\log C}}.
\]
}The parameter vector, $\gamma$ will only have as many unique elements
as there are different blocks in $C$. This number is $(K+1)K/2$,
and we can therefore condense $\gamma$ into a subvector, $\eta$,
such that
\begin{equation}
\gamma=B\eta,\label{eq:FactorGamma}
\end{equation}
where $B$ is a known bit-matrix with a single one in each row and
$\eta\in\mathbb{R}^{K(K+1)/2}$. This factor structure for $\gamma$
was first proposed in \citet{ArchakovHansenLundeMRG}.

For later use, we define the \emph{condensed log-correlation matrix},
$\tilde{C}\in\mathbb{R}^{K\times K}$, whose $(k,l$)-th element is
the off-diagonal element from the $(k,l)$-th block of $\log C$,
$k,l=1,\ldots,K$, and we can set $\eta=\mathrm{vech}(\tilde{C})\in\mathbb{R}^{K(K+1)/2}$.
In the example above, we have
\[
\tilde{C}=\left[\begin{array}{ccc}
\cellcolor{black!10}1.02 & .251 & \cellcolor{black!05}.115\\
.251 & \cellcolor{black!10}.626 & .036\\
\cellcolor{black!05}.115 & .036 & \cellcolor{black!10}.259
\end{array}\right],
\]
such that $\eta=\left[1.02,0.251,0.115,0.626,0.036,0.259\right]^{\prime}$
has dimension six whereas $\gamma$ has dimension 21. Since the block
correlation matrix, $C$, is only a function of $\eta$ we can model
the time-variation in $C$ using a dynamic model for the unrestricted
vector $\eta$. This will be our approach below. 

\subsection{Canonical Form for the Block Correlation Matrix }

Block matrices has a canonical representation that resembles the eigendecomposition
of matrices, see \citet{ArchakovHansen:CanonicalBlockMatrix}. For
a block correlation matrix with block-sizes, $(n_{1},\ldots,n_{K})$,
we have
\begin{equation}
C=QDQ^{\prime},\quad\ensuremath{D=\left[\begin{array}{cccc}
A & 0 & \cdots & 0\\
0 & \lambda_{1}I_{n_{1}-1} & \ddots & \vdots\\
\vdots & \ddots & \ddots & 0\\
0 & \cdots & 0 & \lambda_{K}I_{n_{K}-1}
\end{array}\right],\quad\lambda_{k}=\frac{n_{k}-A_{kk}}{n_{k}-1},}\label{eq:CanonicalC}
\end{equation}
where the upper left block, $A$, is an $K\times K$ matrix with elements
$A_{kl}=\rho_{kl}\sqrt{n_{k}n_{l}}$, for $k\neq l,$and $A_{kk}=1+\left(n_{k}-1\right)\rho_{kk}$.
The matrix $Q$ is a group-specific orthonormal matrix, i.e., $Q^{\prime}Q=QQ^{\prime}=I_{n}$.
Importantly, $Q$ is solely determined by the block sizes, $\ensuremath{(n_{1},\ldots,n_{K}})$,
and does not depend on the elements in $C$. This matrix is given
by
\[
\ensuremath{Q=\left[\begin{array}{cccccccc}
v_{n_{1}} & 0 & \cdots &  & v_{n_{1}}^{\perp} & 0 & \cdots & 0\\
0 & v_{n_{2}} &  &  & 0 & v_{n_{2}}^{\perp} &  & \vdots\\
\vdots &  & \ddots &  &  &  & \ddots\\
0 & \cdots &  & v_{n_{K}} & 0 & \cdots &  & v_{n_{K}}^{\perp}
\end{array}\right]},
\]
where $v_{n_{k}}=(1/\sqrt{n_{k}},\ldots,1/\sqrt{n_{k}})^{\prime}\in\mathbb{R}^{n_{k}}$
and $v_{n_{k}}^{\perp}$ is an $n_{k}\times(n_{k}-1)$ matrix, which
is orthogonal to $v_{n_{k}}$, i.e., $v_{n_{k}}^{\prime}v_{n_{k}}^{\perp}=0$,
and orthonormal, such that $v_{n_{k}}^{\perp\prime}v_{n_{k}}^{\perp}=I_{n_{k}-1}.$\footnote{The Gram-Schmidt process can be used to obtain $v_{n\perp}$ from
$v_{n}$.} The canonical representation enables us to rotate $Z$ with $Q$
and define
\begin{equation}
Y=Q^{\prime}Z,\qquad\text{with }\quad Y=(Y_{0}^{\prime},Y_{1}^{\prime},\ldots,Y_{K}^{\prime})^{\prime},\label{eq:ZtoY}
\end{equation}
where $Y_{0}$ is $K$-dimensional with ${\rm var}(Y_{0})=A$, and
$Y_{k}$ is $n_{k}-1$ dimensional with ${\rm var}(Y_{k})=\lambda_{k}I_{n_{k}-1}$
for $k=1,\ldots,K$. The block-diagonal structure of $D$ implies
that $Y_{0},Y_{1},\ldots$, and $Y_{K}$ are uncorrelated, which simplifies
several expressions. For instance, we have the following identities:
\begin{equation}
|C|=|A|\cdot\prod_{k=1}^{K}\lambda_{k}^{n_{k}-1},\quad Z^{\prime}C^{-1}Z=Y_{0}^{\prime}A^{-1}Y_{0}+\sum_{k=1}^{K}\lambda_{k}^{-1}Y_{k}^{\prime}Y_{k},\label{eq:DetQuaTerm}
\end{equation}
such that the computation of the determinant and any power of $C$
is greatly simplified. The square-root of the $n\times n$ correlation
matrix, $C^{1/2}$, is straight forward to compute. From the eigendecomposition
of $A$, $A=P\Lambda_{a}P^{\prime}$, we define the block diagonal
matrix: $D^{1/2}=\mathrm{diag}(P\Lambda_{a}^{1/2}P^{\prime},\lambda_{1}^{1/2}I_{n_{1}-1},\ldots,\lambda_{K}^{1/2}I_{n_{K}-1})$,
and set $C^{1/2}\equiv QD^{1/2}Q^{\prime}$. It is easy to verify
that $C=C^{1/2}C^{1/2}$ and that $C^{1/2}$ is symmetric. Computing
$C^{1/2}$ therefore only requires an eigendecomposition of the symmetric
and positive definite $K\times K$ matrix, $A$, rather than the eigendecomposition
of $C$, which is $n\times n$. Computing other power of $C$ can
be done similarly.

We can use \citet[corollary 2]{ArchakovHansen:CanonicalBlockMatrix}
to recover the elements of the condensed log-correlation matrix,
\[
\tilde{C}=\ensuremath{\ensuremath{\Lambda_{n}^{-1}W\Lambda_{n}^{-1}}},\quad W=\log A-\log\Lambda_{\lambda},
\]
where
\[
\Lambda_{\lambda}=\left[\begin{array}{ccc}
\lambda_{1} & \cdots & 0\\
\vdots & \ddots & \vdots\\
0 & \cdots & \lambda_{K}
\end{array}\right],\qquad\text{and}\qquad\ensuremath{\Lambda_{n}=\left[\begin{array}{ccc}
\sqrt{n_{1}} &  & 0\\
 & \ddots\\
0 &  & \sqrt{n_{K}}
\end{array}\right]}.
\]
The unique values in $\tilde{C}$, which are the elements in $\eta$,
can be expressed as 
\[
\eta={\rm vech}(\tilde{C})=L_{K}\left(\Lambda_{n}^{-1}\otimes\Lambda_{n}^{-1}\right){\rm vec}(W),
\]
where $L_{K}$ is the elimination matrix, that solves $\mathrm{vech}(A)=L_{k}\mathrm{vec}(A)$.
This parametrization of block correlation matrices does not impose
additional superfluous restrictions, and the canonical representation
facilitates simple computation of the determinant, the matrix inverse,
and any other power, as well as the matrix logarithm and the matrix
exponential. This is very useful for the evaluation of the likelihood
function, especially for the more complicated models with heterogeneous
heavy tails and complex dependencies, which we pursue in the next
section.

\section{Distributions\label{sec:Distributions}}

The next step is to specify a distribution for the $n$-dimensional
random vector $Z$, from which the log-likelihood function, $\ell$,
is defined. We consider several specifications, ranging from the multivariate
normal distribution to convolutions of multivariate $t$-distributions.
The convolution-$t$ distributions by \citet{HansenTong:2024} have
simple log-likelihood functions and the canonical representation of
a block correlation matrix motivates some particular specifications
of the convolution-$t$ distribution.

We define
\[
U=C^{-1/2}Z,
\]
such that $\mathrm{var}(U)=I_{n}$,\footnote{An advantage of having defined $C^{1/2}$ from the eigendecomposition,
is that the normalized variables in $U$ are invariant to reordering
of the elements in $Z$, which would not be the case if a Cholesky
form was used to define $C^{1/2}$.} and a convenient property of any log-likelihood function, $\ell$,
is that
\begin{equation}
\ell\left(Z\right)=-\tfrac{1}{2}\log|C|+\ell\left(U\right).\label{eq:LogZU}
\end{equation}
This shows that the log-likelihood function will be in closed-form
if we adopt a distribution for $U$ with a closed-form expression
for $\ell(U)$, and this is important for obtaining tractable score-driven
models. It is well known that the multivariate $t$-distribution and
the Gaussian distribution have simple expression for $\ell(U)$. Fortunately,
so does the multivariate convolution-$t$ distributions, which has
different and interesting statistical properties for $Z$.

\subsection{Multivariate $t$-Distributions}

We begin with the simplest heavy-tailed distribution, a scaled multivariate
$t$-distribution, which nests the Gaussian distribution as a limited
case. The multivariate $t$-distribution is widely used to model vectors
of returns with heavy tailed distributions, see e.g. \citet{CrealKoopmanLucasJBES:2012},
\citet{OpschoorJanusLucasVanDick:2017}, and \citet{HafnerWang:2023}. 

The $n$-dimensional multivariate $t$-distribution with $\nu$ degrees
of freedom, location $\mu\in\mathbb{R}^{n}$, and scale matrix $\Sigma\in\mathbb{R}^{n\times n}$,
typically written $X\sim t_{\nu}(\mu,\Sigma)$, has density
\[
f_{X}(x)=\tfrac{\Gamma(\tfrac{\nu+n}{2})}{\Gamma(\tfrac{\nu}{2})}[\nu\pi]^{-\frac{n}{2}}|\Sigma|^{-\frac{1}{2}}\left[1+\tfrac{1}{\nu}(x-\mu)^{\prime}\Sigma^{-1}(x-\mu)\right]^{-\frac{\nu+n}{2}}.
\]
The variance is well-defined when $\nu>2$, in which case $\mathrm{var}(X)=\tfrac{\nu}{\nu-2}\Sigma$.
The parameter $\nu$ governs the heaviness of the tail and the multivariate
$t$-distribution converges to the multivariate normal distribution,
$N(\mu,\Sigma)$, as $\nu\rightarrow\infty$.

To simplify the notation, we will use a scaled multivariate $t$-distribution,
denoted $t_{\nu}^{\mathrm{std}}(0,\Sigma)$, which is defined for
$\nu>2$. Its density is given by,
\begin{equation}
f_{Y}(y)=\tfrac{\Gamma(\tfrac{\nu+n}{2})}{\Gamma(\tfrac{\nu}{2})}[(\nu-2)\pi]^{-\frac{n}{2}}|\Sigma|^{-\frac{1}{2}}\left[1+\tfrac{1}{\nu-2}y{}^{\prime}\Sigma^{-1}y\right]^{-\frac{\nu+n}{2}},\qquad\nu>2.\label{eq:Std-t-density}
\end{equation}
The relation between the two distributions is as follows: If $X\sim t_{\nu}(0,\Sigma)$
with $\nu>2$, then $Y=\sqrt{\tfrac{\nu-2}{\nu}}X\sim t_{\nu}^{\mathrm{std}}(0,\Sigma)$.
The main advantage of the scaled $t$-distribution is that $\mathrm{var}(Y)=\Sigma$.
Thus, if $U\sim t_{\nu}^{\mathrm{std}}(0,I_{n})$ then $Z=C^{1/2}U\sim t_{\nu}^{\mathrm{std}}(0,C)$,
and the corresponding log-likelihood function is given by
\begin{equation}
\ensuremath{\ell(Z)=c(\nu,n)-\tfrac{1}{2}\log|C|-\tfrac{\nu+n}{2}\log\left(1+\tfrac{1}{\nu-2}Z^{\prime}C^{-1}Z\right),}\label{eq:LogLstuT}
\end{equation}
where $c(\nu,n)=\log\left(\Gamma(\tfrac{\nu+n}{2})/\Gamma(\tfrac{\nu}{2})\right)-\frac{n}{2}\log\left[\left(\nu-2\right)\pi\right]$
is a normalizing constant that does not depend on the correlation
matrix, $C$. If $C$ has a block structure we can use the identities
in (\ref{eq:DetQuaTerm}), and obtain the following simplified expression,
\begin{equation}
\ensuremath{\begin{aligned}\ell(Z)= & c(\nu,n)-\tfrac{1}{2}\log|A|-\tfrac{1}{2}\sum_{k=1}^{K}\left(n_{k}-1\right)\log\lambda_{k}\\
 & -\tfrac{\nu+n}{2}\log\left(1+\tfrac{1}{\nu-2}\left(Y_{0}^{\prime}A^{-1}Y_{0}+\sum_{k=1}^{K}\lambda_{k}^{-1}Y_{k}^{\prime}Y_{k}\right)\right).
\end{aligned}
}\label{eq:LogLstuTBlock}
\end{equation}

The multivariate $t$-distribution has two implications for all elements
of the vector $Z$. First, all elements of a multivariate $t$-distribution
are dependent, because they share a common random mixing variable.
Second, all elements of $U$ are identically distributed, because
they are $t$-distributed with the same degrees of freedom. Both implications
may be too restrictive in many applications, especially if the dimension,
$n$, is large. Below we consider the convolution-$t$ distribution
proposed in \citet{HansenTong:2024}, which allows for heterogeneity
and cluster structures in the tail properties and the tail dependencies.

\subsection{Multivariate Convolution-$t$ Distributions}

The multivariate convolution-$t$ distribution is a suitable rotations
of a random vector that is made up of independent multivariate $t$-distributions.
More specific, let $V_{1},\ldots,V_{G}$ be mutually independent standardized
multivariate $t$-distributed variables, $V_{g}\sim t_{\nu_{g}}^{\mathrm{std}}(0,I_{m_{g}})$,
with $\nu_{g}>2$ for all $g=1,\ldots,G$ and $n=\sum_{g=1}^{G}m_{g}$.

Then $V=(V_{1}^{\prime},\ldots,V_{G}^{\prime})^{\prime}\in\mathbb{R}^{n}$
has the standardized convolution-$t$ distribution (with zero location
vector and identity scale-rotation matrix) that is denoted by
\[
V\sim\mathrm{CT}_{\boldsymbol{m},\boldsymbol{\nu}}^{{\rm std}}(0,I_{n}),
\]
where $\boldsymbol{\nu}=(\nu_{1},\ldots,\nu_{G})^{\prime}$ is the
vector with degrees of freedom and $\boldsymbol{m}=(m_{1},\ldots,m_{G})^{\prime}$
is the vector with the dimensions for the $G$ multivariate $t$-distributions.
We can think of the partitioning of elements in $V$ as a second cluster
structure, as we discuss below. 

We will model the distribution of $U$ using $U=PV$, where $P\in\mathbb{R}^{n\times n}$
is an orthonormal matrix, i.e. $P^{\prime}P=I_{n}$, and we use the
notation $U\sim\mathrm{CT}_{\boldsymbol{m},\boldsymbol{\nu}}^{{\rm std}}(0,P)$.
While $=\mathrm{var}(U)=\mathrm{var}(V)=I_{n}$, they will not have
the same distribution, unless $P$ has a particular structure, such
as $P=I_{n}$. Similarly, we use the following notation for the distribution
of 
\[
Z=C^{1/2}PV\sim\mathrm{CT}_{\boldsymbol{m},\boldsymbol{\nu}}^{{\rm std}}(0,C^{1/2}P),
\]
which is a convolution-$t$ distribution with location zero and scale-rotation
matrix $C^{1/2}P$. Note that we have $\mathrm{var}(Z)=C$, for any
orthonormal matrix, $P$, but different choices for $P$ lead to different
distributions with distinct non-linear dependencies that arise from
the cluster structure in $V$. 

Conveniently, we have the expression, $V=P^{\prime}C^{-1/2}Z=P^{\prime}U$,
and if we partition the columns in $P$, using the same cluster structure
as in $V$, i.e. $P=(P_{1},\ldots,P_{G})$ with $P_{g}\in\mathbb{R}^{n\times m_{g}}$,
then it follows that $V_{g}=P_{g}^{\prime}U\in\mathbb{R}^{m_{g}}$,
for $g=1,\ldots,G$. Next, $U$ and $V$ have the exact same log-likelihoods,
$\ell(U)=\ell(P^{\prime}U)=\ell(V)$, and we can use (\ref{eq:LogZU})
to express the log-likelihood function for $Z$ as
\begin{align}
\ell(Z) & =-\tfrac{1}{2}\log|C|+\sum_{g=1}^{G}c_{g}-\tfrac{\nu_{g}+m_{g}}{2}\log\left(1+\tfrac{1}{\nu_{g}-2}V_{g}^{\prime}V_{g}\right),\label{eq:LogLGenC_StrucT}
\end{align}
where $c_{g}=c(\nu_{g},m_{g})$. When $C$ has a block structure,
then we also have 
\[
V=P^{\prime}Q\left[\begin{array}{c}
A^{-1/2}Y_{0}\\
\lambda_{1}^{-1/2}Y_{1}\\
\vdots\\
\lambda_{K}^{-1/2}Y_{K}
\end{array}\right],
\]
where $Y=Q^{\prime}Z$, and some interesting special cases emerge
from this structure. 

We have previously used a partitioning of the variables to form a
block correlation structure, which arises from a cluster structure
for the variables. The convolution-$t$ distribution involves a second
partitioning that defines the $G$ independent multivariate $t$-distributions.
This is a cluster structure in the underlying random innovations in
the model. The two cluster structures can be identical, or can be
different, as we illustrate with examples and in our empirical application.
Next, we highlight six distributional properties that are the product
of this model design. 
\begin{enumerate}
\item Each element of $V_{g}\in\mathbb{R}^{m_{g}}$, has the same marginal
$t$-distribution with $\nu_{g}$ degrees of freedom. This does not
carry over to the same elements of $Z$ (even if $P=I$). In general,
the marginal distribution of an element of $Z$, will be a unique
convolution of (as many as) $G$ independent $t$-distributions with
different degrees of freedom. 
\item While the (multivariate) $t$-distributions are independent across
groups, this does not carry over to the corresponding sub-vectors
of $Z$. 
\item The convolution for each element of $Z$ is, in part, defined by the
correlation matrix, $C$. So, time-variation in $C$ will induce time-varying
marginal distributions for the elements of $Z$. 
\item The partitioning of $V=P^{\prime}U$ into $G$ clusters ($G$-clusters)
induces heterogeneity in tail dependencies and the heavyness of the
tails. The $G$-clusters can be entirely different from the $K$-clusters
(partitioning of $Z$ variables) that define the block structure in
the correlation matrix, and the two numbers of clusters can be different. 
\item Increasing the number of $G$-clusters, does not necessarily improve
the empirical fit. While increasing $G$ will increase the number
parameters (degrees of freedom) in the model, it also entails dividing
$V$ into additional subvectors, which eliminates the innate dependence
between elements of $V$, which apply to elements from the same multivariate
$t$-distribution. 
\item Sixth, this model framework nests the conventional multivariate $t$-distribution
as the special case, $G=1$, which facilitates simple comparisons
with a natural benchmark model. 
\end{enumerate}

\subsection{Density and CDF of Convolution-$t$ Distribution}

\begin{figure}
\centering{}\includegraphics[width=1\textwidth]{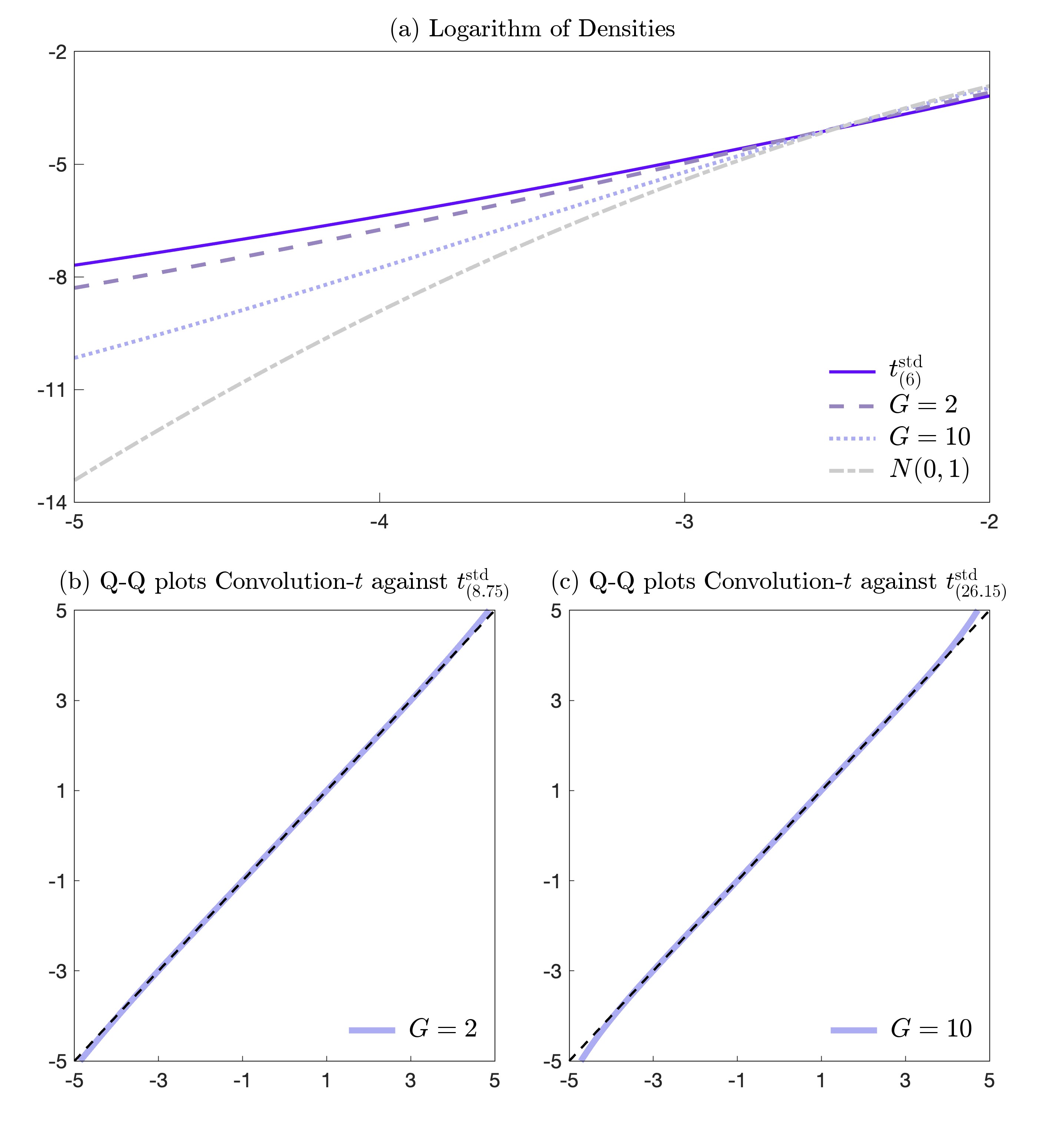}\caption{{\small{}Panel (a) plots the logarithm of marginal distribution for
$\sum_{g=1}^{G}V_{g}/\sqrt{G}$ for $G=1$, $G=2$, and $G=6$, where
$V_{g}$ are independent and identically distributed as $t_{6}^{\mathrm{std}}(0,1)$.
Panels (b) and (c) are Q-Q-plots of the Convolution-$t$ distribution,
$\sum_{g=1}^{G}V_{g}/\sqrt{G}$, with $G=2$ and $G=10$, respectively,
against the best approximating standardized Student's-$t$ distribution,
as defined by the Kullback-Leibler discrepancy.\label{fig:LogDensity-Panel}}}
\end{figure}

The marginal distributions of the elements of $Z$ are convolutions
of independent $t$-distributed variables, and neither their densities
nor their cumulative distribution function have simple expressions.\footnote{Even for the simplest case -- a convolution of two univariate $t$-distributions
-- the resulting density does not have a simple closed-form expression.} However, using \citet[theorem 1]{HansenTong:2024} we obtain the
following semi-analytical expressions, where ${\rm Re}\left[x\right]$
and ${\rm Im}\left[x\right]$ denote the real and imaginary part of
$x\in\mathbb{C}$, respectively, and $e_{j,n}$ is the $j$-th column
of identity matrix $I_{n}$.
\begin{prop}
\label{prop:Convo-t}Suppose $Z\sim\mathrm{CT}_{\boldsymbol{m},\boldsymbol{\nu}}^{{\rm std}}(0,C^{1/2}P)$.
Then the marginal density and cumulative distribution function for
$Z_{j}$, $j=1,\ldots,n$, are given by
\begin{align*}
f_{Z_{j}}(z) & =\frac{1}{\pi}\int_{0}^{\infty}{\rm Re}\left[e^{-isz}\varphi_{Z_{j}}(s)\right]\mathrm{d}s,\quad\quad F_{Z_{j}}(z)=\ensuremath{\frac{1}{2}-\frac{1}{\pi}\int_{0}^{\infty}\frac{{\rm Im}\left[e^{-isz}\varphi_{Z_{j}}(s)\right]}{s}\mathrm{d}s},
\end{align*}
respectively, where $\varphi_{Z_{j}}(s)=\prod_{g=1}^{G}\varphi_{\nu_{g}}^{\mathrm{std}}\left(is\|P_{g}^{\prime}C^{\frac{1}{2}}e_{j,n}\|\right)$
is the characteristic function for $Z_{j}$, and 
\[
\varphi_{\nu}^{\mathrm{std}}(s)=\frac{K_{\frac{\nu}{2}}(\sqrt{\nu-2}|s|)(\sqrt{\nu-2}|s|)^{\frac{1}{2}\nu}}{\Gamma\left(\frac{\nu}{2}\right)2^{\frac{\nu}{2}-1}},
\]
 is the characteristic function of the univariate $t_{\nu}^{\mathrm{std}}$-distribution.
\end{prop}
To gain some insight about convolution-$t$ distributions and the
expressions in Proposition \ref{prop:Convo-t} we present features
of two densities in Figure \ref{fig:LogDensity-Panel}. We specifically
consider convolutions, $\tfrac{1}{\sqrt{G}}\sum_{g=1}^{G}V_{g}$,
for $G=2$ and $G=10$, where $V_{1},\ldots,V_{g}$ are independent
and standardized $t$-distributed with six degrees of freedom.

The upper panel of Figure \ref{fig:LogDensity-Panel}, Panel (a),
shows the log-densities of the (left) tail of the distribution, and
how they compare to those of a standardized $t_{(6)}^{{\rm std}}$-distribution
and a standard normal distribution. As $G\rightarrow\infty$ the convolution-$t$
distribution will approach the normal distribution. So, it is not
surprisingly that the log-densities for the convolutions are between
that of a $t_{(6)}^{{\rm std}}$ and that of a standard normal. Unsurprisingly,
the convolution of $G=10$ standardized $t$-distributions is closer
to the normal distribution than the convolution of $G=2$ distributions.
However, the convolution-$t$ distribution is not a $t$-distribution
for $G>1$. In terms of Kullback-Leibler discrepancy, the best approximating
$t$-distribution to the convolution-$t$ distribution is a $t_{(8.75)}^{{\rm std}}$-distribution
when $G=2$ and a $t_{(26.15)}^{{\rm std}}$-distribution when $G=10$,
see the Q-Q plots in Panels (b) and (c) in Figure \ref{fig:LogDensity-Panel}.

The expression for the marginal density of convolution-$t$ distributions
is particularly useful in our empirical analysis, because it gives
us a factorization of the joint density into marginal densities and
the copula density by Sklar's theorem. This leads to the decomposition
of the log-likelihood, $\ell(Z)=\sum_{j=1}^{n}\ell(Z_{j})+\log(c(Z))$,
where $c(Z)$ denotes the copula density, and we can see if gains
in the log-likelihood are primarily driven by gains in the marginal
distributions or by gains in the copula density.

\subsection{Three Special Types of Convolution-$t$ Distributions}

The convolution-$t$ distributions define a broad class of distributions,
with many possible partitions of $V$ and choices for $P$. Below
we elaborate on som particular details of three special types of convolution-$t$
distributions. For latter use, we use $e_{k}\in\mathbb{R}^{K\times1}$
to denote the $k$-th column of identity matrix $I_{K}$. 

\subsubsection{Special Type 1: Cluster-$t$ Distribution}

The first special type of convolution-$t$ distribution has $P=I$,
such that $U=V$, and a single cluster structure. The cluster structure,
$\boldsymbol{m}$, is imposed on $V$, whereas $C$ can be unrestricted,
or have block structure based on the the same clustering, in which
case $\bm{n}=\bm{m}$ and $G=K$. 

Without a block correlation structure on $C$, we have $V=C^{-1/2}Z$
and the log-likelihood function is simply computed using (\ref{eq:LogLGenC_StrucT}).
If the block structure is imposed on $C$, then we can express the
multivariate $t$-distributed variables as linear combinations on
the canonical variables, $Y_{0},\ldots,Y_{K}$, 
\begin{equation}
U_{k}=V_{k}=v_{n_{k}}e_{k}^{\prime}A^{-\frac{1}{2}}Y_{0}+\lambda_{k}^{-\frac{1}{2}}v_{n_{k}}^{\perp}Y_{k},\qquad\text{for}\quad k=1,\ldots,K,\label{eq:VkExpression}
\end{equation}
We therefore have the expression for the quadratic terms, 
\[
U_{k}^{\prime}U_{k}=Y_{0}^{\prime}A^{-\tfrac{1}{2}}e_{k}e_{k}^{\prime}A^{-\tfrac{1}{2}}Y_{0}+\lambda_{k}^{-1}Y_{k}^{\prime}Y_{k},\quad k=1,\ldots,K,
\]
 and the log-likelihood function simplifies to
\begin{align}
\ell(Z) & =-\tfrac{1}{2}\log|A|+\sum_{k=1}^{K}c_{k}-\tfrac{1}{2}\left(n_{k}-1\right)\log\lambda_{k}-\tfrac{\nu_{k}+n_{k}}{2}\log\left(1+\tfrac{1}{\nu_{k}-2}U_{k}^{\prime}U_{k}\right),\label{eq:LogLGroupT}
\end{align}
where $c_{k}=c(\nu_{k},n_{k})$. The block structure simplifies implementation
of the score-driven model for this specification, and makes it possible
to implement the model with a large number of stocks. 

\subsubsection{Special Type 2: Hetero-$t$ Distribution}

A second special type of convolution-$t$ distributions has $P=I$
and $G=n$. So, the elements of $U$ are made up of $n$ independent
univariate $t$-distributions with degrees of freedom, $\nu_{i}$,
$i=1,\cdots,n$. This distribution can accommodate a high degree of
heterogeneity in the tail properties of $Z_{i}$, $i=1,\ldots,n$,
which are different convolutions of the $n$ independent $t$-distributions.
For this reason, we refer to these distributions as the Hetero-$t$
distributions. The number of degrees of freedom increases from $G$
to $n$, but the additional parameters do not guarantee a better in-sample
log-likelihood, because all dependence between elements of $V$ is
eliminated. The Cluster-$t$ distribution has dependence between $V$-variables
within the same cluster. This has implications the linear combinations
of $U$, including those that define $Z$. 

For the case with a general correlation matrix, the Hetero-$t$ distribution
simplifies the log-likelihood function in (\ref{eq:LogLGenC_StrucT})
to
\[
\ell(Z)=-\tfrac{1}{2}\log|C|+\sum_{i=1}^{n}c_{i}-\tfrac{\nu_{i}+1}{2}\log\left(1+\tfrac{1}{\nu_{i}-2}U_{i}^{2}\right),
\]
where $c_{i}=c(\nu_{i},1)$.\footnote{Note that we can obtain preliminary estimates (starting values) of
the $n$ degrees of freedom parameters, by estimating $\nu_{i}$ from
$e_{i}^{\prime}\tilde{U}_{t}$, where $\tilde{U}_{t}=\tilde{C}^{-\frac{1}{2}}Z_{t}$
and $\tilde{C}$ is an estimate of the unconditional correlation matrix,
for $i=1,\ldots,n$. }

We can combine the heterogenous $t$-distributions with a block correlation
matrix, in which case the log-likelihood function simplifies to
\begin{equation}
\ensuremath{\ell(Z)=c-\tfrac{1}{2}\log|A|-\tfrac{1}{2}\sum_{k=1}^{K}\left(n_{k}-1\right)\log\lambda_{k}-\sum_{k=1}^{K}\sum_{j=1}^{n_{k}}\tfrac{\nu_{k,j}+1}{2}\log\left(1+\tfrac{1}{\nu_{k,j}-2}U_{k,j}^{2}\right)},\label{eq:LogLHeteroT}
\end{equation}
where $c=\sum_{i=1}^{n}c(\nu_{i},1)$ and  $U_{k,j}$ is the $j$-th
element of the vector $U_{k}$ expressed by (\ref{eq:VkExpression}). 

\subsubsection{Special Type 3: Canonical-Block-$t$ Distribution}

A third special type of convolution-$t$ distributions is based on
the canonical canonical variables, as defined by the canonical representation
of the block correlation matrix. The Canonical-Block-$t$ distribution
has $P=Q$ and $\boldsymbol{m}=(K,n_{1}-1,\ldots,n_{K}-1)^{\prime}$,
such that $V=Q^{\prime}U$ is composed of $G=K+1$ independent multivariate
$t$-distributions. So,
\[
Q^{\prime}U=\left(V_{0}^{\prime},V_{1}^{\prime},\cdots,V_{K}^{\prime}\right)^{\prime},\quad{\rm where}\ V_{0}\sim t_{\nu_{0}}(0,I_{K}),\quad{\rm and}\text{ }V_{k}\sim t_{\nu_{k}}(0,I_{n_{k}-1}).
\]
This construction is motivated by the $K+1$ canonical variables,
$Y_{0},\ldots,Y_{K}$, that arises from the canonical representation
of block correlation matrices. Interestingly, this type of convolution-$t$
distribution can be used, regardless of $C$ having a block structure
or not. For a general correlation matrix, $C$, the log-likelihood
function is given by
\begin{align*}
\ell(Z) & =-\tfrac{1}{2}\log|C|+c_{0}-\tfrac{\nu_{0}+K}{2}\log\left(1+\frac{V_{0}^{\prime}V_{0}}{\nu_{0}-2}\right)+\sum_{k=1}^{K}c_{k}-\tfrac{\nu_{k}+n_{k}-1}{2}\log\left(1+\frac{V_{k}^{\prime}V_{k}}{\nu_{k}-2}\right),
\end{align*}
where $V=Q^{\prime}U=Q^{\prime}C^{-1/2}Z$. 

From a practical viewpoint, a more interesting situation is when $C$
has a block structure, such that $C=QDQ^{\prime}$. With this structure,
the log-likelihood function simplifies to
\begin{align}
\ell(Z) & =c_{0}-\tfrac{1}{2}\log|A|-\tfrac{\nu_{0}+K}{2}\log\left(1+\tfrac{1}{\nu_{0}-2}Y_{0}^{\prime}A^{-1}Y_{0}\right)\nonumber \\
 & \quad+\sum_{k=1}^{K}\left[c_{k}-\tfrac{1}{2}\left(n_{k}-1\right)\log\lambda_{k}-\tfrac{\nu_{k}+n_{k}-1}{2}\log\left(1+\tfrac{1}{\nu_{k}-2}Y_{k}^{\prime}Y_{k}\lambda_{k}^{-1}\right)\right],\label{eq:LogLConvT}
\end{align}
which is computationally advantageous, because it does note require
an inverse (nor a determinant) of an $n\times n$ matrix.

The expression for the log-likelihood function shows that this distribution
is equivalent to assuming that $Y_{0},Y_{1},\ldots,Y_{K}$ are independent
and distributed as $\ensuremath{Y_{0}\sim t_{\nu_{0}}^{\mathrm{std}}(0,A)}$
and $\ensuremath{Y_{k}\sim t_{\nu_{k}}^{\mathrm{std}}(0,\lambda_{k}I_{n_{k}-1})}$,
for $k=1,\cdots,K$. This yields insight about the standardized returns
within each block. Let $Z_{k}$ be the $n_{k}$-dimensional subvector
of $Z=(Z_{1}^{\prime},\ldots,Z_{K}^{\prime})^{\prime}$. From $Z=QQ^{\prime}Z=QY$
it follows that 
\[
Z_{k}=v_{n_{k}}Y_{0,k}+v_{n_{k}}^{\perp}Y_{k},
\]
such that a standardized return in the $k$-th block has the same
loading on the common variable $Y_{0,k}$, and orthogonal loadings
on the vector $Y_{k}$.

Additional convolution-$t$ distributions could be bases on this structure.
For instance, we could combine $P=Q$ with heterogeneous univariate
$t$-distributions, for some or all of the canonical variables. For
instance, the canonical variable, $V_{0}$, could be made up of $K$heterogeneous
$t$-distributions, while other canonical variables, $V_{1},\ldots,V_{K}$
have multivariate $t_{\nu_{k}}^{\mathrm{std}}$-distributions.

\section{Score-Driven Models\label{sec:Score-Driven-Models}}

We turn to the dynamic modeling of the conditional correlation matrix
in this section. To this end we adopt the score-drive framework by
\citet{CrealKoopmanLucas:2013}, to model the dynamic properties of
$\gamma_{t}={\rm vecl}(\log C_{t})\in\mathbb{R}^{d}$, with $d=n\left(n-1\right)/2$.
Specifically, we adopt the vector autoregressive model of order one,
VAR(1):
\begin{equation}
\gamma_{t+1}=(I_{d}-\beta)\mu+\beta\gamma_{t}+\alpha\varepsilon_{t},\label{eq:VAR1}
\end{equation}
where $\beta$ and $\alpha$ are $d\times d$ matrices of coefficients,
$\mu=\mathbb{E}(\gamma_{t})$, and $\varepsilon_{t}$ will be defined
by the first-order conditions of the log-likelihood at times $t$.\footnote{It is straightforward to include additional lagged values of $\eta_{t}$,
such that (\ref{eq:VAR1}) has a higher-order VAR(p) structure, and
adding $q$ lagged values of $\varepsilon_{t}$, would generalize
(\ref{eq:VAR1}) to a VARMA(p,q) model, we do not pursue these extensions
in this paper.} The key aspect of a score-driven model is that the score of the predictive
likelihood function is used to define the innovation $\varepsilon_{t}$,
specifically 
\begin{equation}
\varepsilon_{t}=\mathcal{S}_{t}^{-1}\nabla_{t},\quad{\rm where}\quad\nabla_{t}=\frac{\partial\ell_{t-1}(Z_{t})}{\partial\gamma_{t}},\label{eq:DefScore}
\end{equation}
and $\mathcal{S}_{t}$ is a scaling matrix. The score $\nabla_{t}$
is the first-order derivative of log-likelihood with respect to $\gamma_{t}$,
and $\nabla_{t}$ is a martingale difference process if the model
is correctly specified. The Fisher information matrix, $\mathcal{I}_{t}=\mathbb{E}_{t-1}\left(\nabla_{t}\nabla_{t}^{\prime}\right)$,
is often used as the scaling matrix, in which case the time-varying
parameter vector is updated in a manner that resembles a Newton-Raphson
algorithm, see \citet{CrealKoopmanLucas:2013}.\footnote{One exception is \citet{HafnerWang:2023}, who used an unscaled score,
i.e. $\mathcal{S}_{t}=I$, which does not take any curvature of the
log-likelihood into account when parameter values are revised.}

A potential drawback of using $\mathcal{S}_{t}^{-1}$ as the scaling
matrix in (\ref{eq:DefScore}) is that the precision of the inverse
deteriorates as the dimension increases. We will therefore approximate
$\mathcal{S}_{t}^{-1}$ by imposing a diagonal structure, and simply
inverting the diagonal elements of $\mathcal{I}_{t}$. This is equivalent
to using the scaling matrix,
\[
\mathcal{S}_{t}={\rm diag}\left(\mathcal{I}_{t,11},\ldots,\mathcal{I}_{t,dd}\right).
\]
In this manner, each element of the parameter vector is updated with
a scaled version of the corresponding element of the score. Computing
the inverse, $\mathcal{S}_{t}^{-1}$, is now straightforward and simple
to implement.

The score is computed using the following decomposition,
\begin{equation}
\frac{\partial\ell}{\partial\gamma^{\prime}}=\frac{\partial\ell}{\partial{\rm vecl}(C)^{\prime}}\frac{\partial{\rm vecl}(C)}{\partial{\rm vecl}\left(\log C\right)^{\prime}}.\label{eq:GenericDecompos}
\end{equation}
The expression for the last term was derived in \citet{ArchakovHansen:Correlation}
using results from \citet{LintonMcCrorie:1995}. The drawback of this
approach is that it requires an eigendecomposition of $n^{2}\times n^{2}$
matrix and this is impractical and unstable when $n$ is large. Moreover,
the computational burden for the corresponding information matrix
is even worse. Fortunately, when $C$ has a block structure, we have
the following simplified expression,
\[
\frac{\partial\ell}{\partial\eta^{\prime}}=\frac{\partial\ell}{\partial{\rm vec}\left(A\right)^{\prime}}\frac{\partial{\rm vec}\left(A\right)}{\partial{\rm vec}\left(W\right)^{\prime}}\left(\Lambda_{n}\otimes\Lambda_{n}\right)D_{K}.
\]
The first term can be computed very fast for all the variants of the
convolution-$t$ distributions we consider. The second term only requires
an eigendecomposition of $A$ (the upper-left $K\times K$ submatrix
of $D$), and this greatly reduces the computational burden for evaluating
both the score and the information matrix.

For block correlation matrices, we use the vector autoregression of
order one for the subvector,
\begin{equation}
\eta_{t+1}=\left(I_{d}-\beta\right)\mu+\beta\eta_{t}+\alpha\varepsilon_{t},\label{eq:VAR1-1}
\end{equation}
where $\mu=\mathbb{E}(\text{\ensuremath{\eta}}_{t})\in\mathbb{R}^{d}$,
and $\alpha$ and $\beta$ are $d\times d$ matrices with $d=K\left(K+1\right)/2$.

To implement the score-driven model we need to derive the appropriate
score and scaling matrix for each of the log-likelihoods. For this
purpose, we will adopt the following notation involving matrices and
matrix operators, with some notation adopted from \citet{CrealKoopmanLucasJBES:2012}.
Let $A$ and $B$ be two matrices with suitable dimensions. The Kronecker
product is denoted by $A\otimes B$ and we use $A_{\otimes}\equiv A\otimes A$
and $A\oplus B\equiv A\otimes B+B\otimes A$. We let $K_{k}$ denote
the commutation matrix, $D_{k}$ the duplication matrix, and $L_{k}$,
$E_{l}$, $E_{u}$, are $E_{d}$ elimination matrices. These are defined
by the following identities: 
\[
\begin{array}{c}
K_{k}{\rm vec}(B)={\rm vec}\left(B^{\prime}\right),\quad D_{k}{\rm vech}(A)={\rm vec}(A),\quad L_{k}{\rm vec}(B)={\rm vech}(B),\\
\ensuremath{E_{l}{\rm vec}(B)={\rm vecl}(B),\quad E_{u}{\rm vec}(B)}={\rm vecl}\left(B^{\prime}\right),\quad\ensuremath{E_{d}{\rm vec}(B)={\rm diag}(B),}
\end{array}
\]
for any symmetric matrix, $A\in\mathbb{R}^{k\times k}$, and any matrix,
$B\in\mathbb{R}^{k\times k}$.

\subsection{Scores and Information Matrices for a General Correlation Matrix\label{subsec:Scores-and-InformationGeneralC}}

We first derive expressions for $\nabla$ and $\mathcal{I}$ with
a general correlation matrix. Recall that the log-likelihood function,
based on a convolution-$t$ distribution, is given by (\ref{eq:LogLstuT}),
and in the special case with a multivariate $t$-distribution, the
log-likelihood simplifies to the expression in (\ref{eq:LogLGenC_StrucT}).

\subsubsection{Score-Driven Model with Multivariate $t$-Distribution}
\begin{thm}
\label{thm:Derivatives}Suppose that $Z\sim t_{n,\nu}^{\mathrm{std}}(0,C)$.
Then the score vector and information matrix with respect to $\gamma={\rm vecl}\left(\log C\right)$,
are given by:
\begin{align}
\nabla & =\tfrac{1}{2}M^{\prime}C_{\otimes}^{-1}\left[W{\rm vec}\left(ZZ^{\prime}\right)-{\rm vec}\left(C\right)\right],\label{eq:ScoreConvolT}\\
\mathcal{I} & =\tfrac{1}{4}M^{\prime}\left[\phi C_{\otimes}^{-1}H_{n}+(\phi-1){\rm vec}(C^{-1}){\rm vec}(C^{-1})^{\prime}\right]M,\label{eq:InfoConvolT}
\end{align}
respectively, with $H_{n}=I_{n^{2}}+K_{n}$, 
\[
W=\frac{\nu+n}{\nu-2+Z^{\prime}C^{-1}Z},\quad\phi=\frac{\nu+n}{\nu+n+2},
\]
and 
\[
M=\partial{\rm vec}\left(C\right)/\partial\gamma^{\prime}=\left(E_{l}+E_{u}\right)^{\prime}E_{l}\left(I_{n^{2}}-\Gamma E_{d}^{\prime}\left(E_{d}\Gamma E_{d}^{\prime}\right)^{-1}E_{d}\right)\Gamma\left(E_{l}+E_{u}\right)^{\prime},
\]
where the expression for $\Gamma=\ensuremath{\partial{\rm vec}(C)/\partial{\rm vec}\left(\log C\right)^{\prime}}$
is presented in the appendix, see (\ref{eq:ParCparLogC}).
\end{thm}
The expression of $W$ shows that the impact of extreme values (outliers)
is dampened by the degrees of freedom, however this mitigation subsides
as $\nu\rightarrow\infty$. The result for the Gaussian distribution
is obtained by setting $W=\phi=1$, which are their limits as $\nu\rightarrow\infty$. 

\subsubsection{Score-Driven Model with Convolution-$t$ Distributions}
\begin{thm}
\label{thm:Convo}Suppose that $Z\sim\mathrm{CT}_{\boldsymbol{m},\boldsymbol{\nu}}^{{\rm std}}(0,C^{1/2}P)$.
Then the score vector and information matrix with respect to $\gamma={\rm vecl}\left(\log C\right)$,
are given by:
\begin{align*}
\nabla & =M^{\prime}\Omega\left[\sum_{g=1}^{G}W_{g}{\rm vec}\left(P_{g}V_{g}U^{\prime}\right)-{\rm vec}\left(I_{n}\right)\right],\\
\mathcal{I} & =M^{\prime}\Omega\left(K_{n}+\Upsilon_{G}\right)\Omega M,
\end{align*}
respectively, where $M$ is defined in Theorem \ref{thm:Derivatives},
$\Omega=\ensuremath{(I_{n}\otimes C^{-\frac{1}{2}})(C^{\frac{1}{2}}\oplus I_{n})^{-1}}$,
and $\Upsilon_{G}=\sum_{g=1}^{G}\Psi_{g}$ with
\begin{align*}
\Psi_{g} & =\psi_{g}\left(I_{n}\otimes J_{g}\right)+\ensuremath{\left(\phi_{g}-\psi_{g}\right)J_{g\otimes}+\left(\phi_{g}-1\right)\left[J_{g\otimes}K_{n}+{\rm vec}\left(J_{g}\right){\rm vec}\left(J_{g}\right)^{\prime}\right],}
\end{align*}
where $J_{g}=P_{g}P_{g}^{\prime}$,
\[
W_{g}=\frac{\nu_{g}+m_{g}}{\nu_{g}-2+V_{g}^{\prime}V_{g}},\quad\phi_{g}=\frac{\nu_{g}+m_{g}}{\nu_{g}+m_{g}+2},\quad\psi_{g}=\phi_{g}\frac{\nu_{g}}{\nu_{g}-2},
\]
for $g=1,\ldots,G$.
\end{thm}
The inverse of $C^{\frac{1}{2}}\oplus I_{n}$ (an $n^{2}\times n^{2}$
matrix) is available in closed form (see Appendix A) and is computationally
inexpensive because it relies on an eigendecomposition of $C$, which
is already needed for computing $\Gamma$ in the expression of $M$.

Some insight can be gained from considering the case $P=I$. A key
component of $\nabla$ is $\sum_{g=1}^{G}\left(W_{g}P_{g}V_{g}\right)=\left(W_{1}V_{1}^{\prime},W_{2}V_{2}^{\prime},\ldots,W_{G}V_{G}^{\prime}\right)^{\prime}$,
which shows that the impact that $g$-th cluster, $V_{g}$, has one
the score is controlled by the coefficient $W_{g}$.

\subsection{Scores and Information Matrices for a Block Correlation Matrix\label{subsec:Scores-and-InformationBlockC}}

Next, we derive the corresponding expression for the case where $C$
has a block structure. For the score we have the following expression
\[
\nabla^{\prime}=\frac{\partial\ell}{\partial\eta^{\prime}}=\nabla_{A}^{\prime}\Pi_{A},\quad{\rm where}\quad\nabla_{A}=\frac{\partial\ell}{\partial{\rm vec}(A)},
\]
and the expression for $\Pi_{A}$ is given in the following Lemma.
\begin{lem}
\label{lem:dAdEta}Let $\Pi_{A}=\partial{\rm vec}(A)/\partial\eta^{\prime}$,
then
\begin{equation}
\Pi_{A}=\left[\Gamma_{A}-\Gamma_{A}E_{d}^{\prime}\left(\Phi+E_{d}\Gamma_{A}E_{d}^{\prime}\right)^{-1}E_{d}\Gamma_{A}\right]\Lambda_{n\otimes}D_{k},\label{eq:Pi_A}
\end{equation}
where $\Phi$ is a $K\times K$ diagonal matrix with $\Phi_{kk}=\lambda_{k}\left(n_{k}-1\right)$,
$k=1,\ldots,K$, and $\Gamma_{A}=\ensuremath{\partial{\rm vec}(A)/\partial{\rm vec}\left(\log A\right)^{\prime}}$
has the expression given in (\ref{eq:ParlogCparC}). 
\end{lem}
Conveniently, the computation of $\Pi_{A}$ only requires the inverse
of a $K\times K$ matrix. From the results for $\nabla_{A}$ we have
$\nabla=\Pi_{A}^{\prime}\nabla_{A}$ and similarly,
\[
\mathcal{I}=\Pi_{A}^{\prime}\mathcal{I}_{A}\Pi_{A},\quad{\rm where}\quad\mathcal{I}_{A}=\mathbb{E}\left(\nabla_{A}\nabla_{A}^{\prime}\right).
\]

\subsubsection{Score-Driven Model with Block Correlation and Multivariate $t$-Distribution}

With a block correlation structure, we define the standardized canonical
variables
\[
X=\ensuremath{\left(X_{0}^{\prime},X_{1}^{\prime},\ldots,X_{K}^{\prime}\right)^{\prime}=Q^{\prime}U=D^{-\frac{1}{2}}Y},
\]
such that $X_{0}=A^{-\frac{1}{2}}Y_{0}$ with ${\rm var}(X_{0})=I_{K}$
and $X_{k}=\lambda_{k}^{-\frac{1}{2}}Y_{k}$ with ${\rm var}(X_{k})=I_{n_{k}-1}$
for $k=1,\ldots,K$.
\begin{thm}
\label{thm:BlockMultT}Suppose that $Z\sim t_{\nu,n}^{{\rm std}}(0,C)$.
Then the score vector and information matrix with respect to the dynamic
parameters, ${\rm vec}(A)$, are given by:
\begin{align*}
\nabla_{A} & =\tfrac{1}{2}\ensuremath{A_{\otimes}^{-\frac{1}{2}}}\left[W{\rm vec}\left(X_{0}X_{0}^{\prime}\right)-{\rm vec}\left(I_{K}\right)\right]+\tfrac{1}{2}E_{d}^{\prime}S,\\
\mathcal{I}_{A} & =\tfrac{1}{4}\left[\ensuremath{\ensuremath{\phi A_{\otimes}^{-1}H_{K}+(\phi-1){\rm vec}(A^{-1}){\rm vec}(A^{-1})^{\prime}}}\right]+\tfrac{\phi}{2}E_{d}^{\prime}\Xi E_{d}\\
 & \quad+\tfrac{1-\phi}{4}\left[{\rm vec}(A^{-1})\xi^{\prime}E_{d}+E_{d}^{\prime}\xi{\rm vec}(A^{-1})^{\prime}-E_{d}^{\prime}\xi\xi^{\prime}E_{d}\right],
\end{align*}
respectively, where 
\[
\phi=\frac{\nu+n}{\nu+n+2},\quad W=\frac{\nu+n}{\nu-2+X_{0}^{\prime}X_{0}+\sum_{k=1}^{K}X_{k}^{\prime}X_{k}},
\]
and $S\in\mathbb{R}^{K}$, $\xi\in\mathbb{R}^{K}$, and the diagonal
matrix, $\Xi$, are defined by
\[
S_{k}=\frac{1}{\lambda_{k}}-\frac{WX_{k}^{\prime}X_{k}}{\lambda_{k}\left(n_{k}-1\right)},\quad\xi_{k}=\lambda_{k}^{-1},\quad\ensuremath{\Xi_{kk}=\lambda_{k}^{-2}\left(n_{k}-1\right)^{-1},}
\]
for $k=1,\ldots,K$. In the special case where $Z$ has a multivariate
Gaussian distribution ($\nu=\infty$, $\phi=1$), the expression for
the information matrix simplifies to $\mathcal{I}_{A}=\tfrac{1}{4}\ensuremath{A_{\otimes}^{-1}}H_{K}+\tfrac{1}{2}E_{d}^{\prime}\Xi E_{d}$.
\end{thm}

\subsubsection{Score-Driven Model with Block Correlation and Cluster-$t$ Distribution}
\begin{thm}[Cluster-$t$ with Block-$C$]
\label{thm:BlockClusterT}Suppose that $Z\sim\mathrm{CT}_{\boldsymbol{n},\boldsymbol{\nu}}^{{\rm std}}(0,C^{1/2})$
where $C$ has the block structure defined by $\boldsymbol{n}$. Then
the score vector and information matrix with respect to dynamic parameters,
${\rm vec}(A)$, are given by:
\begin{align*}
\nabla_{A} & =\Omega_{A}\left[\sum_{k=1}^{K}W_{k}X_{0,k}{\rm vec}\left(e_{k}X_{0}^{\prime}\right)-{\rm vec}\left(I_{K}\right)\right]+\tfrac{1}{2}E_{d}^{\prime}S,\\
\mathcal{I}_{A} & =\Omega_{A}\left(K_{K}+\Upsilon_{K}\right)\Omega_{A}+\ensuremath{\tfrac{1}{4}E_{d}^{\prime}\Xi E_{d}}+\ensuremath{\tfrac{1}{2}E_{d}^{\prime}\Theta\Omega_{A}}+\ensuremath{\tfrac{1}{2}\Omega_{A}\Theta^{\prime}E_{d}},
\end{align*}
respectively, where $\Omega_{A}=\ensuremath{(I_{K}\otimes A^{-\frac{1}{2}})(A^{\frac{1}{2}}\oplus I_{K})^{-1}}$,
and vector $e_{k}$ is the $k$-th column of the identity matrix $I_{K}$.
The vector $S\in\mathbb{R}^{K}$, the diagonal matrix, $\Xi$, and
$\Theta$ are defined as
\begin{align*}
S_{k} & =\frac{1}{\lambda_{k}}-\frac{W_{k}X_{k}^{\prime}X_{k}}{\lambda_{k}\left(n_{k}-1\right)},\quad\quad\ \ensuremath{W_{k}=\frac{\nu_{k}+n_{k}}{\nu_{k}-2+\ensuremath{X_{0,k}^{2}+X_{k}^{\prime}X_{k}}},}\\
\Xi_{kk} & =\ensuremath{\frac{\phi_{k}-1}{\lambda_{k}^{2}}+\frac{2\phi_{k}}{\lambda_{k}^{2}\left(n_{k}-1\right)}},\quad\Theta=\sum_{k=1}^{K}\lambda_{k}^{-1}\left(1-\phi_{k}\right)e_{k}{\rm vec}\left(J_{k}^{e}\right)^{\prime},
\end{align*}
for $k=1,\ldots,K$. The matrix $\Upsilon_{K}$ is defined analogously
to $\Upsilon_{G}$ in Theorem \ref{thm:Convo}.
\end{thm}

\subsubsection{Score-Driven Model with Block Correlation and Hetero-$t$ Distribution}
\begin{thm}[Heterogeneous-Block Convolution-$t$]
\label{thm:HeteroBlock}Suppose that $Z\sim\mathrm{CT}_{\boldsymbol{n},\boldsymbol{\nu}}^{{\rm std}}(0,C^{1/2})$,
where $C$ has the block structure defined by $\boldsymbol{n}$. Then
the score vector and information matrix with respect to the dynamic
parameters, ${\rm vec}(A)$, are given by:
\begin{align*}
\nabla_{A} & =\Omega_{A}\left[\sum_{k=1}^{K}\sum_{i=1}^{n_{k}}W_{k,i}U_{k,i}{\rm vec}\left(e_{k}X_{0}^{\prime}\right)n_{k}^{-\frac{1}{2}}-{\rm vec}(I_{K})\right]+\tfrac{1}{2}E_{d}^{\prime}S,\\
\mathcal{I}_{A} & =\ensuremath{\Omega_{A}\left(K_{K}+\Upsilon_{K}^{e}\right)\Omega_{A}+\ensuremath{\tfrac{1}{4}E_{d}^{\prime}\Xi E_{d}}+\ensuremath{\tfrac{1}{2}E_{d}^{\prime}\Theta\Omega_{A}}+\ensuremath{\tfrac{1}{2}\Omega_{A}\Theta^{\prime}E_{d}}},
\end{align*}
respectively, where
\[
\ensuremath{S_{k}=\frac{1}{\lambda_{k}}-\frac{\sum_{i=1}^{n_{k}}W_{k,i}U_{k,i}F_{k,i}U_{k}}{\left(n_{k}-1\right)\lambda_{k}}},\qquad k=1,\ldots,K,
\]
with 
\[
W_{k,i}=\frac{\nu_{k,i}+1}{\nu_{k,i}-2+U_{k,i}^{2}},\qquad F_{k,i}=\ensuremath{\tilde{e}_{i}^{\prime}\left(I_{n_{k}}-v_{n_{k}}v_{n_{k}}^{\prime}\right)},
\]
and $\tilde{e}_{i}$ is the $i$-th column of identity matrix $I_{n_{k}}$.
The matrix $\Upsilon_{K}^{e}=\sum_{k=1}^{K}\Psi_{k}^{e}$ is given
by:
\[
\Psi_{k}^{e}=\ensuremath{n_{k}^{-1}\left(3\bar{\phi}_{k}-2-\bar{\psi}_{k}\right)J_{k\otimes}^{e}+\bar{\psi}_{k}\left(I_{K}\otimes J_{k}^{e}\right)},
\]
where $J_{k}^{e}=e_{k}e_{k}^{\prime}$, and 
\[
\ensuremath{\bar{\phi}_{k}=\frac{1}{n_{k}}\sum_{i=1}^{n_{k}}\phi_{k,i}},\quad\bar{\psi}_{k}=\frac{1}{n_{k}}\sum_{i=1}^{n_{k}}\psi_{k,i}.
\]
The diagonal matrix $\Xi$ and $\Theta$ are given by:
\begin{align*}
\Xi_{kk} & =\ensuremath{\lambda_{k}^{-2}n_{k}^{-1}\left[3\bar{\phi}_{k}-1+\left(\bar{\psi}_{k}+1\right)\left(n_{k}-1\right)^{-1}\right]},\\
\Theta & =\sum_{k=1}^{K}\ensuremath{\lambda_{k}^{-1}n_{k}^{-1}\left(\bar{\psi}_{k}+2-3\bar{\phi}_{k}\right)}e_{k}{\rm vec}\left(J_{k}^{e}\right)^{\prime}.
\end{align*}
\end{thm}

\subsubsection{Score-Driven Model with Block Correlation and Canonical-Block-$t$
Distribution}
\begin{thm}[Canonical-Block Convolution-$t$]
Suppose that $Z\sim\mathrm{CT}_{\boldsymbol{m},\boldsymbol{\nu}}^{{\rm std}}(0,C^{1/2}Q)$,
where $C$ has the block structure defined by $\boldsymbol{n}$ and
$\boldsymbol{m}=(K,n_{1}-1,\ldots,n_{K}-1)^{\prime}$. Then the score
vector and information matrix with respect to the dynamic parameters,
${\rm vec}(A)$, are given by:
\begin{align*}
\nabla_{A} & =\tfrac{1}{2}A_{\otimes}^{-\frac{1}{2}}\left[W_{0}{\rm vec}(X_{0}X_{0}^{\prime})-{\rm vec}(I_{K})\right]+\tfrac{1}{2}E_{d}^{\prime}S,\\
\mathcal{I}_{A} & =\tfrac{1}{4}\left[\ensuremath{\ensuremath{\phi_{0}A_{\otimes}^{-1}H_{K}+(\phi_{0}-1){\rm vec}(A^{-1}){\rm vec}(A^{-1})^{\prime}}}+E_{d}^{\prime}\Xi E_{d}\right],
\end{align*}
where the expressions for $S$ and diagonal matrix, $\Xi$, are those
given in Theorem \ref{thm:HeteroBlock} with
\begin{align*}
W_{0} & =\frac{\nu_{0}+K}{\nu_{0}-2+X_{0}^{\prime}X_{0}},\quad W_{k}=\frac{\nu_{k}+n_{k}-1}{\nu_{k}-2+X_{k}^{\prime}X_{k}},\\
\phi_{0} & =\frac{\nu_{0}+K}{\nu_{0}+K+2},\quad\quad\phi_{k}=\frac{\nu_{k}+n_{k}-1}{\nu_{k}+n_{k}+1},
\end{align*}
for $k=1,\ldots,K$.
\end{thm}

\section{Some details about practical implementation}

\subsection{Obtaining the $A$-matrix from the vector $\eta$}

The $K\times K$ matrix, $A=\mathrm{var}(Y_{0})$, plays a central
role in the score models with block-correlation matrices. Below we
show how $A_{t}$ can be computed from $\eta_{t}$. 

In order to obtain $A$ from $\eta$, we adopt the algorithm developed
in \citet[theorem 5]{ArchakovHansenLuo-RandomCorr:2024} to generate
random block correlation matrices. The algorithm has three steps. 
\begin{enumerate}
\item Compute the elements of the $K\times K$ matrix, $\tilde{A}$, using
\[
\tilde{A}_{k,l}=\begin{cases}
\tilde{c}_{kk}\left(n_{k}-1\right) & \text{ for }k=l,\\
\tilde{c}_{kl}\sqrt{n_{k}n_{l}} & \text{ for }k\neq l,
\end{cases}
\]
where $\tilde{c}_{kl}$ are elements of $\eta$, as defined by the
identity, $\eta={\rm vech}(\tilde{C})$.
\item From an arbitrary starting value, $y^{(0)}\in\mathbb{R}^{K}$, e.g.
a vector of zeroes, evaluate the recursion,
\[
y_{k}^{(N+1)}=y_{k}^{(N)}+\log n_{k}-\log\left(\left[\exp\left\{ \tilde{A}+{\rm diag}\left(y^{(N)}\right)\right\} \right]_{kk}+\left(n_{k}-1\right)e^{y_{k}^{(N)}-\tilde{c}_{kk}}\right),
\]
repeatedly, until convergence. Let $y$ denote the final value. (The
convergences tends to be quick because $y$ is a fixed point to a
contraction).
\item Compute $A={\rm exp}\left(\tilde{A}+{\rm diag}(y)\right)$.
\end{enumerate}

\subsection{Correlation/Moment Targeting of Dynamic Parameters\label{subsec:CorrelationTargeting}}

The dimension of $\eta$ in the score-driven model with $K$ groups
is $d=K\left(K+1\right)/2$. For this model we adopt the following
dynamic model 
\[
\eta_{t+1}=\left(I_{d}-\beta\right)\mu+\beta\eta_{t}+\alpha s_{t},
\]
where $\beta$ and $\alpha$ are diagonal matrices. This makes the
total number of parameters to be estimated $K\left(K+1\right)/2\times3$
when we use the Gaussian specification. Specifications with $t$-distributions
will have additional degrees of freedom parameters.

So-called \emph{variance targeting} is often used when estimating
multivariate GARCH models, where the expected value of the conditional
covariance matrix is estimated in an initial step.\footnote{Targeting is often found to be beneficial for prediction but can have
drawbacks, e.g. for inference, see \citet{Pedersen:2016}.} This idea can also be applied to the transformed correlations with
an estimate of $\mu=\mathbb{E}(\eta_{t})$ as the target. In the present
context, it would be more appropriate to call it \emph{correlation
targeting}, or \emph{moment targeting} that encompasses many variations
of this method. For the initial estimation of the target, $\mathbb{E}(\eta_{t})$,
we follow \citet{ArchakovHansen:CanonicalBlockMatrix} and estimate
the unconditional sample block-correlation matrix with
\[
\hat{C}=Q\hat{D}Q^{\prime},\quad\hat{D}=\left[\begin{array}{cccc}
\hat{A} & 0 & \cdots & 0\\
0 & \hat{\lambda}_{1}I_{n_{1}-1} & \ddots & \vdots\\
\vdots & \ddots & \ddots & 0\\
0 & \cdots & 0 & \hat{\lambda}_{K}I_{n_{K}-1}
\end{array}\right],
\]
where
\[
\ensuremath{Y_{t}=Q^{\prime}X_{t}=}\ensuremath{\left(Y_{0,t}^{\prime},Y_{1,t}^{\prime},\ldots,Y_{K,t}^{\prime}\right)^{\prime}}\ensuremath{,\quad\hat{A}=\sum_{t=1}^{T}\ensuremath{Y_{0,t}Y_{0,t}^{\prime}},\quad\hat{\lambda}_{k}=\frac{n_{k}-\hat{A}_{kk}}{n_{k}-1}}.
\]
We then proceed to compute $\hat{\mu}=\gamma(\hat{C})$. Because $\gamma(C)$
is  non-linear, $\hat{\mu}$ is only a first-order approximation of
$\mu$, but our empirical results suggest that it is a good approximation.

\subsection{Benchmark Correlation Model: The DCC Model}

The original DCC model was proposed by \citet{Engle2002}, see also
\citet{EngleSheppard:2001}. The original form of variance targeting
could result in inconsistencies, see \citet{Aielli:2013}, who proposed
a modification that resolves this issue. This model is known as cDCC
model and is given by:
\begin{align*}
C_{t} & =\Lambda_{Q_{t}}^{-1/2}Q_{t}\Lambda_{Q_{t}}^{-1/2},
\end{align*}
where $Q_{t}$ is a symmetric positive definite matrix (whose dynamic
properties are defined below) and $\Lambda_{Q_{t}}$ is the diagonal
matrix with the same diagonal elements as $Q_{t}$. This structure
ensures that $C_{t}$ is a valid correlation matrix. The dynamic properties
of $C_{t}$ are defined from those of $Q_{t}$, which are defined
by 
\begin{align}
\ensuremath{Q_{t+1}} & =\left(\iota\iota^{\prime}-\alpha-\beta\right)\odot\bar{C}+\beta\odot Q_{t}+\alpha\odot\left(\Lambda_{Q_{t}}^{1/2}Z_{t}Z_{t}^{\prime}\Lambda_{Q_{t}}^{1/2}\right),\label{eq:cDCCmodel}
\end{align}
where $\iota$ is the vector of ones, $Z_{t}$ is a $n\times1$ vector
with standardized return shocks, $\odot$ is the Hadamard product
(element by element multiplication), and $\bar{C}$, $\beta$ and
$\alpha$ are unknown $n\times n$ matrices. Here $\bar{C}$ is the
unconditional correlation matrix, which can be parametrized as $\mu={\rm vecl}(\log\bar{C})$.
Note that this model has $n(n+1)/2$ time-varying parameters, as defined
by the unique elements of ${\rm vech}(Q_{t})$. However, $C_{t}$
only has $n(n-1)/2$ distinct correlations, so there are $n$ redundant
variable in $Q_{t}$.

\section{Empirical Analysis}

We estimate and evaluate the models using nine stocks (small universe)
as well as 100 stocks (large universe). We will use industry sectors,
as defined by the Global Industry Classification Standard (GICS),
to form block structures in the correlation matrix and/or the heavy
tail index. The ticker symbols for all 100 stocks are listed in Table
\ref{tab:NameAssets}, organized by industry sectors. The nine stocks
in the small universe are highlighted with bold font.

\begin{table}
\caption{List of 100 stocks}

\begin{centering}
\vspace{0.2cm}
\begin{small}
\begin{tabularx}{\textwidth}{YYYYYYYYYYYY}
\toprule
\midrule
    Energy & Materials & Industrials & \multicolumn{1}{c}{Consumer} & \multicolumn{1}{c}{Consumer } \\
          &        &       & \multicolumn{1}{c}{Discretionary} & \multicolumn{1}{c}{Staples} \\
    \midrule
          &        &       &       &  \\
    APA   & APD    & BA    & \multicolumn{1}{c}{AMZN} & \multicolumn{1}{c}{CL} \\
    BKR   & DD     & CAT   & \multicolumn{1}{c}{EBAY} & \multicolumn{1}{c}{COST} \\
    COP   & FCX    & EMR   & \multicolumn{1}{c}{F} & \multicolumn{1}{c}{CPB} \\
    CVX   & IP     & FDX   & \multicolumn{1}{c}{HD} & \multicolumn{1}{c}{KO} \\
\bf DVN   & SHW    & GD    & \multicolumn{1}{c}{LOW} & \multicolumn{1}{c}{MDLZ} \\
    HAL   &        & GE    & \multicolumn{1}{c}{MCD} & \multicolumn{1}{c}{MO} \\
\bf MRO   &        & HON   & \multicolumn{1}{c}{NKE} & \multicolumn{1}{c}{PEP} \\
    NOV   &        & LMT   & \multicolumn{1}{c}{SBUX} & \multicolumn{1}{c}{PG} \\
\bf OXY   &        & MMM   & \multicolumn{1}{c}{TGT} & \multicolumn{1}{c}{WBA} \\
    SLB   &        & NSC   &       & \multicolumn{1}{c}{WMT} \\
    WMB   &        & UNP   &       &  \\
    XOM   &        & UPS   &       &  \\
          &        &       &       &  \\
    \midrule
    Healthcare & Financials & Information & \multicolumn{1}{c}{Telecom.} & \multicolumn{1}{c}{Utilities} \\
          &        & \multicolumn{1}{c}{Technology} & \multicolumn{1}{c}{Services} &  \\
    \midrule
          &        &          &       &  \\
    ABT   &    ALL &    AAPL  & \multicolumn{1}{c}{CMCSA} & \multicolumn{1}{c}{AEE} \\
    AMGN  &    AXP &    ACN   & \multicolumn{1}{c}{DIS} & \multicolumn{1}{c}{AEP} \\
    BAX   &\bf BAC &    ADBE  & \multicolumn{1}{c}{DISH} & \multicolumn{1}{c}{DUK} \\
    BMY   &    BK  &    CRM   & \multicolumn{1}{c}{GOOGL} & \multicolumn{1}{c}{ETR} \\
    DHR   &\bf C   &\bf CSCO  & \multicolumn{1}{c}{OMC} & \multicolumn{1}{c}{EXC} \\
    GILD  &    COF &    IBM   & \multicolumn{1}{c}{T} & \multicolumn{1}{c}{NEE} \\
    JNJ   &    GS  &\bf INTC  & \multicolumn{1}{c}{VZ} & \multicolumn{1}{c}{SO} \\
    LLY   &\bf JPM &\bf MSFT  &       &  \\
    MDT   &    MET &    NVDA  &       &  \\
    MRK   &    RF  &    ORCL  &       &  \\
    PFE   &    USB &    QCOM  &       &  \\
    TMO   &    WFC &    TXN   &       &  \\
    UNH   &        &    XRX   &       &  \\
\\[0.0cm]
\\[-0.5cm]
\midrule
\bottomrule
\end{tabularx}
\end{small}

\par\end{centering}
{\small{}Note: Ticker symbols for 100 stocks that define the Large
Universe, listed by sector according to their Global Industry Classification
Standard (GICS) codes. The nine stocks in the Small Universe are highlighted
with bold font.\label{tab:NameAssets}}{\small\par}
\end{table}
The sample period spans the period from January 3, 2005 to December
31, 2021, with a total of $T=4,280$ trading days. We obtained daily
close-to-close returns from the CRSP daily stock files in the WRDS
database.

The focus of this paper concerns the dynamic modeling of correlations,
but in practice we also need to estimate the conditional variances.
In our empirical analysis, we estimated each of the univariate time
series of conditional variances using the EGARCH models by \citet{Nelson91},
where the conditional mean has an AR(1) structure, as is common in
this literature. Thus, the model for the $i$-th asset return on day
$t$, $r_{i,t}$, is given by:
\begin{align}
r_{i,t} & =\kappa_{i}+\phi_{i}r_{i,t-1}+\sqrt{h_{i,t}}z_{i,t},\quad z_{i,t}\sim\left(0,1\right),\nonumber \\
\log h_{i,t+1} & =\xi_{i}+\theta_{i}\log h_{i,t}+\tau_{i}z_{i,t}+\delta_{i}|z_{i,t}|.\label{eq:EGARCH}
\end{align}
The parameter, $\tau_{i}$, is related to the well-known leverage
effect, whereas $\theta_{i}$ is tied to the degree of volatility
clustering. By modeling the logarithm of conditional volatility, the
estimated volatility paths are guaranteed to be positive, which in
conjunction with the parametrization of the correlation matrix, $C(\gamma)$,
guarantees a positive definite conditional covariance matrix. At this
stage of the estimation, we do not want to select a particular type
of heavy tail distributions for $z_{i,t}$. So, we simply estimate
the EGARCH models by quasi maximum likelihood estimation using a Gaussian
specification. From the estimated time series for $h_{i,t}$, we obtain
the vector of standardized returns, $Z_{t}=\left[z_{1,t},z_{2,t},\cdots,z_{n,t}\right]$,
which are common to all the multivariate models we consider below.
\begin{table}[t]
\caption{Small Universe: Sample Correlation Matrix, $\hat{C}$, and $\log\hat{C}$
(full sample)}

\begin{centering}
\vspace{0.2cm}
\begin{footnotesize}
\begin{tabularx}{\textwidth}{XXYYYYYYYYY}
\toprule
\midrule
  & &  \multicolumn{3}{c}{Energy} & \multicolumn{3}{c}{Financial} & \multicolumn{3}{c}{Information Tech.} \\
\cmidrule(l){3-5}
\cmidrule(l){6-8}
\cmidrule(l){9-11}
  &  & MRO & OXY  & DVN & BAC & C & JPM & MSFT & INTC & CSCO \\
\midrule
\\[-0.2cm]
\\[-0.2cm]
 \parbox[c]{1cm}{\multirow{5}{*}{\rotatebox[origin=c]{90}{\footnotesize Energy}}} & MRO &  &  0.758 &  0.855 & \cellcolor{black!10}0.152 & \cellcolor{black!10}0.149 & \cellcolor{black!10}0.180 &  0.139 &  0.137 &  0.133 \\
  &  &  &  &  & \cellcolor{black!10} & \cellcolor{black!10} & \cellcolor{black!10} &  &   &  
\\[0.0cm]
  & OXY & 0.790 &  & 0.709 & \cellcolor{black!10}0.152 & \cellcolor{black!10}0.153 & \cellcolor{black!10}0.199 &  0.117 &  0.153 &  0.163 \\
  &  &   & &  & \cellcolor{black!10} & \cellcolor{black!10} & \cellcolor{black!10} &   &    & 
\\[0.0cm]
  & DVN &  0.814 &  0.775 &  & \cellcolor{black!10}0.145 & \cellcolor{black!10}0.153 & \cellcolor{black!10}0.126 &   0.125 &  0.145 &  0.136 \\
\\[0.0cm]
 \parbox[c]{1cm}{\multirow{5}{*}{\rotatebox[origin=c]{90}{\footnotesize Financial}}} & BAC & \cellcolor{black!10}0.439 & \cellcolor{black!10}0.442 & \cellcolor{black!10}0.424 &  &  0.859 &  0.873 & \cellcolor{black!10}0.143 & \cellcolor{black!10}0.152 & \cellcolor{black!10}0.200 \\
  &  & \cellcolor{black!10} & \cellcolor{black!10} & \cellcolor{black!10} &  &  &   & \cellcolor{black!10} & \cellcolor{black!10} & \cellcolor{black!10}
\\[0.0cm]
  & C & \cellcolor{black!10}0.429 & \cellcolor{black!10}0.433 & \cellcolor{black!10}0.418 &  0.819 &  &  0.608 & \cellcolor{black!10}0.151 & \cellcolor{black!10}0.151 & \cellcolor{black!10}0.195 \\
  &  & \cellcolor{black!10} & \cellcolor{black!10} & \cellcolor{black!10} &  &  &  & \cellcolor{black!10} & \cellcolor{black!10} & \cellcolor{black!10}
\\[0.0cm]
  & JPM & \cellcolor{black!10}0.459 & \cellcolor{black!10}0.466 & \cellcolor{black!10}0.435 &  0.829 &  0.762 &  & \cellcolor{black!10}0.222 & \cellcolor{black!10}0.245 & \cellcolor{black!10}0.251 \\
\\[0.0cm]
 \parbox[c]{1cm}{\multirow{5}{*}{\rotatebox[origin=c]{90}{\footnotesize Info Tech.}}} & MSFT & 0.372 & 0.367 & 0.361 & \cellcolor{black!10}0.422 & \cellcolor{black!10}0.412 & \cellcolor{black!10}0.467 &  &  0.494 &  0.455 \\
  &  &  &   &  & \cellcolor{black!10} & \cellcolor{black!10} & \cellcolor{black!10} &  &  & 
\\[0.0cm]
  & INTC &  0.386 &  0.392 &  0.381 & \cellcolor{black!10}0.435 & \cellcolor{black!10}0.422 & \cellcolor{black!10}0.484 & 0.576 &  & 0.426 \\
  &  &   &   &   & \cellcolor{black!10} & \cellcolor{black!10} & \cellcolor{black!10} &  & & 
\\[0.0cm]
  & CSCO &  0.391 &  0.401 &  0.384 & \cellcolor{black!10}0.471 & \cellcolor{black!10}0.457 & \cellcolor{black!10}0.506 & 0.584 & 0.576 &  \\
\\[0.0cm]
\\[-0.5cm]
\midrule
\bottomrule
\end{tabularx}
\end{footnotesize}

\par\end{centering}
{\small{}Note: The sample correlation matrix estimated for the nine
assets (Small Universe) over the full sample period, January 3, 2005,
to December 31, 2020. The elements of $\hat{C}$ are given below the
diagonal and elements of $\log\hat{C}$ are given above the diagonal.
The block structure is illustrated with shaded regions. \label{tab:Summary}}{\small\par}
\end{table}

\subsection{Small Universe: Dynamic Correlations for Nine Stocks}

We begin by analyzing nine stocks and we refer to this data set as
the \emph{small universe}. The nine stocks are: Marathon Oil (MRO),
Occidental Petroleum (OXY), and Devon Energy (DVN) from the energy
sector, Bank of America (BAC), Citigroup (C), and JPMorgan Chase \&
Co (JPM) from the Financial sector, and Microsoft (MSFT), Intel (INTC),
and Cisco (CSCO) from the Information Technology sector. Table \ref{tab:Summary}
reports the full-sample unconditional correlation matrix (lower triangle)
and its logarithm (upper-triangle) with the sector-based block structure
illustrated with the shaded regions. Note that the estimated unconditional
correlations within each of the blocks have similar averages. The
assets within the Energy sector and Financial sector are highly correlated,
with an average correlation of about 0.80. Within-sector correlations
for Information Technology stock returns tend to be smaller, with
an average of about 0.58. The between-sector correlations tend to
be smaller and range from 0.36 to 0.51. A similar pattern is observed
for the corresponding elements of the logarithm of the unconditional
correlation matrix, as the logarithm transformation preserves the
block structure.
\begin{sidewaystable}
\caption{Small Universe Estimation Results: $C_{t}$ Unrestricted}

\begin{centering}
\vspace{0.2cm}
\begin{footnotesize}
\begin{tabularx}{\textwidth}{p{0.5cm}p{0.3cm}p{-0.2cm}Yp{-0.2cm}Yp{-0.2cm}Yp{-0.2cm}Yp{-0.2cm}Yp{-0.2cm}Yp{-0.2cm}Yp{-0.2cm}Yp{-0.2cm}Yp{-0.2cm}Y}
\toprule
\midrule
          &       &       & \multicolumn{9}{c}{Score-Driven Model}                                               &       & \multicolumn{9}{c}{DCC Model} \\
\\[-0.2cm]          
          &       &       & Gaussian     &       & Multiv.-$t$     &       & Canon-$t$ &       & Cluster-$t$ &       & Hetero-$t$ &       & Gaussian     &       & Multiv.-$t$     &       & Canon-$t$ &       & Cluster-$t$ &       & Hetero-$t$ \\
\cmidrule{4-12}\cmidrule{14-22}          
\\[-0.2cm]
    \multirow{6}[0]{*}{$\mu$} & Mean  &       & 0.248 &       & 0.265 &       & 0.268 &       & 0.269 &       & 0.268 &       & 0.239 &       & 0.255 &       & 0.250 &       & 0.266 &       & 0.266 \\
          & Min   &       & 0.022 &       & 0.027 &       & 0.091 &       & 0.076 &       & 0.037 &       & 0.077 &       & 0.097 &       & 0.093 &       & 0.088 &       & 0.095 \\
          & $Q_{25}$   &       & 0.127 &       & 0.124 &       & 0.124 &       & 0.123 &       & 0.129 &       & 0.106 &       & 0.116 &       & 0.116 &       & 0.117 &       & 0.129 \\
          & $Q_{50}$   &       & 0.164 &       & 0.169 &       & 0.166 &       & 0.165 &       & 0.169 &       & 0.136 &       & 0.144 &       & 0.149 &       & 0.150 &       & 0.155 \\
          & $Q_{75}$   &       & 0.227 &       & 0.327 &       & 0.307 &       & 0.325 &       & 0.306 &       & 0.240 &       & 0.331 &       & 0.261 &       & 0.317 &       & 0.305 \\
          & Max   &       & 0.816 &       & 0.777 &       & 0.805 &       & 0.807 &       & 0.815 &       & 0.874 &       & 0.783 &       & 0.829 &       & 0.848 &       & 0.856 \\
\\[-0.2cm]
    \multirow{6}[0]{*}{$\beta$} & Mean  &       & 0.917 &       & 0.962 &       & 0.952 &       & 0.970 &       & 0.970 &       & 0.967 &       & 0.973 &       & 0.974 &       & 0.971 &       & 0.972 \\
          & Min   &       & 0.503 &       & 0.601 &       & 0.502 &       & 0.803 &       & 0.817 &       & 0.937 &       & 0.945 &       & 0.938 &       & 0.938 &       & 0.939 \\
          & $Q_{25}$ &       & 0.881 &       & 0.976 &       & 0.951 &       & 0.966 &       & 0.964 &       & 0.964 &       & 0.972 &       & 0.973 &       & 0.967 &       & 0.968 \\
          & $Q_{50}$ &       & 0.980 &       & 0.990 &       & 0.988 &       & 0.988 &       & 0.985 &       & 0.968 &       & 0.975 &       & 0.977 &       & 0.975 &       & 0.976 \\
          & $Q_{75}$ &       & 0.995 &       & 0.996 &       & 0.994 &       & 0.994 &       & 0.994 &       & 0.972 &       & 0.979 &       & 0.980 &       & 0.978 &       & 0.978 \\
          & Max   &       & 0.999 &       & 0.999 &       & 0.999 &       & 0.999 &       & 0.999 &       & 0.979 &       & 0.984 &       & 0.991 &       & 0.986 &       & 0.981 \\
\\[-0.2cm]
    \multirow{6}[0]{*}{$\alpha$} & Mean  &       & 0.019 &       & 0.014 &       & 0.016 &       & 0.015 &       & 0.016 &       & 0.016 &       & 0.013 &       & 0.011 &       & 0.012 &       & 0.012 \\
          & Min   &       & 0.001 &       & 0.001 &       & 0.001 &       & 0.001 &       & 0.001 &       & 0.011 &       & 0.008 &       & 0.005 &       & 0.007 &       & 0.008 \\
          & $Q_{25}$ &       & 0.007 &       & 0.005 &       & 0.004 &       & 0.005 &       & 0.005 &       & 0.012 &       & 0.010 &       & 0.008 &       & 0.009 &       & 0.011 \\
          & $Q_{50}$ &       & 0.014 &       & 0.011 &       & 0.009 &       & 0.009 &       & 0.011 &       & 0.016 &       & 0.013 &       & 0.012 &       & 0.012 &       & 0.012 \\
          & $Q_{75}$ &       & 0.025 &       & 0.019 &       & 0.023 &       & 0.023 &       & 0.024 &       & 0.019 &       & 0.015 &       & 0.014 &       & 0.014 &       & 0.014 \\
          & Max   &       & 0.071 &       & 0.069 &       & 0.071 &       & 0.057 &       & 0.063 &       & 0.028 &       & 0.021 &       & 0.016 &       & 0.022 &       & 0.019 \\
\\[-0.2cm]
    $\nu_{0}$    &       &       &       &       & 6.232 &       & 6.391 &       &       &       &       &       &       &       & 6.029 &       & 6.501 &       &       &       &  \\
\\[0.0cm]
          &       &       &       &       &       &       &       &       &       &       & 6.193 &       &       &       &       &       &       &       &       &       & 5.907 \\
    $\nu_{1}$    &       &       &       &       &       &       & 5.517 &       & 6.022 &       & 5.320 &       &       &       &       &       & 5.097 &       & 5.839 &       & 5.162 \\
          &       &       &       &       &       &       &       &       &       &       & 4.597 &       &       &       &       &       &       &       &       &       & 4.533 \\
          &       &       &       &       &       &       &       &       &       &       & 4.552 &       &       &       &       &       &       &       &       &       & 4.330 \\
    $\nu_{2}$    &       &       &       &       &       &       & 4.919 &       & 4.871 &       & 4.614 &       &       &       &       &       & 4.397 &       & 4.671 &       & 4.340 \\
          &       &       &       &       &       &       &       &       &       &       & 5.165 &       &       &       &       &       &       &       &       &       & 4.985 \\
          &       &       &       &       &       &       &       &       &       &       & 3.289 &       &       &       &       &       &       &       &       &       & 3.237 \\
    $\nu_{3}$    &       &       &       &       &       &       & 3.714 &       & 4.175 &       & 3.807 &       &       &       &       &       & 3.315 &       & 4.046 &       & 3.740 \\
          &       &       &       &       &       &       &       &       &       &       & 4.023 &       &       &       &       &       &       &       &       &       & 3.910 \\
\\[0.0cm]
    $p$ &       &       & 108   &       & 109   &       & 112   &       & 111   &       & 117   &       & 126   &       & 127   &       & 130   &       & 129   &       & 135 \\
\\[-0.2cm]
    $\ell$    &       &       & -42603 &       & -39971 &       & {-39762} &       & \textbf{-39282} &       & -39348 &       & -42675 &       & -40054 &       & -39827 &       & \textbf{-39370} &       & -39465 \\
    $\ell_m$ &       &       & -54653 &       & -52866 &       & -52893 &       & -52774 &       & \textbf{-52770} &       & -54653 &       & -52857 &       & -52858 &       & -52767 &       & \textbf{-52746} \\
    $\ell_c$ &       &       & 12050 &       & 12895 &       & 13131 &       & \textbf{13492} &       & 13422 &       & 11978 &       & 12803 &       & 13031 &       & \textbf{13397} &       & 13281 \\
\\[-0.2cm]
    AIC   &       &       & 85422 &       & 80160 &       & {79748} &       & \textbf{78786} &       & 78930 &       & 85602 &       & 80362 &       & 79914 &       & \textbf{78998} &       & 79200 \\
    BIC   &       &       & 86109 &       & 80853 &       & {80461} &       & \textbf{79492} &       & 79674 &       & 86404 &       & 81170 &       & 80741 &       & \textbf{79819} &       & 80059 \\
\\[-0.2cm]
\\[-0.5cm]
\midrule
\bottomrule
\end{tabularx}
\end{footnotesize}

\par\end{centering}
{\small{}Note: Parameter estimates for the full sample period, January
2005 to December 2021. The Score-Driven model and DCC model are both
estimated with five distributional specifications, without imposing
a block structure on $C_{t}$. We report summary statistics for the
estimates of $\mu$, $\alpha$, and $\beta$, and report all estimates
of the degrees of freedom. $p$ is the number of parameters and we
report the maximized log-likelihoods, $\ell=\ell_{m}+\ell_{c}$, and
its two components: the log-likelihoods for the nine marginal distributions,
$\ell_{m}$, and the corresponding log-copula density, $\ell_{c}$.
We also report the AIC $=-2\ell+2p$ and BIC $=-2\ell+p\ln T$. Bold
font is used to identify the ``best performing'' specification in
each row among Score-Driven models and among DCC models. \label{tab:JointEstimation9Full}}{\small\par}
\end{sidewaystable}

We estimate three types of dynamic correlation models using five different
distributions. The first type of model is the DCC model, see (\ref{eq:cDCCmodel}).
The second model is the new score-driven model for $C_{t}$, which
we introduced in Section \ref{subsec:Scores-and-InformationGeneralC}.
The third model is the score-driven model for a block correlation
matrix, see Section \ref{subsec:Scores-and-InformationBlockC}. We
consider five distributional specifications for $U$, for each of
these models. The distributions are: Gaussian, multivariate $t$,
Canonical-Block-$t$, Cluster-$t$, and Hetero-$t$ distributions.
We impose a diagonal structure on the matrices, $\alpha$ and $\beta$.
In Tables \ref{tab:JointEstimation9Full} and \ref{tab:JointEstimation100Assets}
we report means and quantiles for the estimated parameters, $\mu$,
${\rm diag}\left(\beta\right)$, ${\rm diag}\left(\alpha\right)$
for score-driven model, and $\mu$, ${\rm vech}\left(\beta\right)$,
${\rm vech}\left(\alpha\right)$ for DCC model, i.e. the DCC models
have more parameters. We denote $p$ as the number of parameters,
$\ell$ is the full log-likelihood function, $\ell_{m}$ and $\ell_{c}$
are the log-likelihood for marginal densities and copula functions.
We also report the Akaike and Bayesian information criteria (AIC and
BIC) to compare the performance of models with different number of
parameters.

Table \ref{tab:JointEstimation9Full} reports the estimation results
for the DCC model and score-driven models for general correlation
matrix (Score-Full model). There are several interesting findings:
First, the score-driven model provides superior performance relative
to the simple DCC model for all five specifications of distributions.
Second, the models with heavy-tailed distributions perform better
than the corresponding model with a Gaussian distribution. For the
score-driven models we see that persistence parameter, $\beta$, is
larger for with heavy tailed specifications, as the existence of $W$
would mitigate the effect from extreme value in updating interested
parameters. Third, introducing the structured heavy tails greatly
improve the model performances, as indicated by higher likelihood
values $\ell$. That this improves the empirical fit is supported
by the estimated degree of freedoms, which are different for different
asset groups. The Information Tech sector is estimated to have the
heaviest tails, follow by the Financial and Energy sectors. Fourth,
the degree of freedoms estimated from Cluster-$t$ distribution is
larger than the averages of each group from Hetero-$t$ distribution,
as we have explained earlier. Fifth, from the decomposition of $\ell$,
we could observe that the improvements of Canonical-Block-$t$ relative
to the multivariate $t$-distribution are all driven by the copula
part. This is also the case for comparing Hetero-$t$ and Cluster-$t$
distributions. Although the former provides more flexibility in fitting
marginal distribution of individual asset, it doesn't necessarily
lead to a better dependence structure. In this dataset, the Cluster-$t$
provides the largest copula functions, as it allows for a common $\chi^{2}$
shock among assets within the same group.
\begin{table}
\caption{Small Universe Estimation Results: $C_{t}$ with Block Structure}

\begin{centering}
\vspace{0.2cm}
\begin{footnotesize}
\begin{tabularx}{\textwidth}{p{1cm}Yp{0.5cm}Yp{0.5cm}Yp{0.5cm}Yp{0.5cm}Yp{0.5cm}YYYYY}
\toprule
\midrule
          & Gaussian &       & Multiv.-$t$ &       & Canon-$t$&     &Cluster-$t$&     & Hetero-$t$\\
    \midrule
\\[-0.2cm]
    $\mu_{11}$    & 0.663 &       & 0.666 &       & 0.687 &       & 0.697 &       & 0.689 \\
    $\mu_{12}$    & 0.138 &       & 0.139 &       & 0.140 &       & 0.140 &       & 0.141 \\
    $\mu_{13}$    & 0.089 &       & 0.110 &       & 0.108 &       & 0.111 &       & 0.112 \\
    $\mu_{22}$    & 0.793 &       & 0.778 &       & 0.802 &       & 0.811 &       & 0.810 \\
    $\mu_{23}$    & 0.169 &       & 0.188 &       & 0.179 &       & 0.178 &       & 0.185 \\
    $\mu_{33}$    & 0.302 &       & 0.435 &       & 0.380 &       & 0.434 &       & 0.405 \\
          &       &       &       &       &       &       &       &       &  \\
    $\beta_{11}$    & 0.918 &       & 0.988 &       & 0.964 &       & 0.982 &       & 0.974 \\
    $\beta_{12}$    & 0.990 &       & 0.987 &       & 0.988 &       & 0.990 &       & 0.990 \\
    $\beta_{13}$    & 0.988 &       & 0.990 &       & 0.988 &       & 0.987 &       & 0.990 \\
    $\beta_{22}$    & 0.868 &       & 0.985 &       & 0.920 &       & 0.954 &       & 0.956 \\
    $\beta_{23}$    & 0.921 &       & 0.912 &       & 0.921 &       & 0.942 &       & 0.936 \\
    $\beta_{33}$    & 0.930 &       & 0.950 &       & 0.956 &       & 0.956 &       & 0.958 \\
          &       &       &       &       &       &       &       &       &  \\
    $\alpha_{11}$    & 0.082 &       & 0.035 &       & 0.058 &       & 0.044 &       & 0.052 \\
    $\alpha_{12}$    & 0.025 &       & 0.034 &       & 0.031 &       & 0.031 &       & 0.031 \\
    $\alpha_{13}$    & 0.025 &       & 0.029 &       & 0.030 &       & 0.033 &       & 0.030 \\
    $\alpha_{22}$    & 0.129 &       & 0.052 &       & 0.123 &       & 0.093 &       & 0.093 \\
    $\alpha_{23}$    & 0.041 &       & 0.064 &       & 0.056 &       & 0.051 &       & 0.054 \\
    $\alpha_{33}$    & 0.067 &       & 0.045 &       & 0.047 &       & 0.050 &       & 0.053 \\
            &       &       &       &       &       &       &       &       &  \\
    $\nu_0$   &       &       & 6.323 &       & 6.476 &       &       &       &  \\
            &       &       &       &       &       &       &       &       &  \\
            &       &       &       &       &       &       &       &       & 6.496 \\
    $\nu_1$   &       &       &       &       & 5.465 &       & 6.098 &       & 5.287 \\
            &       &       &       &       &       &       &       &       & 4.797 \\
            &       &       &       &       &       &       &       &       & 4.568 \\
    $\nu_2$   &       &       &       &       & 4.771 &       & 4.918 &       & 4.691 \\
            &       &       &       &       &       &       &       &       & 4.977 \\
            &       &       &       &       &       &       &       &       & 3.254 \\
    $\nu_3$   &       &       &       &       & 3.637 &       & 4.182 &       & 3.816 \\
            &       &       &       &       &       &       &       &       & 4.094 \\
            &       &       &       &       &       &       &       &       &  \\
    $p$     & 18    &       & 19    &       & 22    &       & 21    &       & 27 \\
\\[-0.2cm]
    $\ell$     & -42696 &       & -40068 &       & -39814 &       & \textbf{-39352} &       & -39428 \\
    $\ell_m$   & -54653 &       & -52870 &       & -52883 &       & -52770 &       & -52769 \\
    $\ell_c$   &  11957 &       &  12803 &       &  13070 &       & 13417 &       & 13341 \\
    \\[-0.2cm]            
    AIC   & 85428 &       & 80174 &       & 79672 &       & \textbf{78746} &       & 78910 \\
    BIC   & 85543 &       & 80295 &       & 79812 &       & \textbf{78880} &       & 79082 \\
\\[0.0cm]
\\[-0.5cm]
\midrule
\bottomrule
\end{tabularx}
\end{footnotesize}

\par\end{centering}
{\small{}Note: Parameter estimates for the full sample period, January
2005 to December 2021. Score-Driven models with a block correlation
structure and five distributional specifications are estimated. The
parameter estimates are reported with subscript that refer to (within/between)
clusters, where Energy=1, Financial=2, and Information Tech=3. $p$
is the number of parameters and we report the maximized log-likelihoods,
$\ell=\ell_{m}+\ell_{c}$, and its two components: the log-likelihoods
for the nine marginal distributions, $\ell_{m}$, and the corresponding
log-copula density, $\ell_{c}$. We also report the AIC $=-2\ell+2p$
and BIC $=-2\ell+p\ln T$. Bold font is used to identify the ``best
performing'' specification in each row among Score-Driven models.\label{tab:JointEstimation9Block}}{\small\par}
\end{table}

\begin{sidewaysfigure}[ph]
\centering{}\includegraphics[scale=0.34]{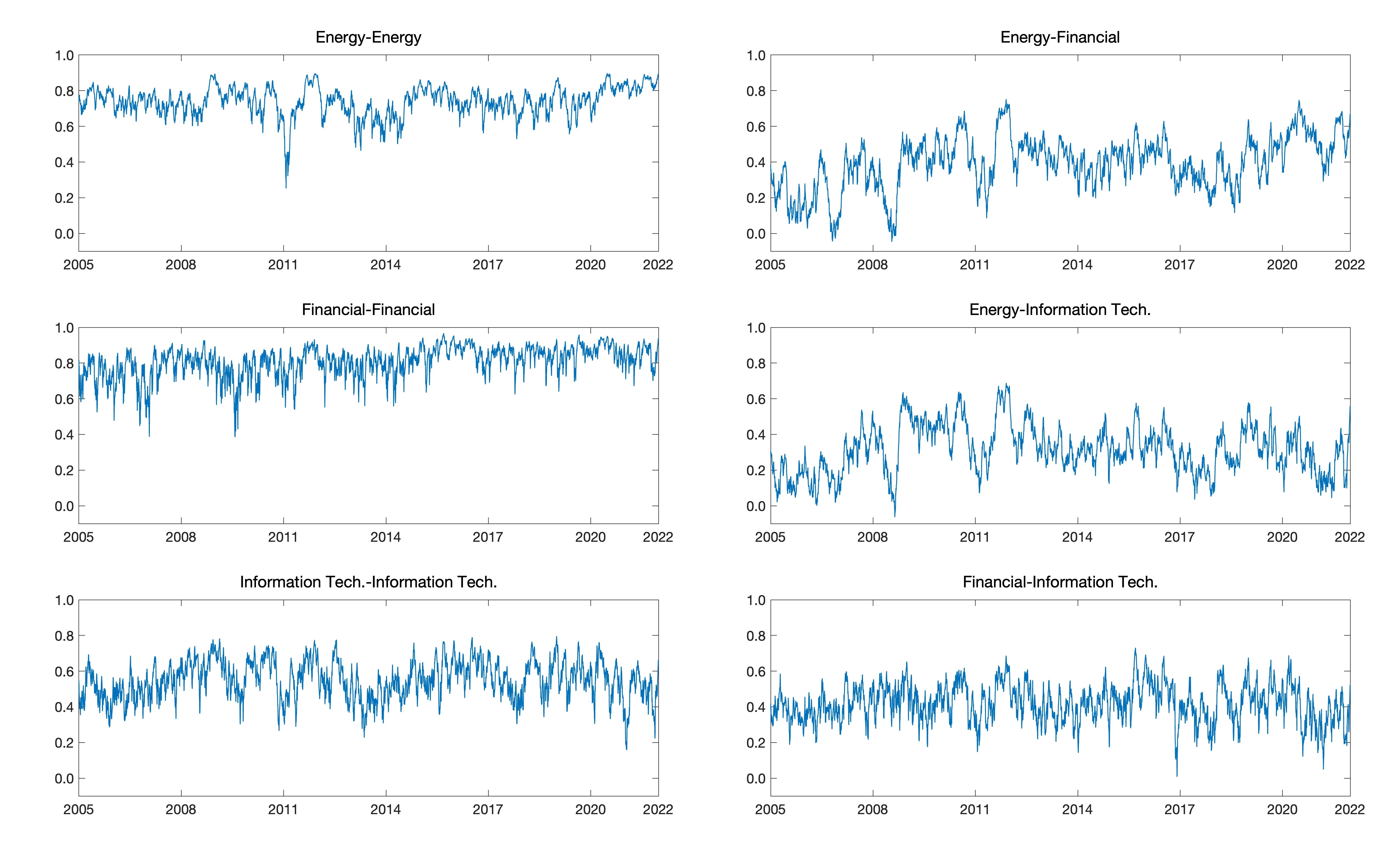}\caption{{\small{}Within-sector and between-sector conditional correlations
implied by the estimated Score-Driven model with a sector-based cluster
structure in correlations and the Convolution-$t$ distribution (Cluster-$t$
with block correlation matrix).\label{fig: withinCorr}}}
\end{sidewaysfigure}

Table \ref{tab:JointEstimation9Block} presents the estimation results
for the score-driven models for block correlation matrix (Score-Block
model). We report all the estimated coefficients with subscripts referring
to the parameters for within/between groups with \textquotedblleft Energy=1,
Financial=2, Information Tech=3\textquotedblright . Results are similar
to the Table \ref{tab:JointEstimation9Full}. When compare with the
\ref{tab:JointEstimation9Full}, we could find although the DCC models
the general correlation matrix, the restricted Score-Block models
provide superior performances with the last three convolution-$t$
specifications. And compared with the Score-Full models, the Score-Block
models delivery smaller BIC for all specifications, and smaller AIC
for the last three cases. We plot the time series of correlations
in Figure \ref{fig: withinCorr} filtered by Cluster-$t$ distributions.
Several heterogenous patterns are observed: First, expect for the
within correlations for financial sector, other correlations have
a sharp decline in late 2010 and increase in early 2011. Second, the
inter-group correlations that involves Energy sector have a evident
decline in late 2008 and the recovered.

\subsection{Large Universe: Dynamic Correlation Matrix for $100$ Assets}

Next, we estimate the model with the large universe, where $C_{t}$
has dimension $100\times100$. We use the sector classification, see
Table \ref{tab:NameAssets}, to define the block structure on $C_{t}$.
Ten (of the eleven) sectors represented in the Large Universe, such
that $K=10$, and the number of unique correlations in $C_{t}$ is
reduced from 4,950 to 55. We estimate the score-driven model with
and without correlation targeting, see Section \ref{subsec:CorrelationTargeting}.
With correlation targeting, the intercept, $\mu$, is estimated first,
and the remaining parameters are estimated in a second stage.
\begin{sidewaystable}
\caption{Large Universe Estimation Results: $C_{t}$ with Block Structure}

\begin{centering}
\vspace{0.2cm}
\begin{footnotesize}
\begin{tabularx}{\textwidth}{p{0.5cm}p{0.3cm}p{-0.2cm}Yp{-0.2cm}Yp{-0.2cm}Yp{-0.2cm}Yp{-0.2cm}Yp{-0.2cm}Yp{-0.2cm}Yp{-0.2cm}Yp{-0.2cm}Yp{-0.2cm}Y}
\toprule
\midrule
          &       &       & \multicolumn{9}{c}{Score-Block Model with Full Parametrization}                                               &       & \multicolumn{9}{c}{Score-Block Model with Correlation Targeting} \\
\\[-0.2cm]          
          &       &       & Gaussian     &       & Multiv.-$t$     &       & Canon-$t$ &       & Cluster-$t$ &       & Hetero-$t$ &       & Gaussian     &       & Multiv.-$t$     &       & Canon-$t$ &       & Cluster-$t$ &       & Hetero-$t$ \\
\cmidrule{4-12}\cmidrule{14-22} 
\\[-0.2cm]
    \multirow{6}[0]{*}{$\mu$} & Mean  &       & 0.053 &       & 0.051 &       & 0.056 &       & 0.056 &       & 0.055 &       & 0.053 &       & 0.053 &       & 0.053 &       & 0.053 &       & 0.053 \\
          & Min   &       & 0.004 &       & 0.001 &       & 0.008 &       & 0.008 &       & 0.007 &       & 0.002 &       & 0.002 &       & 0.002 &       & 0.002 &       & 0.002 \\
          & $Q_{25}$   &       & 0.026 &       & 0.025 &       & 0.026 &       & 0.027 &       & 0.026 &       & 0.025 &       & 0.025 &       & 0.025 &       & 0.025 &       & 0.025 \\
          & $Q_{50}$   &       & 0.038 &       & 0.037 &       & 0.038 &       & 0.039 &       & 0.040 &       & 0.036 &       & 0.036 &       & 0.036 &       & 0.036 &       & 0.036 \\
          & $Q_{75}$   &       & 0.048 &       & 0.049 &       & 0.048 &       & 0.048 &       & 0.050 &       & 0.049 &       & 0.049 &       & 0.049 &       & 0.049 &       & 0.049 \\
          & Max   &       & 0.333 &       & 0.327 &       & 0.343 &       & 0.348 &       & 0.342 &       & 0.349 &       & 0.349 &       & 0.349 &       & 0.349 &       & 0.349 \\
          &       &       &       &       &       &       &       &       &       &       &       &       &       &       &       &       &       &       &       &       &  \\
    \multirow{6}[0]{*}{$\beta$} & Mean  &       & 0.842 &       & 0.886 &       & 0.888 &       & 0.903 &       & 0.887 &       & 0.850 &       & 0.891 &       & 0.902 &       & 0.915 &       & 0.905 \\
          & Min   &       & 0.443 &       & 0.432 &       & 0.612 &       & 0.654 &       & 0.619 &       & 0.401 &       & 0.415 &       & 0.614 &       & 0.651 &       & 0.621 \\
          & $Q_{25}$ &       & 0.794 &       & 0.859 &       & 0.817 &       & 0.853 &       & 0.808 &       & 0.799 &       & 0.854 &       & 0.827 &       & 0.861 &       & 0.846 \\
          & $Q_{50}$ &       & 0.897 &       & 0.983 &       & 0.935 &       & 0.940 &       & 0.944 &       & 0.898 &       & 0.983 &       & 0.941 &       & 0.951 &       & 0.954 \\
          & $Q_{75}$ &       & 0.975 &       & 0.996 &       & 0.989 &       & 0.985 &       & 0.989 &       & 0.976 &       & 0.996 &       & 0.991 &       & 0.989 &       & 0.990 \\
          & Max   &       & 0.999 &       & 0.999 &       & 0.999 &       & 0.999 &       & 0.999 &       & 0.999 &       & 1.000 &       & 0.999 &       & 0.999 &       & 0.999 \\
          &       &       &       &       &       &       &       &       &       &       &       &       &       &       &       &       &       &       &       &       &  \\
    \multirow{6}[0]{*}{$\alpha$} & Mean  &       & 0.030 &       & 0.023 &       & 0.042 &       & 0.041 &       & 0.041 &       & 0.030 &       & 0.023 &       & 0.041 &       & 0.039 &       & 0.041 \\
          & Min   &       & 0.001 &       & 0.003 &       & 0.005 &       & 0.006 &       & 0.004 &       & 0.001 &       & 0.004 &       & 0.008 &       & 0.005 &       & 0.003 \\
          & $Q_{25}$ &       & 0.013 &       & 0.007 &       & 0.015 &       & 0.014 &       & 0.016 &       & 0.013 &       & 0.007 &       & 0.013 &       & 0.015 &       & 0.015 \\
          & $Q_{50}$ &       & 0.028 &       & 0.015 &       & 0.035 &       & 0.038 &       & 0.036 &       & 0.028 &       & 0.015 &       & 0.032 &       & 0.029 &       & 0.036 \\
          & $Q_{75}$ &       & 0.045 &       & 0.029 &       & 0.057 &       & 0.058 &       & 0.055 &       & 0.045 &       & 0.028 &       & 0.058 &       & 0.055 &       & 0.055 \\
          & Max   &       & 0.102 &       & 0.108 &       & 0.137 &       & 0.127 &       & 0.140 &       & 0.103 &       & 0.106 &       & 0.138 &       & 0.128 &       & 0.139 \\
          &       &       &       &       &       &       &       &       &       &       &       &       &       &       &       &       &       &       &       &       &  \\
    $\nu_0$    &       &       &       &       & 10.06 &       & 12.25 &       &       &       &       &       &       &       & 10.11 &       & 12.20 &       &       &       &  \\
    $\nu_1$    &       &       &       &       &       &       & 7.111 &       & 7.320 &       & 4.852$^\dagger$ &       &       &       &       &       & 7.158 &       & 7.341 &       & 4.882$^\dagger$ \\
    $\nu_2$    &       &       &       &       &       &       & 5.778 &       & 6.012 &       & 4.723$^\dagger$ &       &       &       &       &       & 5.489 &       & 5.812 &       & 4.635$^\dagger$ \\
    $\nu_3$    &       &       &       &       &       &       & 6.640 &       & 6.903 &       & 4.451$^\dagger$ &       &       &       &       &       & 6.397 &       & 6.644 &       & 4.368$^\dagger$ \\
    $\nu_4$    &       &       &       &       &       &       & 5.373 &       & 5.579 &       & 3.884$^\dagger$ &       &       &       &       &       & 5.306 &       & 5.546 &       & 3.849$^\dagger$ \\
    $\nu_5$    &       &       &       &       &       &       & 5.657 &       & 5.896 &       & 4.049$^\dagger$ &       &       &       &       &       & 5.627 &       & 5.847 &       & 4.018$^\dagger$ \\
    $\nu_6$    &       &       &       &       &       &       & 6.024 &       & 6.263 &       & 4.070$^\dagger$ &       &       &       &       &       & 5.983 &       & 6.183 &       & 4.021$^\dagger$ \\
    $\nu_7$    &       &       &       &       &       &       & 6.018 &       & 6.159 &       & 4.280$^\dagger$ &       &       &       &       &       & 6.008 &       & 6.127 &       & 4.289$^\dagger$ \\
    $\nu_8$    &       &       &       &       &       &       & 5.581 &       & 5.784 &       & 3.829$^\dagger$ &       &       &       &       &       & 5.516 &       & 5.706 &       & 3.779$^\dagger$ \\
    $\nu_9$    &       &       &       &       &       &       & 7.022 &       & 7.286 &       & 5.347$^\dagger$ &       &       &       &       &       & 7.043 &       & 7.283 &       & 5.337$^\dagger$ \\
    $\nu_{10}$   &       &       &       &       &       &       & 5.693 &       & 6.007 &       & 4.366$^\dagger$ &       &       &       &       &       & 5.658 &       & 5.945 &       & 4.334$^\dagger$ \\
          &       &       &       &       &       &       &       &       &       &       &       &       &       &       &       &       &       &       &       &       &  \\
    $p$     &       &       & 165   &       & 166   &       & 176   &       & 175   &       & 265   &       & 110   &       & 111   &       & 121   &       & 120   &       & 210 \\
\\[-0.2cm]
    $\ell$    &       &       & -481966 &       & -464247 &       & -448572 &       & -446633 &       & -\textbf{436613} &       & -482019 &       & -464307 &       & -448640 &       & -446711 &       & -\textbf{436704} \\
    $\ell_m$   &       &       & -607256 &       & -588971 &       & -589777 &       & -587713 &       & -586615 &       & -607256 &       & -589006 &       & -589625 &       & -587506      &       & -586446 \\
    $\ell_c$   &       &       & 125292 &       & 124724 &       & 141204 &       & 141079 &       & 150002 &       & 125236 &       & 124699 &       & 140986 &       & 140795      &       & 149742  \\
\\[-0.2cm]
    AIC   &       &       & 964262 &       & 928826 &       & 897496 &       & 893616 &       & \textbf{873756} &       & 964258 &       & 928836 &       & 897522 &       & 893662 &       & \textbf{873828} \\
    BIC   &       &       & 965312 &       & 929882 &       & 898616 &       & 894729 &       & \textbf{875442} &       & 964958 &       & 929542 &       & 898292 &       & 894425 &       & \textbf{875164} \\
\\[-0.2cm]
\\[-0.5cm]
\midrule
\bottomrule
\end{tabularx}
\end{footnotesize}

\par\end{centering}
{\small{}Note: Parameter estimates for the full sample period, January
2005 to December 2021. Score-Driven models with a block correlation
structure and five distributional specifications are estimated without
correlation targeting (left panel) and with correlation targeting
(right panel). We report summary statistics for the estimates of $\mu$,
$\alpha$, and $\beta$, and all estimates of the degrees of freedom,
except for the Heterogeneous Convolution-$t$ specifications where
we report the average estimate within each cluster., as identified
with the $\dagger$-superscript. $p$ is the number of parameters
and we report the maximized log-likelihoods, $\ell=\ell_{m}+\ell_{c}$,
and its two components: the log-likelihoods for the nine marginal
distributions, $\ell_{m}$, and the corresponding log-copula density,
$\ell_{c}$. We also report the AIC $=-2\ell+2p$ and BIC $=-2\ell+p\ln T$.
Bold font is used to identify the ``best performing'' specification
in each row for models with and without correlation targeting.\label{tab:JointEstimation100Assets}}{\small\par}
\end{sidewaystable}

Table \ref{tab:JointEstimation100Assets} reports the estimation results
for the the score-driven models with block correlation matrices. The
left panel has estimation results for models without correlation targeting,
and the right panel has the estimation results based on correlation
targeting. The estimates identified with a $\dagger$-superscript,
are the average degrees of freedom within each cluster. These are
used for specifications with heterogeneous Convolution-$t$ specifications
(Hetero-$t$), which estimates 100 degrees of freedom parameters.
Compared with the results for the Small Universe, we note some interesting
difference. First, different from the results on small universe, the
model with hetero-$t$ distribution now provides the best fitting
performance, and compared with Cluster-$t$ distribution, its improvement
concentrates on the copula part. This may due to the high level of
heterogeneity across the large dataset, and the simple classification
based GICS is poor.\footnote{One could estimates the group structure by using the method in \citet{OhPatton2023},
here we only focus such simple classification to assess our score-driven
model in modeling high-dimensional assets.} Second, the models estimated with targeting perform well and have
the smallest BIC across all distributional specifications. 

\subsection{Out-of-sample Results}

We next compare the out-of-sample (OOS) performance of the different
models/specifications. We estimate all models (once) using data from
2005-2014 and evaluate the estimated models with (out-of-sample) data
that spans the years: 2015-2021. 

The OOS results for the Small Universe are shown in Panel A of Table
\ref{tab:OOSResults}. We decompose the predicted log-likelihood,
$\ell$, into the marginal, $\ell_{m}$, and copula, $\ell_{c}$,
components. For each of the five distributional specifications, we
have highlighted the largest predicted log-likelihood, which is the
Score-Driven model without a block structure on $C_{t}$, for all
five distributions. This is consistent with our in-sample results,
where this model also had the largest (in-sample) log-likelihood for
each of the five distributional specifications, see Tables \ref{tab:JointEstimation9Full}
and \ref{tab:JointEstimation9Block}. Overall, the Convolution-$t$
distribution with a sector-based cluster structure, Cluster-$t$,
has the largest predictive log-likelihood. We also note that the DCC
model is has the worst performance across all distributional specifications.
In sample, the DCC model was slightly better than the Score-Driven
model with a block correlation matrix, for two of the five distributions
(Gaussian and multivariate $t$). This suggests that the DCC suffer
from an overfitting problem.
\begin{table}
\caption{Out-of-sample Results}

\begin{centering}
\vspace{0.2cm}
\begin{footnotesize}
\begin{tabularx}{\textwidth}{p{1cm}Yp{0.5cm}Yp{0.5cm}Yp{0.5cm}Yp{0.5cm}Yp{0.5cm}YYYYY}
\toprule
\midrule
\\[-0.3cm]
        \multicolumn{10}{c}{Panel A: Out-of-sample Results for 9 Assets} \\
\\[-0.2cm]      
          & Gaussian &       & Multiv.-$t$ &       & Canon-$t$&     &Cluster-$t$&     & Hetero-$t$\\
    \midrule
           \multicolumn{10}{c}{DCC Model} \\
\\[-0.2cm]
    $p$     & 126   &       & 127   &       & 130   &       & 129   &       & 135 \\
\\[-0.3cm]          
    $\ell$     & -17148 &       & -15887 &       & -15719 &       & -15538 &       & -15648 \\
    $\ell_m$    & -22721&		&-21825	&	     &-21849&		  &-21770 &		   &-21784 \\
    $\ell_c$    & 5573	&		&	5938&		&6130	&	    	&	6232	 &		&	6136 \\
\\[-0.3cm]
\midrule          
           \multicolumn{10}{c}{Score-Full Model} \\
\\[-0.2cm]
    $p$     & 108   &       & 109   &       & 112   &       & 111   &       & 117 \\
\\[-0.3cm] 
    $\ell$    & \textbf{-17139} &       & \textbf{-15812} &       & \textbf{-15672} &       & \textbf{-15459} &       & \textbf{-15572} \\
    $\ell_m$    & -22721&		&-21839	&		&	-21864 	&		&	-21782	&		&	-21804 \\
    $\ell_c$    & 5582	&		&	6027&		&		6192&		&		6323		&		&6232 \\    
\\[-0.3cm]
\midrule
          \multicolumn{10}{c}{Score-Block Model} \\
\\[-0.2cm]
    $p$     & 18    &       & 19    &       & 22    &       & 21    &       & 27 \\
\\[-0.3cm] 
    $\ell$    & -17142 &       & -15832 &       & -15698 &       & -15484 &       & -15591 \\
    $\ell_m$    & -22721&		&		-21842&		&		-21865	&		&	-21783&		&		-21805 \\
    $\ell_c$    & 5579	&		&	6010&		&		6167&		&		6299		&		&6214 \\     
\\[-0.3cm]
\midrule
\\[-0.3cm]
          \multicolumn{10}{c}{Panel B: Out-of-sample Results for 100 Assets} \\
\\[-0.2cm]          
          & Gaussian &       & Student-$t$ &       & Convo-$t$&     &Group-$t$&     & Hetero-$t$\\
    \midrule
          \multicolumn{10}{c}{Score-Block Model} \\
\\[-0.2cm]
    $p$     & 165   &       & 166   &       & 176   &       & 175   &       & 265 \\
\\[-0.3cm] 
    $\ell$    & -202633 &       & -192366 &       & -184318 &       & -183362 &       & -179509 \\
    $\ell_m$    & -251946	&       &	-242925	&       &	-243415	&       &	-242198		&       &-241225 \\
    $\ell_c$    & 49313		&       & 50559	&       &	59098	&       &	58836	&       &	61716 \\         
\\[-0.3cm]
    \midrule
         \multicolumn{10}{c}{Score-Block Model with Correlation Targeting} \\
\\[-0.2cm]
    $p$     & 110   &       & 111   &       & 121   &       & 120   &       & 210 \\
\\[-0.3cm] 
    $\ell$    & \textbf{-201910} &       & \textbf{-192041} &       & \textbf{-184080} &       & \textbf{-183145} &       & \textbf{-179205} \\
    $\ell_m$    & -251946	&       &	-242891	&       &	-243364	&       &	-242005	&       &	-241072 \\
    $\ell_c$    & 50036		&       & 50850		&       & 59284	&       &	58860	&       &	61867 \\      
\\[-0.2cm]
\\[-0.5cm]
\midrule
\bottomrule
\end{tabularx}
\end{footnotesize}

\par\end{centering}
{\small{}Note: Out-of-sample results for the sample period (January
2015 to December 2021). $p$ is the number of parameters, $\ell$
is the log-likelihood function. The Akaike and Bayesian information
criteria are respectively computed as AIC $=-2\ell+2p$, and BIC $=-2\ell+p\ln T$.
We include the Score-driven log Group-correlation model with several
distribution assumptions. The highest log-likelihood and smallest
AIC and BIC in each row are highlighted in bold.\label{tab:OOSResults}}{\small\par}
\end{table}

We report the OOS results for the Large Universe in Panel B of Table
\ref{tab:OOSResults}, where all model-specifications employ a block
structure on $C_{t}$. The empirical results favor correlation targeting,
because the Score-Driven model with correlation targeting has the
largest predicted log-likelihood for each of the five distributions.
Across the five distributions, the Convolution-$t$ distribution based
on $100$, independent $t$-distributions, Hetero-$t$, has the largest
predictive log-likelihood.

\section{Summary\label{sec:Summery}}

We have introduced the Cluster GARCH model, which is a novel multivariate
GARCH model, with two types of cluster structures. One that relates
to the correlation structure and one that define non-linear dependencies.
The Cluster GARCH framework combines several useful components from
the existing literature. For instance, we incorporate the block correlation
structure by \citet{EngleKelly:2012}, the correlation parametrization
by \citet{ArchakovHansen:Correlation}, and the convolution-$t$ distributions
by \citet{HansenTong:2024}. A convolution-$t$ distribution is a
multivariate heavy-tailed distribution with cluster structures, flexible
nonlinear dependencies, and heterogeneous marginal distributions.
We also adopted the score-driven framework by \citet{CrealKoopmanLucas:2013}
to model the dynamic variation in the correlation structure. The convolution-$t$
distributions are well-suited for score-driven models, because their
density functions are sufficiently tractable, allowing us to derive
closed-form expressions for the key ingredients in score-driven models:
the score and the Hessian. We derived detailed results for three special
types of convolution-$t$ distributions. These are labelled Canonical-Block-$t$,
Cluster-$t$, and Hetero-$t$, and their score functions and Fisher
informations are all available in closed-form.

Applying the model to high-dimensional systems is possible when the
block correlation structure is imposed. This was pointed out in \citet{ArchakovHansenLundeMRG},
but the present paper is first to demonstrate this empirically with
$n=100$. This was achieved with $K=10$ sector-based clusters that
was used to define the block structure on the correlation matrix.
The block structure is advantages for several reason. First, it reduces
the number of distinct correlations in $C_{t}$ from 4,950 to 55 ($n(n-1)/2$
to $K(K+1)/2$). Second, many likelihood computations are greatly
simplified due to the canonical representation of block correlation
matrix, see \citet{ArchakovHansen:CanonicalBlockMatrix}. An important
implication for the dynamic model is that computations only involve
inverses, determinants, square-roots of $K\times K$ matrices rather
than $n\times n$ matrices.

We conduct an extensive empirical investigation on the performance
of our dynamic model for correlation matrices. And we consider a ``small
universe'' with $n=9$ assets and a ``large universe'' with $n=100$
assets. The empirical results find strong support for convolution-$t$
distributions that outperforms conventional distributions, in-sample
as well as out-of-sample. Moreover, the score-driven framework out-performs
the standard DCC model in all cases (dimensions and choice of distribution).
The score-driven model with a sector-based block correlation matrix
has the smallest BIC.

\bibliographystyle{apalike}
\bibliography{prh}

\clearpage{}

\appendix

\section{Proofs}

\setcounter{equation}{0}\renewcommand{\theequation}{A.\arabic{equation}}
\setcounter{lem}{0}\renewcommand{\thelem}{A.\arabic{lem}}

\noindent \textbf{Proof of Proposition \ref{prop:Convo-t}.} Let
$X\sim t_{\nu}^{\mathrm{std}}(0,I_{n})$ and consider $X_{\alpha}\equiv\alpha^{\prime}X$,
for some $\alpha\in\mathbb{R}^{n}$. It follows that $X_{\alpha}=aY$
where $a=\left\Vert \alpha\right\Vert =\sqrt{\alpha^{\prime}\alpha}$
and $Y$ is a univariate random variable with distribution, $Y\sim t_{\nu}^{\mathrm{std}}(0,1)$.
The characteristics function for the conventional Student's $t$-distribution
with $\nu$ degrees of freedom, see \citet{Hurst:1995} and \citet{Joarder:1995},
is given by:
\[
\varphi_{\nu}(s)=\frac{K_{\frac{\nu}{2}}(\sqrt{\nu}|s|)(\sqrt{\nu}|s|)^{\frac{1}{2}\nu}}{\Gamma\left(\frac{\nu}{2}\right)2^{\frac{\nu}{2}-1}},
\]
where $K_{\frac{\nu}{2}}(\cdot)$ is the modified Bessel function
of the second kind, such that the characteristic function for $Y$
is given by,
\[
\varphi_{\nu}^{\mathrm{std}}(s)=\varphi_{\nu}(\sqrt{\tfrac{\nu-2}{\nu}}s)=\frac{K_{\frac{\nu}{2}}(\sqrt{\nu-2}|s|)(\sqrt{\nu-2}|s|)^{\frac{1}{2}\nu}}{\Gamma\left(\frac{\nu}{2}\right)2^{\frac{\nu}{2}-1}},
\]
and the characteristic function for $X_{\alpha}$ is simply $\varphi_{X_{\alpha}}(s)=\varphi_{\nu}^{\mathrm{std}}(s\left\Vert \alpha\right\Vert )$.

Next, the $j$-th element of $Z=C^{\frac{1}{2}}U$ can be expresses
as
\[
Z_{j}=e_{j,n}^{\prime}C^{\frac{1}{2}}U=\sum_{g=1}^{G}\left(e_{j,n}^{\prime}C^{\frac{1}{2}}P_{g}\right)V_{g}=\sum_{g=1}^{G}\alpha_{jg}^{\prime}V_{g},
\]
where $\alpha_{jg}=P_{g}^{\prime}C^{\frac{1}{2}}e_{j,n}\in\mathbb{R}^{m_{g}}$
and $e_{j,n}$ is the $j$-th column of identity matrix $I_{n}$.
From the independence of $V_{1},\ldots,V_{G}$ it now follows that
the characteristic function for $Z_{j}$ is given by
\begin{align*}
\varphi_{Z_{j}}(s) & =\prod_{g=1}^{G}\mathbb{E}\left(\exp\left(is\alpha_{jg}^{\prime}V_{g}\right)\right)=\prod_{g=1}^{G}\varphi_{\nu}^{\mathrm{std}}\left(s\|\alpha_{jg}\|\right).
\end{align*}
Finally, from the inverse Fourier transform, we can recover the probability
and cumulative density functions from the characteristic functions
of $Z_{j}$, given by 
\[
\ensuremath{f_{Z_{j}}(z)}=\frac{1}{\pi}\int_{0}^{\infty}{\rm Re}\left[e^{-isz}\varphi_{Z_{j}}(s)\right]\mathrm{d}s,
\]
and
\[
F_{Z_{j}}(z)=\frac{1}{2}-\frac{1}{\pi}\int_{0}^{\infty}\frac{{\rm Im}\left[e^{-isz}\varphi_{Z_{j}}(s)\right]}{s}\mathrm{d}s,
\]
respectively. $\square$

\subsection{Proofs of Results for Score Model (Section 4)}

First some notation. Let $A$ and $B$ be $k\times k$ matrices, then
$A\otimes B$ denotes the Kronecker product. We use $A_{\otimes}$
to denote $A\otimes A$ and $A\oplus B$ for $A\otimes B+B\otimes A$
as in \citet{CrealKoopmanLucasJBES:2012}. The ${\rm vec}(A)$ stacks
the columns of matrix $A$ into a $k^{2}\times1$ column vector, while
${\rm vech}(A)$ stacks the lower triangular part (including diagonal
elements) into a $k^{*}\times1$ column vector, where $k^{*}=k(k+1)/2$.
The $k\times k$ identity matrix is denoted by $I_{k}$. 

From the eigendecomposition, $C=Q\Lambda Q^{\prime}$, we have from
\citet[Theorem 13.16]{LaubAlan:2004} that $C\oplus I=(Q\otimes Q)(\Lambda\oplus I)(Q^{\prime}\otimes Q^{\prime}).$
The inverse is therefore given by 
\[
\left(C\oplus I\right)^{-1}=\left(Q\otimes Q\right)\left(\Lambda^{-1}\oplus I\right)\left(Q^{\prime}\otimes Q^{\prime}\right).
\]

From \citet{LintonMcCrorie:1995}, the expression for $\Gamma=\partial{\rm vec}\left(C\right)/\partial{\rm vec}\left(\log C\right)^{\prime}$
is
\begin{equation}
\Gamma=(Q\otimes Q)\varXi(Q\otimes Q)^{\prime},\label{eq:ParCparLogC}
\end{equation}
where $Q$ is an orthonormal matrix from the eigenvectors of $\log A$
with eigenvalues, $\lambda_{1},\ldots,\lambda_{n}$, and $\varXi$
is a $n^{2}\times n^{2}$ diagonal matrix with elements $\delta_{ij}$,
for $i,j=1,\ldots,n$
\[
\delta_{ij}=\varXi_{(i-1)n+j,(i-1)n+j}=\begin{cases}
e^{\lambda_{i}}, & \text{ if }\lambda_{i}=\lambda_{j},\\
\frac{e^{\lambda_{i}}-e^{\lambda_{j}}}{\lambda_{i}-\lambda_{j}}, & \text{ if }\lambda_{i}\neq\lambda_{j},
\end{cases}
\]
Note that the the expression for $\partial{\rm vec}\left(\log C\right)/\partial{\rm vec}\left(C\right)^{\prime}$
is just the inverse of $\Gamma$, given by
\begin{equation}
\Gamma^{-1}=(Q\otimes Q)\varXi^{-1}(Q\otimes Q)^{\prime},\label{eq:ParlogCparC}
\end{equation}
where $\varXi^{-1}$ is a $n^{2}\times n^{2}$ diagonal matrix with
elements $\delta_{ij}^{-1}$, for $i,j=1,\ldots,n$.

Next, we presents expectations of some quantities involving the $t_{\nu}^{\mathrm{std}}(0,I_{n})$
distribution, involving the following constant,
\[
\zeta_{p,q}=\left(\tfrac{\nu+n}{\nu-2}\right)^{\frac{p}{2}}\left(\tfrac{\nu-2}{2}\right)^{\frac{q}{2}}\frac{\Gamma(\frac{\nu+n}{2})}{\Gamma(\tfrac{\nu}{2})}\frac{\Gamma(\tfrac{\nu+p-q}{2})}{\Gamma(\tfrac{\nu+p+n}{2})}.
\]

\begin{lem}
\label{thm:qHomogeneous}Suppose that $X\sim t_{\nu}^{\mathrm{std}}(0,I_{n})$
and define
\[
W=\frac{\nu+n}{\nu-2+X^{\prime}X}.
\]

(i) For any integrable function $g$ and any $p>2-\nu$, it holds
that
\[
\mathbb{E}\left[W^{\frac{p}{2}}g(X)\right]=\zeta_{p,0}\mathbb{E}\left[g(Y)\right],
\]
where $Y\sim t_{\nu+p}^{\mathrm{std}}\left(0,\ensuremath{\frac{v-2}{v+p-2}}I_{n}\right)$.

(ii) Moreover, if $g$ is homogeneous of degree $q<\nu+p$, then
\[
\mathbb{E}\left[W^{\frac{p}{2}}g(X)\right]=\zeta_{p,q}\mathbb{E}\left[g(Z)\right],
\]
where $Z\sim N\left(0,I_{n}\right)$.
\end{lem}
By integrable function, $g$, the requirement is $\mathbb{E}|g(Y)|<\infty$
and $\mathbb{E}|g(Z)|<\infty$ in parts (i) and (ii), respectively.
Note that $p$ is allowed to be negative, since $2-\nu<0$. Also,
if $p/2$ is a positive integer, then
\begin{eqnarray*}
\zeta_{p,q} & = & \left(\tfrac{\nu+n}{\nu-2}\right)^{\frac{p}{2}}\left(\tfrac{\nu-2}{2}\right)^{\frac{q}{2}}\frac{\tfrac{\nu+q}{2}}{\tfrac{\nu+n}{2}}\frac{\tfrac{\nu+q}{2}+1}{\tfrac{\nu+n}{2}+1}\cdots\frac{\tfrac{\nu+q}{2}+\tfrac{p}{2}-1}{\tfrac{\nu+n}{2}+\tfrac{p}{2}-1}\\
 & = & \left(\tfrac{\nu+n}{\nu-2}\right)^{\frac{p}{2}}\left(\tfrac{\nu-2}{2}\right)^{\frac{q}{2}}\frac{\nu+q}{\nu+n}\frac{\nu+q+2}{\nu+n+2}\cdots\frac{\nu+q+p-2}{\nu+n+p-2}\\
 & = & \left(\tfrac{\nu+n}{\nu-2}\right)^{\frac{p}{2}}\left(\tfrac{\nu-2}{2}\right)^{\frac{q}{2}}\prod_{k=0}^{\tfrac{p}{2}-1}\frac{\nu+q+2k}{\nu+n+2k}.
\end{eqnarray*}
where we used $\Gamma(x+1)=x\Gamma(x)$, repeatedly. This simplifies
the terms we use to derive the Fisher information matrix in several
score models
\[
\begin{array}{rclcrcl}
\zeta_{2,0} & = & \frac{\nu}{\nu-2}, & \qquad\qquad & \zeta_{2,2} & = & 1,\\
\zeta_{4,0} & = & \frac{(\nu+n)}{(\nu+n+2)}\frac{(\nu+2)\nu}{(\nu-2)^{2}}, &  & \zeta_{4,2} & = & \frac{(\nu+n)\nu}{(\nu+n+2)(\nu-2)},\\
 &  &  &  & \zeta_{4,4} & = & \frac{(\nu+n)}{(\nu+n+2)}.
\end{array}
\]

\noindent\textbf{Proof of Lemma }\ref{thm:qHomogeneous}. Let $\kappa_{\nu,n}=\Gamma(\tfrac{\nu+n}{2})/\Gamma(\tfrac{\nu}{2})$
and the density for $X\sim t_{\nu}^{\mathrm{std}}(0,I_{n})$ is 
\[
f_{x}(x)=\kappa_{\nu,n}[(\nu-2)\pi]^{-\frac{n}{2}}\left(1+\tfrac{x^{\prime}x}{\nu-2}\right)^{-\frac{\nu+n}{2}},
\]
whereas the density for $Y\sim t_{\nu+p}^{\mathrm{std}}(0,\ensuremath{\frac{v-2}{v+p-2}}I_{n})$
is
\begin{eqnarray*}
f_{y}(y) & = & \kappa_{\nu+p,n}[(\nu+p-2)\pi]^{-\frac{n}{2}}\left(\tfrac{\nu+p-2}{\nu-2}\right)^{\frac{n}{2}}\left(1+\tfrac{1}{\nu+p-2}x^{\prime}\left[\tfrac{\nu-2}{\nu+p-2}I_{n}\right]^{-1}x\right)^{-\frac{\nu+p+n}{2}}\\
 & = & \kappa_{\nu+p,n}[(\nu-2)\pi]^{-\frac{n}{2}}\left(1+\tfrac{x^{\prime}x}{\nu-2}\right)^{-\frac{\nu+p+n}{2}}.
\end{eqnarray*}
The expected value we seek is
\begin{eqnarray*}
\mathbb{E}\left[W^{\frac{p}{2}}g(X)\right] & = & \int\left(\tfrac{\nu+n}{\nu-2+x^{\prime}x}\right)^{\frac{p}{2}}g(x)\kappa_{\nu,n}[(\nu-2)\pi]^{-\frac{n}{2}}\left(1+\tfrac{x^{\prime}x}{\nu-2}\right)^{-\frac{\nu+n}{2}}\ensuremath{\mathrm{d}x}\\
 & = & \left(\tfrac{\nu+n}{\nu-2}\right)^{\frac{p}{2}}\int g(x)\kappa_{\nu,n}[(\nu-2)\pi]^{-\frac{n}{2}}\left(1+\tfrac{x^{\prime}x}{\nu-2}\right)^{-\frac{\nu+p+n}{2}}\ensuremath{\mathrm{d}x}\\
 & = & \left(\tfrac{\nu+n}{\nu-2}\right)^{\frac{p}{2}}\frac{\kappa_{\nu,n}}{\kappa_{\nu+p,n}}\int g(x)f_{y}(x)\ensuremath{\mathrm{d}x,}
\end{eqnarray*}
and the results for part (i) follows, since
\[
\zeta_{p,0}=\left(\tfrac{\nu+n}{\nu-2}\right)^{\frac{p}{2}}\frac{\Gamma(\tfrac{\nu+n}{2})/\Gamma(\tfrac{\nu}{2})}{\Gamma(\tfrac{\nu+p+n}{2})/\Gamma(\tfrac{\nu+p}{2})}=\left(\tfrac{\nu+n}{\nu-2}\right)^{\frac{p}{2}}\frac{\kappa_{\nu,n}}{\kappa_{\nu+p,n}}.
\]

To prove (ii) we use that $Y\sim t_{\nu+p}^{\mathrm{std}}\left(0,\ensuremath{\frac{v-2}{v+p-2}}I_{n}\right)$
can be expressed as $Y=Z/\sqrt{\xi/(\nu-2)}$ where $Z\sim N(0,I_{n})$
and $\xi$ is an independent $\chi^{2}$-distributed random variable
with $\nu+p$ degrees of freedom. Hence, $Y=\psi Z$, with $\psi=1/\sqrt{\xi/(\nu-2)}$,
such that $\psi^{q}=\left(\tfrac{\nu-2}{\xi}\right)^{\frac{q}{2}}.$
Now using part (i) and that $g$ is homogeneous, we find
\begin{eqnarray*}
\mathbb{E}\left[W^{\frac{p}{2}}g(X)\right] & = & \zeta_{p,0}\mathbb{E}\left[g(Y)\right]=\zeta_{p,0}\mathbb{E}\left[\psi^{q}g(Z)\right]\\
 & = & \zeta_{p,0}\left(\nu-2\right)^{\frac{q}{2}}\mathbb{E}[\xi^{-\frac{q}{2}}]\mathbb{E}\left[g(Z)\right],\\
 & = & \zeta_{p,0}\left(\nu-2\right)^{\frac{q}{2}}\frac{\Gamma(\tfrac{\nu+p-q}{2})}{\Gamma(\tfrac{\nu+p}{2})}\left(\tfrac{1}{2}\right)^{\frac{q}{2}}\mathbb{E}\left[g(Z)\right]\\
 & = & \left(\tfrac{\nu+n}{\nu-2}\right)^{\frac{p}{2}}\frac{\Gamma(\tfrac{\nu+n}{2})}{\Gamma(\tfrac{\nu}{2})}\frac{\Gamma(\tfrac{\nu+p-q}{2})}{\Gamma(\tfrac{\nu+p+n}{2})}\left(\tfrac{\nu-2}{2}\right)^{\frac{q}{2}}\mathbb{E}\left[g(Z)\right],
\end{eqnarray*}
where we used that $\xi$ and $Z$ are independent, and \citet[Results 2]{CrealKoopmanLucasJBES:2012},
which states that if $\xi\sim\chi_{\nu+p}^{2}$, then
\[
\ensuremath{\mathbb{E}\left(\xi^{-\frac{q}{2}}\right)=\frac{\Gamma(\tfrac{\nu+p-q}{2})}{\Gamma(\tfrac{\nu+p}{2})}\left(\tfrac{1}{2}\right)^{\frac{q}{2}}},\qquad\text{for}\quad q<\nu+p.
\]
This completes the proof.\hfill{}$\square$

\noindent \textbf{Proof of Theorem \ref{thm:Derivatives}.} The log-likelihood
function for a vector, $Z$, with the multivariate $t$-distribution,
is given by
\begin{align*}
\ell(Z) & =c_{\nu,n}-\tfrac{1}{2}\log\left|C\right|-\tfrac{\nu+n}{2}\log\left(1+\tfrac{1}{\nu-2}Z^{\prime}C^{-1}Z\right).
\end{align*}
So, we define $W=\left(\nu+n\right)/\left(\nu-2+Z^{\prime}C^{-1}Z\right)$,
we have
\begin{align*}
\frac{\partial\ell}{\partial{\rm vec}(C)^{\prime}} & =-\tfrac{1}{2}{\rm vec}\left(C^{-1}\right)^{\prime}-\tfrac{1}{2}\frac{\nu+n}{\nu-2+Z^{\prime}C^{-1}Z}\frac{\partial\left(Z^{\prime}C^{-1}Z\right)}{\partial{\rm vec}(C^{-1})^{\prime}}\frac{\partial{\rm vec}(C^{-1})}{\partial{\rm vec}(C)^{\prime}}\\
 & =-\tfrac{1}{2}\left[{\rm vec}\left(C^{-1}\right)^{\prime}+W{\rm vec}\left(ZZ^{\prime}\right)^{\prime}C_{\otimes}^{-1}\right]\\
 & =\tfrac{1}{2}\left[W{\rm vec}\left(ZZ^{\prime}\right)^{\prime}-{\rm vec}\left(C\right)^{\prime}\right]C_{\otimes}^{-1},
\end{align*}
such that the score is given by
\[
\ensuremath{\nabla}^{\prime}=\frac{\partial\ell}{\partial\gamma^{\prime}}=\frac{\partial\ell}{\partial{\rm vec}(C)^{\prime}}\frac{\partial{\rm vec}(C)^{\prime}}{\partial\gamma^{\prime}}=\tfrac{1}{2}\left[W{\rm vec}\left(ZZ^{\prime}\right)^{\prime}-{\rm vec}\left(C\right)^{\prime}\right]C_{\otimes}^{-1}M.
\]
From \citet[Proposition 3]{ArchakovHansen:Correlation} we have the
expression 
\begin{equation}
M=\ensuremath{\frac{\partial{\rm vec}\left(C\right)}{\partial\gamma^{\prime}}=\left(E_{l}+E_{u}\right)^{\prime}E_{l}\left(I-\Gamma E_{d}^{\prime}\left(E_{d}\Gamma E_{d}^{\prime}\right)^{-1}E_{d}\right)\Gamma\left(E_{l}+E_{u}\right)^{\prime},}\label{eq:ParCparGamma}
\end{equation}
which uses the fact that $\partial{\rm vec}\left(C\right)/\partial{\rm vecl}\left(C\right)=E_{l}+E_{u}$,
where $E_{l}$, $E_{u}$, $E_{d}$ are elimination matrices, and the
expression $\Gamma=\partial{\rm vec}\left(C\right)/\partial{\rm vec}\left(\log C\right)^{\prime}$
is given in (\ref{eq:ParCparLogC}).

Next we rewrite $\nabla$ as
\begin{align*}
\nabla^{\prime} & =\tfrac{1}{2}\left[W{\rm vec}\left(ZZ^{\prime}\right)^{\prime}-{\rm vec}\left(C\right)^{\prime}\right]C_{\otimes}^{-1}M=\tfrac{1}{2}\left[W{\rm vec}\left(UU^{\prime}\right)^{\prime}-{\rm vec}\left(I\right)^{\prime}\right]C_{\otimes}^{-\tfrac{1}{2}}M,
\end{align*}
where $U=C^{-\frac{1}{2}}Z\sim t_{\nu}^{\mathrm{std}}\left(0,I_{n}\right)$,
such that
\[
\mathcal{I}=\mathbb{E}\left[\nabla\nabla^{\prime}\right]=\frac{1}{4}M^{\prime}C_{\otimes}^{-\tfrac{1}{2}}\left[\mathbb{E}\left(W^{2}{\rm vec}\left(UU^{\prime}\right){\rm vec}\left(UU^{\prime}\right)^{\prime}\right)-{\rm vec}\left(I\right){\rm vec}\left(I\right)^{\prime}\right]C_{\otimes}^{-\tfrac{1}{2}}M.
\]
From Lemma \ref{thm:qHomogeneous} with $\phi=\zeta_{42}=(v+n)/(v+n+2)$,
we have
\[
\mathbb{E}\left[W^{2}{\rm vec}\left(UU^{\prime}\right){\rm vec}\left(UU^{\prime}\right)^{\prime}\right]=\phi\mathbb{E}\left[{\rm vec}\left(\tilde{Z}\tilde{Z}^{\prime}\right){\rm vec}\left(\tilde{Z}\tilde{Z}^{\prime}\right)^{\prime}\right]=\phi\left[H_{n}+{\rm vec}\left(I\right){\rm vec}\left(I\right)^{\prime}\right],
\]
where $\tilde{Z}\sim N\left(0,I_{n}\right)$. The expression for last
expectation follows from \citet[Theorem 4.1]{MagnusNeudecker:1979},
which states that
\[
\mathbb{E}\left[{\rm vec}\left(\tilde{Z}\tilde{Z}^{\prime}\right){\rm vec}\left(\tilde{Z}\tilde{Z}^{\prime}\right)^{\prime}\right]=H_{n}+{\rm vec}\left(I\right){\rm vec}\left(I\right)^{\prime},
\]
if $\tilde{Z}\sim N\left(0,I_{n}\right)$, where $H_{n}=I_{n^{2}}+K_{n}$,
and $K_{n}$ is the commutation matrix. Finally,
\begin{align}
\mathcal{I} & =\tfrac{1}{4}M^{\prime}C_{\otimes}^{-\tfrac{1}{2}}\left[\phi H_{n}+\left(\phi-1\right){\rm vec}\left(I_{n}\right){\rm vec}\left(I_{n}\right)^{\prime}\right]C_{\otimes}^{-\tfrac{1}{2}}M\nonumber \\
 & =\tfrac{1}{4}M^{\prime}\left[\ensuremath{\phi C_{\otimes}^{-1}H_{n}+(\phi-1){\rm vec}\left(C^{-1}\right){\rm vec}\left(C^{-1}\right)^{\prime}}\right]M.\label{eq:FisherInf_t}
\end{align}
This completes the proof.\hfill{}$\square$

\noindent \textbf{Proof of Theorem \ref{thm:Derivatives}.} For this
case we have the log-likelihood function
\begin{align*}
\ell\left(Z\right) & =-\log|C^{\frac{1}{2}}|+\sum_{g=1}^{G}c_{g}-\tfrac{\nu_{g}+m_{g}}{2}\log\left(1+\tfrac{1}{\nu_{g}-2}V_{g}^{\prime}V_{g}\right),
\end{align*}
where $V_{g}=P_{g}^{\prime}U=P_{g}^{\prime}C^{-\frac{1}{2}}Z$, and
$J_{g}=P_{g}P_{g}^{\prime}$. Because we have
\begin{align*}
\frac{\partial\left(V_{g}^{\prime}V_{g}\right)}{\partial{\rm vec}(C^{\frac{1}{2}})^{\prime}} & =\frac{\partial\left(V_{g}^{\prime}V_{g}\right)}{\partial V_{g}^{\prime}}\frac{\partial{\rm vec}(P_{g}^{\prime}C^{-\frac{1}{2}}Z)}{\partial{\rm vec}(C^{-\frac{1}{2}})^{\prime}}\frac{\partial{\rm vec}(C^{-\frac{1}{2}})}{\partial{\rm vec}(C^{\frac{1}{2}})^{\prime}}\\
 & =-2V_{g}^{\prime}\left(Z^{\prime}\otimes P_{g}^{\prime}\right)C_{\otimes}^{-\frac{1}{2}}\\
 & =-2V_{g}^{\prime}\left(U^{\prime}\otimes P_{g}^{\prime}C^{-\frac{1}{2}}\right)\\
 & =-2{\rm vec}\left(C^{-\frac{1}{2}}P_{g}V_{g}U^{\prime}\right)^{\prime}.
\end{align*}
Define $W_{g}=\left(\nu_{g}+m_{g}\right)/\left(\nu_{g}-2+V_{g}^{\prime}V_{g}\right)$,
then we have
\begin{align*}
\frac{\partial\ell}{\partial{\rm vec}(C^{\frac{1}{2}})} & =\sum_{g=1}^{G}W_{g}{\rm vec}\left(C^{-\frac{1}{2}}P_{g}V_{g}U^{\prime}\right)-{\rm vec}\left(C^{-\frac{1}{2}}\right)=\left(I_{n}\otimes C^{-\frac{1}{2}}\right)\nabla_{s},
\end{align*}
where $\nabla_{s}=\sum_{g=1}^{G}W_{g}{\rm vec}\left(P_{g}V_{g}U^{\prime}\right)-{\rm vec}\left(I_{n}\right)$.
So, we have the formula for the score
\begin{align*}
\nabla^{\prime} & =\frac{\partial\ell}{\partial\gamma^{\prime}}=\frac{\partial\ell}{\partial{\rm vec}(C^{\frac{1}{2}})}\frac{\partial{\rm vec}(C^{\frac{1}{2}})}{\partial{\rm vec}(C)^{\prime}}\frac{\partial{\rm vec}(C)}{\partial\gamma^{\prime}}=\nabla_{s}^{\prime}\Omega M
\end{align*}
where the matrix $M$ is defined in (\ref{eq:ParCparGamma}) and $\Omega=(I_{n}\otimes C^{-\frac{1}{2}})(C^{\frac{1}{2}}\oplus I_{n})^{-1}$,
which is based on
\[
\frac{\partial{\rm vec}(C^{\frac{1}{2}})}{\partial{\rm vec}(C)^{\prime}}=\left(\frac{\partial{\rm vec}(C)}{\partial{\rm vec}(C^{\frac{1}{2}})^{\prime}}\right)^{-1}=\left(C^{\frac{1}{2}}\oplus I_{n}\right)^{-1}.
\]
This proves (\ref{eq:ScoreConvolT}). Next, the inverse of the $n^{2}\times n^{2}$
matrix $C^{\frac{1}{2}}\oplus I$ is available in closed form, see
Appendix A, based on the eigendecomposition $C^{\frac{1}{2}}=Q\Lambda^{\frac{1}{2}}Q^{\prime}$.
This does not add to  the computation burden additionally, because
the eigendecomposition of $C^{\frac{1}{2}}$ is available from that
of $\log C=Q\log\Lambda Q^{\prime}$, which was needed for computing
$\Theta$ from $M$.

\subsubsection*{The Information Matrix}

Next we turn to the information matrix. Note that$\mathcal{I}=M^{\prime}\Omega\mathbb{E}\left(\nabla_{s}\nabla_{s}^{\prime}\right)\Omega M$,
with $\mathbb{E}\left(\nabla_{s}\nabla_{s}^{\prime}\right)$ given
by
\begin{align*}
\mathbb{E}\left(\nabla_{s}\nabla_{s}^{\prime}\right) & =\mathbb{E}\left[\sum_{k=1}^{G}\sum_{l=1}^{G}W_{k}W_{l}{\rm vec}\left(P_{k}V_{k}U^{\prime}\right){\rm vec}\left(P_{l}V_{l}U^{\prime}\right)^{\prime}-{\rm vec}\left(I_{n}\right){\rm vec}\left(I_{n}\right)^{\prime}\right]
\end{align*}
For later use, we define $\psi_{k}=\zeta_{42}$, and $\phi_{k}=\zeta_{44}$,
for $k=1,\ldots,G$, where the constants are given from Lemma \ref{thm:qHomogeneous},
given by
\[
\phi_{k}=\frac{\nu_{k}+m_{k}}{\nu_{k}+m_{k}+2},\quad\text{and}\quad\psi_{k}=\phi_{k}\frac{\nu_{k}}{\nu_{k}-2},
\]
and define the function $\varphi\left(k,l\right)$ as
\[
\varphi\left(k,l\right)=W_{k}W_{l}{\rm vec}\left(P_{k}V_{k}U^{\prime}\right){\rm vec}\left(P_{l}V_{l}U^{\prime}\right)^{\prime}.
\]
Note that we will use the following preliminary results in later analysis
\[
U=\sum_{g=1}^{G}P_{g}V_{g},\quad J_{g}=P_{g}P_{g}^{\prime},\quad\sum_{g=1}^{G}J_{g}=I_{n}.
\]

\subsubsection*{Expectation of $\varphi\left(k,l\right)$ when $k=l$}

We have the expectation for $\varphi\left(k,k\right)$ given by
\begin{align*}
\mathbb{E}\left[\varphi\left(k,k\right)\right]= & \mathbb{E}\left[W_{k}^{2}\sum_{p=1}^{G}{\rm vec}\left(P_{k}V_{k}V_{p}^{\prime}P_{p}^{\prime}\right)\sum_{q=1}^{G}{\rm vec}\left(P_{k}V_{k}V_{q}^{\prime}P_{q}^{\prime}\right)^{\prime}\right]\\
= & \mathbb{E}\left[W_{k}^{2}\sum_{p\neq k}^{G}{\rm vec}\left(P_{k}V_{k}V_{p}^{\prime}P_{p}^{\prime}\right){\rm vec}\left(P_{k}V_{k}V_{p}^{\prime}P_{p}^{\prime}\right)^{\prime}+W_{k}^{2}{\rm vec}\left(P_{k}V_{k}V_{k}^{\prime}P_{k}^{\prime}\right){\rm vec}\left(P_{k}V_{k}V_{k}^{\prime}P_{k}^{\prime}\right)^{\prime}\right].
\end{align*}
Based on Lemma \ref{thm:qHomogeneous} with $\psi_{k}=\zeta_{42}$,
we have
\begin{align*}
 & \sum_{p\neq k}^{G}\mathbb{E}\left[W_{k}^{2}{\rm vec}\left(P_{k}V_{k}V_{p}^{\prime}P_{p}^{\prime}\right){\rm vec}\left(P_{k}V_{k}V_{p}^{\prime}P_{p}^{\prime}\right)^{\prime}\right]\\
= & \sum_{p\neq k}^{G}\left(P_{p}\otimes P_{k}\right)\mathbb{E}\left[W_{k}^{2}{\rm vec}\left(V_{k}V_{p}^{\prime}\right){\rm vec}\left(V_{k}V_{p}^{\prime}\right)^{\prime}\right]\left(P_{p}^{\prime}\otimes P_{k}^{\prime}\right)\\
= & \psi_{k}\sum_{p\neq k}^{G}\left(P_{p}\otimes P_{k}\right)\left(P_{p}^{\prime}\otimes P_{k}^{\prime}\right)\\
= & \psi_{k}\sum_{p\neq k}^{G}\left(J_{p}\otimes J_{k}\right)\\
= & \psi_{k}\left(I_{n}-J_{k}\right)\otimes J_{k},
\end{align*}
and from Lemma \ref{thm:qHomogeneous} we have $\phi_{k}=\zeta_{44}$,
such that
\begin{align*}
 & \mathbb{E}\left[W_{k}^{2}{\rm vec}\left(P_{k}V_{k}V_{k}^{\prime}P_{k}^{\prime}\right){\rm vec}\left(P_{k}V_{k}V_{k}^{\prime}P_{k}^{\prime}\right)^{\prime}\right]\\
= & \left(P_{k}\otimes P_{k}\right)\mathbb{E}\left[W_{k}^{2}{\rm vec}\left(V_{k}V_{k}^{\prime}\right){\rm vec}\left(V_{k}V_{k}^{\prime}\right)^{\prime}\right]\left(P_{k}^{\prime}\otimes P_{k}^{\prime}\right)\\
= & \phi_{k}\left(P_{k}\otimes P_{k}\right)\left[H_{n_{k}}+{\rm vec}(I_{n_{k}}){\rm vec}(I_{n_{k}})^{\prime}\right]\left(P_{k}^{\prime}\otimes P_{k}^{\prime}\right)\\
= & \phi_{k}\left[J_{k\otimes}H_{n}+{\rm vec}(J_{k}){\rm vec}(J_{k})^{\prime}\right].
\end{align*}
Finally we arrive at the expression,
\begin{align*}
\mathbb{E}\left[\varphi\left(k,k\right)\right] & =\psi_{k}\left(I_{n}-J_{k}\right)\otimes J_{k}+\phi_{k}\left[J_{k\otimes}H_{n}+{\rm vec}(J_{k}){\rm vec}(J_{k})^{\prime}\right].
\end{align*}

\subsubsection*{Expectation of $\varphi\left(k,l\right)$ when $k\protect\neq l$}

When $k\neq l$, we have
\begin{align*}
\mathbb{E}\left[\varphi\left(k,l\right)\right]= & \mathbb{E}\left[W_{k}W_{l}\sum_{p=1}^{G}{\rm vec}\left(P_{k}V_{k}V_{p}^{\prime}P_{p}^{\prime}\right)\sum_{q=1}^{G}{\rm vec}\left(P_{l}V_{l}V_{q}^{\prime}P_{q}^{\prime}\right)^{\prime}\right]\\
= & \mathbb{E}\left[W_{k}W_{l}{\rm vec}\left(P_{k}V_{k}V_{k}^{\prime}P_{k}^{\prime}\right){\rm vec}\left(P_{l}V_{l}V_{l}^{\prime}P_{l}^{\prime}\right)^{\prime}+W_{k}W_{l}{\rm vec}\left(P_{k}V_{k}V_{l}^{\prime}P_{l}^{\prime}\right){\rm vec}\left(P_{l}V_{l}V_{k}^{\prime}P_{k}^{\prime}\right)^{\prime}\right].
\end{align*}
From Lemma \ref{thm:qHomogeneous} with $p=2$ and $q=2$, we have
\begin{align*}
 & \mathbb{E}\left[W_{k}W_{l}{\rm vec}\left(P_{k}V_{k}V_{k}^{\prime}P_{k}^{\prime}\right){\rm vec}\left(P_{l}V_{l}V_{q}^{\prime}P_{q}^{\prime}\right)^{\prime}\right]\\
= & {\rm vec}\left(P_{k}\mathbb{E}\left[W_{k}V_{k}V_{k}^{\prime}\right]P_{k}^{\prime}\right){\rm vec}\left(P_{l}\mathbb{E}\left[W_{l}V_{l}V_{l}^{\prime}\right]P_{l}^{\prime}\right)^{\prime}\\
= & {\rm vec}\left(J_{k}\right){\rm vec}\left(J_{l}\right)^{\prime},
\end{align*}
and we also have
\begin{align*}
 & \mathbb{E}\left[W_{k}W_{l}{\rm vec}\left(P_{k}V_{k}V_{l}^{\prime}P_{l}^{\prime}\right){\rm vec}\left(P_{l}V_{l}V_{k}^{\prime}P_{k}^{\prime}\right)^{\prime}\right]\\
= & \mathbb{E}\left[W_{k}W_{l}{\rm vec}\left(P_{k}V_{k}V_{l}^{\prime}P_{l}^{\prime}\right){\rm vec}\left(P_{k}V_{k}V_{l}^{\prime}P_{l}^{\prime}\right)^{\prime}K_{n}\right]\\
= & \left(P_{l}\otimes P_{k}\right)\mathbb{E}\left[{\rm vec}\left(\tilde{V}_{k}\tilde{V}_{l}^{\prime}\right){\rm vec}\left(\tilde{V}_{k}\tilde{V}_{l}^{\prime}\right)^{\prime}\right]\left(P_{l}^{\prime}\otimes P_{k}^{\prime}\right)K_{n}\\
= & \left(P_{l}\otimes P_{k}\right)\left(P_{l}^{\prime}\otimes P_{k}^{\prime}\right)K_{n}\\
= & \left(J_{l}\otimes J_{k}\right)K_{n},
\end{align*}
where $\tilde{V}_{k}\sim N(0,I_{n_{k}})$. Finally we arrive at
\[
\mathbb{E}\left[\varphi\left(k,l\right)\right]={\rm vec}\left(J_{k}\right){\rm vec}\left(J_{l}\right)^{\prime}+\left(J_{l}\otimes J_{k}\right)K_{n}.
\]

\subsubsection*{The Expression for $\mathbb{E}\left(\nabla_{s}\nabla_{s}^{\prime}\right)$}

We have the following expression,
\begin{align*}
\mathbb{E}\left(\nabla_{s}\nabla_{s}^{\prime}\right) & =\mathbb{E}\left[\sum_{k=1}^{G}\sum_{l=1}^{G}W_{k}W_{l}{\rm vec}\left(P_{k}V_{k}U^{\prime}\right){\rm vec}\left(P_{l}V_{l}U^{\prime}\right)^{\prime}\right]-{\rm vec}\left(I_{n}\right){\rm vec}\left(I_{n}\right)^{\prime}\\
 & =\sum_{k=1}^{G}\sum_{l=1}^{G}\left[{\rm vec}\left(J_{k}\right){\rm vec}\left(J_{l}\right)^{\prime}+\left(J_{l}\otimes J_{k}\right)K_{n}\right]-{\rm vec}\left(I_{n}\right){\rm vec}\left(I_{n}\right)^{\prime}\\
 & \quad+\sum_{k=1}^{G}\left[\mathbb{E}\left[\varphi\left(k,k\right)\right]-{\rm vec}\left(J_{k}\right){\rm vec}\left(J_{k}\right)^{\prime}+J_{k\otimes}K_{n}\right]\\
 & =K_{n}+\Upsilon_{G}.
\end{align*}
where $\Upsilon_{G}=\sum_{k=1}^{G}\Psi_{k}$ with $\Psi_{k}$ given
by
\begin{align*}
\Psi_{k} & =\psi_{k}\left(I_{n}-J_{k}\right)\otimes J_{k}+\ensuremath{\phi_{k}\left[J_{k\otimes}H_{n}+{\rm vec}\left(J_{k}\right){\rm vec}\left(J_{k}\right)^{\prime}\right]-{\rm vec}\left(J_{k}\right){\rm vec}\left(J_{k}\right)^{\prime}-J_{k\otimes}K_{n}}\\
 & =\psi_{k}\left(I_{n}-J_{k}\right)\otimes J_{k}+\phi_{k}J_{k\otimes}+\ensuremath{\left(\phi_{k}-1\right)\left[J_{k\otimes}K_{n}+{\rm vec}\left(J_{k}\right){\rm vec}\left(J_{k}\right)^{\prime}\right]}\\
 & =\psi_{k}\left(I_{n}\otimes J_{k}\right)+\ensuremath{\left(\phi_{k}-\psi_{k}\right)J_{k\otimes}+\left(\phi_{k}-1\right)\left[J_{k\otimes}K_{n}+{\rm vec}\left(J_{k}\right){\rm vec}\left(J_{k}\right)^{\prime}\right].}
\end{align*}
So, the final formula for information matrix $\mathcal{I}$ is given
by
\begin{align*}
\mathcal{I} & =M^{\prime}\Omega\mathbb{E}\left(\nabla_{s}\nabla_{s}^{\prime}\right)\Omega M=M^{\prime}\Omega\left(K_{n}+\Upsilon_{G}\right)\Omega M,
\end{align*}
as stated in (\ref{eq:InfoConvolT}). This completes the proof.\hfill{}$\square$

\subsection{Block Correlation Matrix with Multivariate $t$-Distribution\label{sec:AppBlockCorr_t}}

Next, we prove the results in Section 4.2. For latter use, we define
the following variables:
\[
Q_{Y}=Y_{0}^{\prime}A^{-1}Y_{0}+\sum_{k=1}^{K}\lambda_{k}^{-1}Y_{k}^{\prime}Y_{k},\quad W=\frac{\nu+n}{\nu-2+Q_{Y}},\quad\nabla_{A}=\frac{\partial\ell}{\partial{\rm vec}\left(A\right)},\quad\Pi_{A}=\frac{\partial{\rm vec}\left(A\right)}{\partial{\rm vec}\left(W\right)^{\prime}}.
\]
By (\ref{eq:DetQuaTerm}) we have following form of log-likelihood
function
\[
\ell(Z)=c-\tfrac{1}{2}\log|A|-\frac{1}{2}\sum_{k=1}^{K}\left(n_{k}-1\right)\log\lambda_{k}-\tfrac{\nu+n_{k}}{2}\log\left(1+\tfrac{1}{\nu-2}Q_{Y}\right).
\]
Because $\tilde{C}=\ensuremath{\ensuremath{\Lambda_{n}^{-1}W\Lambda_{n}^{-1}}}$,
we have
\[
\eta={\rm vech}(\tilde{C})=L_{K}\Lambda_{n\otimes}^{-1}{\rm vec}\left(W\right),
\]
and the score is given by
\begin{align*}
\nabla^{\prime}=\frac{\partial\ell}{\partial\eta^{\prime}} & =\underset{=\nabla_{A}^{\prime}}{\underbrace{\frac{\partial\ell}{\partial{\rm vec}\left(A\right)^{\prime}}}}\underset{=\Pi_{A}}{\underbrace{\frac{\partial{\rm vec}\left(A\right)}{\partial{\rm vec}\left(W\right)^{\prime}}\frac{\partial{\rm vec}\left(W\right)}{\partial\eta^{\prime}}}}.
\end{align*}

\noindent \textbf{Proof of Lemma \ref{lem:dAdEta}.} We have $\Pi_{A}=\frac{\partial{\rm vec}(A)}{\partial{\rm vec}(W)^{\prime}}\frac{\partial{\rm vec}\left(W\right)}{\partial\eta^{\prime}}$,
where $\frac{\partial{\rm vec}\left(W\right)}{\partial\eta^{\prime}}=\Lambda_{n\otimes}D_{K}$
and
\[
\frac{\partial{\rm vec}\left(W\right)}{\partial{\rm vec}\left(A\right)^{\prime}}=\frac{\partial{\rm vec}\left(\log A\right)}{\partial{\rm vec}\left(A\right)^{\prime}}-\frac{\partial{\rm vec}\left(\log\Lambda_{\lambda}\right)}{\partial{\rm vec}\left(A\right)^{\prime}}=\frac{\partial{\rm vec}\left(\log A\right)}{\partial{\rm vec}\left(A\right)^{\prime}}-E_{d}^{\prime}\frac{\partial{\rm diag}\left(\log\Lambda_{\lambda}\right)}{\partial{\rm diag}\left(A\right)^{\prime}}E_{d},
\]
where the matrix $\tilde{\Phi}=\partial{\rm diag}\left(\log\Lambda_{\lambda}\right)/\partial{\rm diag}\left(A\right)^{\prime}$
is a diagonal matrix with diagonal elements $\left(A_{kk}-n_{k}\right)^{-1}=\lambda_{k}^{-1}\left(1-n_{k}\right)^{-1}$
for $k=1,\ldots,K$. The formula for $d{\rm vec}(\log A)/d{\rm vec}(A)$
is given by $\Gamma_{A}^{-1}$ in (\ref{eq:ParlogCparC}). So, we
have
\[
\Pi_{A}=\frac{\partial{\rm vec}\left(A\right)}{\partial{\rm vec}\left(W\right)^{\prime}}\Lambda_{n\otimes}D_{K}=\left(\frac{\partial{\rm vec}\left(W\right)}{\partial{\rm vec}\left(A\right)^{\prime}}\right)^{-1}\Lambda_{n\otimes}D_{K}.
\]
Using the Woodbury formula, we simplify the inverse of the $K^{2}\times K^{2}$
matrix, 
\[
\left(\frac{\partial{\rm vec}\left(W\right)}{\partial{\rm vec}\left(A\right)^{\prime}}\right)^{-1}=\ensuremath{\left(\Gamma_{A}^{-1}+E_{d}^{\prime}\tilde{\Phi}E_{d}\right)^{-1}=\Gamma_{A}-\Gamma_{A}E_{d}^{\prime}\left(\tilde{\Phi}^{-1}+E_{d}\Gamma_{A}E_{d}^{\prime}\right)^{-1}E_{d}\Gamma_{A}},
\]
which only requires the inverse of the low dimension, $K\times K$
matrix, $\tilde{\Phi}^{-1}+E_{d}\Gamma_{A}E_{d}^{\prime}$ to be evaluated.
Moreover, because $\tilde{\Phi}$ is a diagonal matrix with elements
$\tilde{\Phi}_{kk}=\lambda_{k}^{-1}\left(n_{k}-1\right)^{-1}$, we
define the diagonal matrix $\Phi$ with diagonal elements $\Phi_{kk}=\lambda_{k}\left(n_{k}-1\right)$,
such that $\Phi=\tilde{\Phi}^{-1}$. This proves (\ref{eq:Pi_A})
and completes the proof of Lemma \ref{lem:dAdEta}.\hfill{}$\square$

\noindent \textbf{Proof of Theorem \ref{thm:BlockMultT}.} The expression
for $\nabla_{A}$ is given by,
\[
\nabla_{A}^{\prime}=-\tfrac{1}{2}{\rm vec}\left(A^{-1}\right)^{\prime}-\tfrac{1}{2}\sum_{k=1}^{K}\frac{n_{k}-1}{\lambda_{k}}\frac{\partial\lambda_{k}}{\partial{\rm vec}\left(A\right)^{\prime}}-\tfrac{1}{2}W\frac{\partial Q_{Y}}{\partial{\rm vec}(A)^{\prime}},
\]
with
\[
\frac{\partial\lambda_{k}}{\partial{\rm vec}\left(A\right)^{\prime}}=\frac{\partial\lambda_{k}}{\partial{\rm diag}\left(A\right)^{\prime}}\frac{\partial{\rm diag}\left(A\right)}{\partial{\rm vec}\left(A\right)^{\prime}}=\left(1-n_{k}\right)^{-1}e_{k,K}^{\prime}E_{d},
\]
where $e_{k,K}$ is the $k$-th column of the $K\times K$ identity
matrix $I_{K}$. Then we obtain
\[
\frac{\partial Q_{Y}}{\partial{\rm vec}(A)^{\prime}}=-{\rm vec}\left(Y_{0}Y_{0}^{\prime}\right)^{\prime}A_{\otimes}^{-1}-\sum_{k=1}^{K}\lambda_{k}^{-2}\left(1-n_{k}\right)^{-1}\left(Y_{k}^{\prime}Y_{k}\right)e_{k,K}^{\prime}E_{d},
\]
which leads to
\begin{align}
\nabla_{A}^{\prime} & =\tfrac{1}{2}\left[W{\rm vec}\left(Y_{0}Y_{0}^{\prime}\right)^{\prime}-{\rm vec}\left(A\right)^{\prime}\right]A_{\otimes}^{-1}+\tfrac{1}{2}S^{\prime}E_{d}\nonumber \\
 & =\tfrac{1}{2}\left[W{\rm vec}\left(X_{0}X_{0}^{\prime}\right)^{\prime}-{\rm vec}\left(I_{K}\right)^{\prime}\right]A_{\otimes}^{-\frac{1}{2}}+\tfrac{1}{2}S^{\prime}E_{d},\label{eq:DelA}
\end{align}
where $S$ is a $K\times1$ vector defined by
\[
S=\sum_{k=1}^{K}\left(\lambda_{k}^{-1}-W\lambda_{k}^{-1}\left(n_{k}-1\right)^{-1}X_{k}^{\prime}X_{k}\right)e_{k}^{\prime},\quad{\rm with}\quad S_{k}=\frac{\left(n_{k}-1\right)-WX_{k}^{\prime}X_{k}}{\lambda_{k}\left(n_{k}-1\right)}.
\]

\subsubsection*{The Information Matrix}

First, from the formula of score, we have following form of information
matrix, 
\[
\mathcal{I}=\Pi_{A}^{\prime}\mathcal{I}_{A}\Pi_{A}.
\]
So we need to compute the matrix $\mathcal{I}_{A}=\mathbb{E}\left(\nabla_{A}\nabla_{A}^{\prime}\right)$.
From, (\ref{eq:DelA}), we could find its first term is a function
of $X_{0}$ and the second term is a function of $X_{k}^{\prime}X_{k},k=1,2,..K$.
So, for the first term, we have
\begin{eqnarray*}
\nabla_{A}^{(1)} & \equiv & \tfrac{1}{2}A_{\otimes}^{-\frac{1}{2}}\left[W{\rm vec}\left(X_{0}X_{0}^{\prime}\right)-{\rm vec}\left(I_{K}\right)\right]\\
\mathcal{I}_{A}^{(1)} & = & \tfrac{1}{4}A_{\otimes}^{-1/2}\left(\mathbb{E}\left[W^{2}{\rm vec}\left(X_{0}X_{0}^{\prime}\right){\rm vec}\left(X_{0}X_{0}^{\prime}\right)^{\prime}\right]-{\rm vec}\left(I_{K}\right){\rm vec}\left(I_{K}\right)^{\prime}\right)A_{\otimes}^{-1/2}.
\end{eqnarray*}
Similar to (\ref{eq:FisherInf_t}), we have
\begin{align*}
\mathcal{I}_{A}^{(1)} & =\tfrac{1}{4}\left[\phi A_{\otimes}^{-1}H_{K}+\left(\phi-1\right){\rm vec}\left(A^{-1}\right){\rm vec}\left(A^{-1}\right)^{\prime}\right].
\end{align*}
For the second term, we first define
\begin{align*}
\nabla_{A}^{(2)} & \equiv\tfrac{1}{2}E_{d}^{\prime}S=\tfrac{1}{2}E_{d}^{\prime}\Lambda_{\lambda}\bar{S},\quad\mathcal{I}_{A}^{(2)}=\tfrac{1}{4}E_{d}^{\prime}\Lambda_{\lambda}\mathbb{E}\left(\bar{S}\bar{S}^{\prime}\right)\Lambda_{\lambda}E_{d},
\end{align*}
where the element in vector $\bar{S}$ and diagonal matrix $\Lambda_{\lambda}$
are define by $\bar{S}_{k}=WX_{k}^{\prime}X_{k}-\left(n_{k}-1\right)$
and $\left[\Lambda_{\lambda}\right]_{kk}=\left[-\lambda_{k}\left(n_{k}-1\right)\right]^{-1}$.
We know that
\begin{align*}
\mathbb{E}\left[W^{2}\left(X_{k}^{\prime}X_{k}\right)\left(X_{l}^{\prime}X_{l}\right)\right] & =\phi\mathbb{E}\left[\left(\tilde{Z}_{k}^{\prime}\tilde{Z}_{k}\right)\left(\tilde{Z}_{l}^{\prime}\tilde{Z}_{l}\right)\right]=\begin{cases}
\phi\left(n_{k}-1\right)\left(n_{l}-1\right) & k\neq l,\\
\phi\left[\left(n_{k}-1\right)^{2}+2\left(n_{k}-1\right)\right] & k=l,
\end{cases}
\end{align*}
where $\tilde{Z}_{k}\sim N\left(0,I_{n_{k}}\right)$. So $\mathbb{E}\left(\bar{S}_{k}\bar{S}_{l}^{\prime}\right)=\mathbb{E}\left[W^{2}\left(X_{k}^{\prime}X_{k}\right)\left(X_{l}^{\prime}X_{l}\right)\right]-\left(n_{k}-1\right)\left(n_{l}-1\right)$
is given by
\[
\mathbb{E}\left(\bar{S}_{k}\bar{S}_{l}^{\prime}\right)=\begin{cases}
\left(\phi-1\right)\left(n_{k}-1\right)\left(n_{l}-1\right) & k\neq l,\\
\left(\phi-1\right)\left(n_{k}-1\right)^{2}+2\phi\left(n_{k}-1\right) & k=l,
\end{cases}
\]
and along with the following $K\times1$ vector $\xi$ and diagonal
matrix $\Xi$, we have
\[
\Lambda_{\lambda}\mathbb{E}\left(\bar{S}\bar{S}^{\prime}\right)\Lambda_{\lambda}=\ensuremath{\left(\phi-1\right)}\ensuremath{\xi\xi^{\prime}}+2\phi\Xi,\quad\xi_{k}=\lambda_{k}^{-1},\quad\Xi_{kk}=\lambda_{k}^{-2}\left(n_{k}-1\right)^{-1}.
\]
So,
\begin{align*}
\mathcal{I}_{A}^{(2)} & =\tfrac{1}{4}E_{d}^{\prime}\left[\ensuremath{\left(\phi-1\right)}\ensuremath{\xi\xi^{\prime}}+2\phi\Xi\right]E_{d}=\tfrac{\phi-1}{4}E_{d}^{\prime}\ensuremath{\xi\xi^{\prime}}E_{d}+\tfrac{\phi}{2}E_{d}^{\prime}\Xi E_{d},
\end{align*}
and we also have
\begin{align*}
\mathbb{E}\left(W{\rm vec}\left(X_{0}X_{0}^{\prime}\right)-{\rm vec}(I_{K})\right)\bar{S}_{k} & =\mathbb{E}\left[W^{2}{\rm vec}\left(X_{0}X_{0}^{\prime}\right)\left(X_{k}^{\prime}X_{k}\right)\right]-\left(n_{k}-1\right){\rm vec}(I_{K})\\
 & =\left(\phi-1\right)\left(n_{k}-1\right){\rm vec}\left(I_{K}\right).
\end{align*}
Hence,
\begin{align*}
\mathcal{I}_{A}^{(12)}=\mathbb{E}\left(\nabla_{A}^{(1)}\nabla_{A}^{(2)\prime}\right)= & -\tfrac{1}{4}A_{\otimes}^{-1/2}\left(\phi-1\right){\rm vec}\left(I_{K}\right)\xi^{\prime}E_{d}=\tfrac{1-\phi}{4}{\rm vec}\left(A_{\otimes}^{-1}\right)\xi^{\prime}E_{d}.
\end{align*}
Finally, we have
\[
\mathcal{I}_{A}=\mathcal{I}_{A}^{(1)}+\mathcal{I}_{A}^{(2)}+\mathcal{I}_{A}^{(12)}+\mathcal{I}_{A}^{(21)},
\]
which gives the expression in Theorem \ref{thm:BlockMultT}. Finally,
in the limited case, $\nu\rightarrow\infty$, which corresponds to
the multivariate normal distribution, we have $\phi\rightarrow1$,
and the information matrix simplifies to $\mathcal{I}_{A}=\tfrac{1}{4}\ensuremath{A_{\otimes}^{-1}}H_{K}+\tfrac{1}{2}E_{d}^{\prime}\Xi E_{d}.$
This completes the proof.\hfill{}$\square$

\noindent \textbf{Proof of Theorem \ref{thm:BlockClusterT}.} For
this case we have 
\[
\nabla_{A}^{\prime}=\tfrac{1}{2}\left[W_{0}{\rm vec}\left(X_{0}X_{0}^{\prime}\right)^{\prime}-{\rm vec}\left(I_{K}\right)^{\prime}\right]A_{\otimes}^{-\frac{1}{2}}+\tfrac{1}{2}S^{\prime}E_{d},
\]
where $W_{0}$ and $S\in\mathbb{R}^{K}$ are given by
\[
W_{0}=\frac{\nu_{0}+K}{\nu_{0}-2+X_{0}^{\prime}X_{0}},\quad S_{k}=\frac{(n_{k}-1)-W_{k}X_{k}^{\prime}X_{k}}{\lambda_{k}(n_{k}-1)},\quad\text{with}\quad\ensuremath{W_{k}=\frac{\nu_{k}+n_{k}-1}{\nu_{k}-2+X_{k}^{\prime}X_{k}}}.
\]
The covariance of the first part was derived in Theorem \ref{thm:BlockMultT}
and is given by
\[
\mathcal{I}_{A}^{(1)}=\tfrac{1}{4}\left(\phi_{0}A_{\otimes}^{-1}H_{K}+\left(\phi-1\right){\rm vec}\left(A^{-1}\right){\rm vec}\left(A^{-1}\right)^{\prime}\right),\quad\phi_{0}=\frac{\nu_{0}+K}{\nu_{0}+K+2}.
\]
For the second part, we have $\mathbb{E}\left(\bar{S}_{k}\bar{S}_{l}\right)=0$
for $k\neq l$, and
\[
\mathbb{E}\left(\bar{S}_{k}^{2}\right)=\left(\phi_{k}-1\right)\left(n_{k}-1\right)^{2}+2\phi_{k}\left(n_{k}-1\right),
\]
with $\phi_{k}=\left(\nu_{k}+n_{k}-1\right)/\left(\nu_{k}+n_{k}+1\right)$.
Therefore, we have
\[
\mathcal{I}_{A}^{(2)}=E_{d}^{\prime}\Lambda_{\lambda}\mathbb{E}\left(\bar{S}\bar{S}^{\prime}\right)\Lambda_{\lambda}E_{d}=\Xi,\quad\Xi_{kk}=\ensuremath{\frac{2\phi_{k}}{\lambda_{k}^{2}\left(n_{k}-1\right)}+\frac{\phi_{k}-1}{\lambda_{k}^{2}}},
\]
and $\mathcal{I}_{A}^{(12)}=\mathbb{E}\left(\nabla_{A}^{(1)}\nabla_{A}^{(2)\prime}\right)=0$.
Finally, we obtain
\[
\ensuremath{\mathcal{I}_{A}=\tfrac{1}{4}\left[\phi_{0}A_{\otimes}^{-1}H_{K}+(\phi-1){\rm vec}\left(A^{-1}\right){\rm vec}\left(A^{-1}\right)^{\prime}+E_{d}^{\prime}\Xi E_{d}\right]}.
\]

\section{Block Correlation Matrix with Cluster-$t$ Distribution}

\setcounter{equation}{0}\renewcommand{\theequation}{F.\arabic{equation}}

Because $P=I_{n}$, and $\bm{n}=\bm{m}$, we have $V=U$, and $V_{k}=U_{k}$.
Then the log-likelihood function is given by
\begin{align*}
\ell\left(Z\right) & =-\tfrac{1}{2}\log|C|+\sum_{k=1}^{K}c_{k}-\tfrac{\nu_{k}+n_{k}}{2}\log\left(1+\tfrac{1}{\nu_{k}-2}U_{k}^{\prime}U_{k}\right),
\end{align*}
by using the canonical representation of block correlation matrix
$C=QDQ^{\prime}$, we define the vectors $X$ and $Y$ as $Y=Q^{\prime}Z$
and $X=Q^{\prime}U$, so we have $X_{0}=A^{-\frac{1}{2}}Y_{0}$, $X_{k}=\lambda_{k}^{-\frac{1}{2}}Y_{k}$.
From $U=QX$ and the structure of $Q$, we have
\begin{align*}
U_{k,i} & =n_{k}^{-1/2}X_{0,k}+\left(\tilde{e}_{i}^{\prime}v_{n_{k}}^{\perp}\right)X_{k}\\
U_{k,i}^{2} & =X_{0,k}^{2}n_{k}^{-1}+X_{k}^{\prime}\left(v_{n_{k}}^{\perp\prime}\tilde{e}_{i}\tilde{e}_{i}^{\prime}v_{n_{k}}^{\perp}\right)X_{k}+2X_{0,k}\left(v_{n_{k}}^{\prime}\tilde{e}_{i}\tilde{e}_{i}^{\prime}v_{n_{k}}^{\perp}\right)X_{k},
\end{align*}
for $i=1,\ldots,n_{k}$, and $\tilde{e}_{i}\in\mathbb{R}^{n_{k}\times1}$
denote the $i$-th column of $I_{n_{k}}$. So we obtain
\[
U_{k}^{\prime}U_{k}=\sum_{i=1}^{n_{k}}U_{k,i}^{2}=X_{0,k}^{2}+X_{k}^{\prime}X_{k}=Y_{0}^{\prime}\tilde{A}_{k}^{-1}Y_{0}+\lambda_{k}^{-1}Y_{k}^{\prime}Y_{k},
\]
where $\tilde{A}_{k}^{-1}=A^{-\tfrac{1}{2}}J_{k}^{e}A^{-\tfrac{1}{2}}$
, and $J_{k}^{e}=e_{k}e_{k}^{\prime}$ with $e_{k}$ the $k$-th column
of the $K\times K$ identity matrix $I_{K}$. This leads to the the
simplified expression for log-likelihood function
\begin{align*}
\ell\left(Z\right) & =-\tfrac{1}{2}\log|A|+\sum_{k=1}^{K}\left[c_{v_{k},n_{k}}-\tfrac{n_{k}-1}{2}\log\lambda_{k}-\tfrac{\nu_{k}+n_{k}}{2}\log\left(1+\tfrac{1}{\nu_{k}-2}\left(X_{0,k}^{2}+X_{k}^{\prime}X_{k}\right)\right)\right].
\end{align*}
Note that we have
\[
X_{0,k}=v_{n_{k}}^{\prime}U_{k},\quad X_{k}=\ensuremath{v_{n_{k}}^{\perp\ \prime}U_{k}}\qquad\text{for}\quad k=1,\ldots,K,
\]
such that $X_{0,k}$ and $X_{k}$ are simply linear combinations of
$U_{k}$. From the structure of $Q$ it follows that $X_{0,k}$ and
$X_{0,l}$ are independent for $k\neq l$, just as it the case for
$X_{k}$ and $X_{l}$ (by their definition). We also have that $X_{0,k}$
and $X_{k}$ are uncorrelated, but they are not independent, because
they have $t$-distributed shocks in common. 

\subsection{The Form of the Score}

By using $X_{0,k}=e_{k,K}^{\prime}A^{-\frac{1}{2}}Y_{0}$, we first
have
\begin{align*}
\frac{\partial X_{0,k}^{2}}{\partial{\rm vec}\left(A\right)^{\prime}} & =\frac{\partial X_{0,k}^{2}}{\partial X_{0,k}}\frac{\partial\left(e_{k}^{\prime}A^{-\frac{1}{2}}Y_{0}\right)}{\partial{\rm vec}\left(A^{-\frac{1}{2}}\right)^{\prime}}\frac{\partial{\rm vec}\left(A^{-\frac{1}{2}}\right)^{\prime}}{\partial{\rm vec}\left(A^{\frac{1}{2}}\right)^{\prime}}\frac{\partial{\rm vec}\left(A^{\frac{1}{2}}\right)^{\prime}}{\partial{\rm vec}\left(A\right)^{\prime}}\\
 & =-2X_{0,k}\left(Y_{0}^{\prime}\otimes e_{k}^{\prime}\right)A_{\otimes}^{-\frac{1}{2}}\ensuremath{\left(A^{\frac{1}{2}}\oplus I\right)^{-1}}\\
 & =-2X_{0,k}\left(X_{0}^{\prime}\otimes e_{k}^{\prime}A^{-\frac{1}{2}}\right)\ensuremath{\left(A^{\frac{1}{2}}\oplus I\right)^{-1}}\\
 & =-2X_{0,k}{\rm vec}\left(A^{-\frac{1}{2}}e_{k}X_{0}^{\prime}\right)^{\prime}\ensuremath{\left(A^{\frac{1}{2}}\oplus I\right)^{-1}}\\
 & =-2X_{0,k}{\rm vec}\left(e_{k}X_{0}^{\prime}\right)^{\prime}\Omega,
\end{align*}
where $\Omega=\left(I\otimes A^{-\frac{1}{2}}\right)\ensuremath{\left(A^{\frac{1}{2}}\oplus I\right)^{-1}}$.
It follows that
\[
\nabla_{A}^{\prime}=\frac{\partial\ell}{\partial{\rm vec}\left(A\right)^{\prime}}=\left[\sum_{k=1}^{K}W_{k}X_{0,k}{\rm vec}\left(e_{k}X_{0}^{\prime}\right)^{\prime}-{\rm vec}\left(I_{K}\right)^{\prime}\right]\Omega+\tfrac{1}{2}E_{d}^{\prime}S
\]
where $S$ is a $K\times1$ vector defined by
\[
S_{k}=\frac{\left(n_{k}-1\right)-W_{k}X_{k}^{\prime}X_{k}}{\left(n_{k}-1\right)\lambda_{k}},\quad W_{k}=\frac{\nu_{k}+n_{k}}{\nu_{k}-2+X_{0,k}^{2}+X_{k}^{\prime}X_{k}}.
\]

\subsection{The Form of the Information Matrix}

The information matrix of $\nabla_{A}$ can be decompose into four
components, given by
\[
\mathcal{I}_{A}=\mathcal{I}_{A}^{(1)}+\mathcal{I}_{A}^{(2)}+\mathcal{I}_{A}^{(21)}+\mathcal{I}_{A}^{(12)}.
\]

\subsubsection{The Form of Matrix $\mathcal{I}_{A}^{(1)}$}

Similar to previous proof, the covariance of the first part of $\partial\ell/\partial{\rm vec}\left(A\right)$
is given by
\begin{align*}
\mathcal{I}_{A}^{(1)} & =\Omega\mathbb{E}\left[\sum_{k=1}^{K}\sum_{l=1}^{K}W_{k}W_{l}X_{0,k}X_{0,l}{\rm vec}\left(e_{k}X_{0}^{\prime}\right){\rm vec}\left(e_{l}X_{0}^{\prime}\right)^{\prime}-{\rm vec}\left(I_{K}\right){\rm vec}\left(I_{K}\right)^{\prime}\right]\Omega,\\
 & =\Omega\left(K_{K}+\Psi^{e}\right)\Omega
\end{align*}
where $\Psi^{e}=\sum_{k=1}^{K}\Psi_{k}^{e}$ with
\begin{align*}
\Psi_{k}^{e} & =\psi_{k}\left(I-J_{k}^{e}\right)\otimes J_{k}^{e}+\phi_{k}J_{k\otimes}^{e}+\left(\phi_{k}-1\right)\left[J_{k\otimes}^{e}K_{K}+{\rm vec}\left(J_{k}^{e}\right){\rm vec}\left(J_{k}^{e}\right)^{\prime}\right],
\end{align*}
with $\phi_{k}=(v_{k}+n_{k})/(v_{k}+n_{k}+2)$.

\subsubsection{The Form of Matrix $\mathcal{I}_{A}^{(2)}$}

As for the second part, we have
\[
\mathcal{I}_{A}^{(2)}=\tfrac{1}{4}E_{d}^{\prime}\mathbb{E}\left(SS^{\prime}\right)E_{d}=\tfrac{1}{4}E_{d}^{\prime}\Xi E_{d},
\]
where $\Xi_{kk}=\mathbb{E}\left(S_{k}^{2}\right)$ given by
\begin{align*}
\mathbb{E}\left(S_{k}^{2}\right) & =\left[\mathbb{E}\left(W_{k}^{2}\left(X_{k}^{\prime}X_{k}\right)^{2}\right)-\left(n_{k}-1\right)^{2}\right]/\left[\lambda_{k}^{2}\left(n_{k}-1\right)^{2}\right]\\
 & =\left[\phi_{k}\left[\left(n_{k}-1\right)^{2}+2\left(n_{k}-1\right)\right]-\left(n_{k}-1\right)^{2}\right]/\left[\lambda_{k}^{2}\left(n_{k}-1\right)^{2}\right]\\
 & =\left(\phi_{k}-1\right)\lambda_{k}^{-2}+2\phi_{k}\lambda_{k}^{-2}\left(n_{k}-1\right)^{-1},
\end{align*}
and $\Xi_{kl}=\mathbb{E}\left(S_{k}S_{l}\right)=0$ for $k\neq l$,
so $\Xi$ is a $K\times K$ diagonal matrix.

\subsubsection{The Form of Matrix $\mathcal{I}_{A}^{(12)}$}

As for the interaction term, we have $\mathcal{I}_{A}^{(21)}=\left[\mathcal{I}_{A}^{(12)}\right]^{\prime}$,
and
\[
\mathcal{I}_{A}^{(12)}=\tfrac{1}{2}E_{d}^{\prime}\mathbb{E}\left[S\left(\sum_{k=1}^{K}W_{k}X_{0,k}{\rm vec}\left(e_{k}X_{0}^{\prime}\right)^{\prime}-{\rm vec}\left(I\right)^{\prime}\right)\right]\Omega=\tfrac{1}{2}E_{d}^{\prime}\Theta\Omega,
\]
where the $k$-th row of the $K\times K^{2}$ matrix $\Theta$ is
given by
\begin{align*}
e_{k}^{\prime}\Theta & =\mathbb{E}\left[S_{k}\left(\sum_{l=1}^{K}W_{l}X_{0,l}{\rm vec}\left(e_{l}X_{0}^{\prime}\right)^{\prime}-{\rm vec}\left(I_{K}\right)^{\prime}\right)\right]\\
 & =\left(n_{k}-1\right)^{-1}\lambda_{k}^{-1}\left[\mathbb{E}\left(\sum_{l=1}^{K}W_{k}W_{l}\left(X_{k}^{\prime}X_{k}\right)X_{0,l}{\rm vec}\left(e_{l}X_{0}^{\prime}\right)\right)-\left(n_{k}-1\right){\rm vec}\left(I_{K}\right)\right],
\end{align*}
when $k=l$, we have
\begin{align*}
\mathbb{E}\left[W_{k}^{2}\left(X_{k}^{\prime}X_{k}\right)X_{0,k}{\rm vec}\left(e_{k}X_{0}^{\prime}\right)\right]= & \mathbb{E}\left[\sum_{p=1}^{K}W_{k}^{2}\left(X_{k}^{\prime}X_{k}\right)X_{0,k}{\rm vec}\left(e_{k}X_{0,p}e_{p}^{\prime}\right)\right]\\
= & \mathbb{E}\left[W_{k}^{2}\left(X_{k}^{\prime}X_{k}\right)X_{0,k}^{2}{\rm vec}\left(J_{k}^{e}\right)\right]\\
= & \phi_{k}\left(n_{k}-1\right){\rm vec}\left(J_{k}^{e}\right),
\end{align*}
when $k\neq l$, we have
\begin{align*}
\mathbb{E}\left[W_{k}W_{l}\left(X_{k}^{\prime}X_{k}\right)X_{0,l}{\rm vec}\left(e_{l}X_{0}^{\prime}\right)\right]= & \mathbb{E}\left[\sum_{p=1}^{K}W_{k}W_{l}\left(X_{k}^{\prime}X_{k}\right)X_{0,l}{\rm vec}\left(e_{l}X_{0,p}e_{l}^{\prime}\right)\right]\\
= & \mathbb{E}\left[W_{k}W_{l}\left(X_{k}^{\prime}X_{k}\right)X_{0,l}^{2}\right]{\rm vec}\left(J_{l}^{e}\right)\\
= & \left(n_{k}-1\right){\rm vec}\left(J_{l}^{e}\right).
\end{align*}
Thus, we have
\begin{align*}
 & \mathbb{E}\left[\sum_{l=1}^{K}W_{k}W_{l}\left(X_{k}^{\prime}X_{k}\right)X_{0,l}{\rm vec}\left(e_{l}X_{0}^{\prime}\right)\right]-\left(n_{k}-1\right){\rm vec}\left(I_{K}\right)\\
= & \sum_{l}\left(n_{k}-1\right){\rm vec}\left(J_{l}^{e}\right)-\left(n_{k}-1\right){\rm vec}\left(J_{k}^{e}\right)+\phi_{k}\left(n_{k}-1\right){\rm vec}\left(J_{k}^{e}\right)-\left(n_{k}-1\right){\rm vec}\left(I_{K}\right)\\
= & \left(n_{k}-1\right){\rm vec}\left(I_{K}\right)-\left(n_{k}-1\right){\rm vec}\left(J_{k}^{e}\right)+\phi_{k}\left(n_{k}-1\right){\rm vec}\left(J_{k}^{e}\right)-\left(n_{k}-1\right){\rm vec}\left(I_{K}\right)\\
= & \left(n_{k}-1\right)\left(\phi_{k}-1\right){\rm vec}\left(J_{k}^{e}\right).
\end{align*}
Finally, the $k$-th row of matrix $M$ is $e_{k}^{\prime}\Theta=-\lambda_{k}^{-1}\left(\phi_{k}-1\right){\rm vec}\left(J_{k}^{e}\right)^{\prime}$,
so we have
\[
\Theta=\sum_{k=1}^{K}e_{k}\left(e_{k}^{\prime}\Theta\right)=\sum_{k=1}^{K}\lambda_{k}^{-1}\left(1-\phi_{k}\right)e_{k}{\rm vec}\left(J_{k}^{e}\right)^{\prime}.
\]

\section{Block Correlation Matrix with Hetero-$t$ Distribution}

\setcounter{equation}{0}\renewcommand{\theequation}{A.\arabic{equation}}

Because $P=I$, we have $U=PV=V$. By using $C=QDQ^{\prime}$, the
log-likelihood function is now given by
\[
\ensuremath{\ell(Z)=c-\tfrac{1}{2}\log|A|-\tfrac{1}{2}\sum_{k=1}^{K}\left(n_{k}-1\right)\log\lambda_{k}-\sum_{k=1}^{K}\sum_{i=1}^{n_{k}}\tfrac{\nu_{k,i}+1}{2}\log\left(1+\tfrac{1}{\nu_{k,i}-2}U_{k,i}^{2}\right)},
\]
where $c=\sum_{i=1}^{n}c\left(\nu_{i},1\right)$, and $U_{k,i}$ is
the $i$-th innovation of $U_{k}$. The cluster structure is implied
by the block correlation matrix. To simplify the notation we let $\tilde{e}_{i}\in\mathbb{R}^{n_{k}\times1}$
denote the $i$-th column of $I_{n_{k}}$. The identity $U_{k,i}=n_{k}^{-1/2}X_{0,k}+\left(e_{i}^{\prime}v_{n_{k}}^{\perp}\right)X_{k}$,
which means that
\begin{align*}
\frac{\partial\left(U_{k,i}^{2}\right)}{\partial{\rm vec}\left(A\right)} & =\frac{\partial\left(U_{k,i}^{2}\right)}{\partial U_{k,i}}\frac{\partial\left(n_{k}^{-1/2}X_{0,k}+\left(\tilde{e}_{i}^{\prime}v_{n_{k}}^{\perp}\right)X_{k}\right)}{\partial{\rm vec}\left(A\right)}\\
 & =2U_{k,i}\left[n_{k}^{-1/2}{\rm vec}\left(e_{k}X_{0}^{\prime}\right)^{\prime}\Omega-\frac{1}{2}\left(e_{i}^{\prime}v_{n_{k}}^{\perp}\right)X_{k}\lambda_{k}^{-1}\ensuremath{\left(1-n_{k}\right)^{-1}e_{k}^{\prime}E_{d}}\right]\\
 & =2U_{k,i}{\rm vec}\left(e_{k}X_{0}^{\prime}\right)^{\prime}n_{k}^{-1/2}\Omega-U_{k,i}F_{k,i}U_{k}e_{k}^{\prime}E_{d}/\left[\lambda_{k}\ensuremath{\left(1-n_{k}\right)}\right],
\end{align*}
where $e_{k}$ is $k$-th column of the $K\times K$ identity matrix
$I_{K}$, $\tilde{e}_{i}$ is the $i$-th column of the $n_{k}\times n_{k}$
identity matrix $I_{n_{k}}$, $X_{k}=\ensuremath{v_{n_{k}}^{\perp\ \prime}U_{k}}$
and $F_{k,i}=\tilde{e}_{i}^{\prime}v_{n_{k}}^{\perp}v_{n_{k}}^{\perp\prime}$.
So we have
\begin{align*}
\nabla_{A}^{\prime}=\frac{\partial\ell}{\partial{\rm vec}\left(A\right)^{\prime}} & =\ensuremath{\left[\sum_{k=1}^{K}\sum_{i=1}^{n_{k}}W_{k,i}U_{k,i}{\rm vec}\left(e_{k}X_{0}^{\prime}\right)^{\prime}n_{k}^{-\frac{1}{2}}-{\rm vec}(I_{K})^{\prime}\right]}\Omega+\tfrac{1}{2}S^{\prime}E_{d},
\end{align*}
with the $W_{k,i}$ and $S_{k}$ defined by
\[
W_{k,i}=\frac{\nu_{k,i}+1}{\nu_{k,i}-2+U_{k,i}^{\prime}U_{k,i}},\quad S_{k}=\frac{\left(n_{k}-1\right)-\sum_{i=1}^{n_{k}}W_{k,i}U_{k,i}F_{k,i}U_{k}}{\left(n_{k}-1\right)\lambda_{k}}.
\]
The information matrix of $\nabla_{A}$ can be expressed by $\mathcal{I}_{A}=\mathcal{I}_{A}^{(1)}+\mathcal{I}_{A}^{(2)}+\mathcal{I}_{A}^{(21)}+\mathcal{I}_{A}^{(12)}$.

\subsection{The Form of Matrix $\mathcal{I}_{A}^{(1)}$}

\begin{align*}
\mathcal{I}_{A}^{(1)}= & \mathbb{E}\left[\sum_{k=1}^{K}\sum_{i=1}^{n_{k}}W_{k,i}U_{k,i}{\rm vec}\left(e_{k}X_{0}^{\prime}\right)^{\prime}n_{k}^{-\frac{1}{2}}\right]\left[\sum_{l=1}^{K}\sum_{j=1}^{n_{k}}W_{l,j}U_{l,j}{\rm vec}\left(e_{l}X_{0}^{\prime}\right)^{\prime}n_{l}^{-\frac{1}{2}}\right]-{\rm vec}(I_{K}){\rm vec}(I_{K})^{\prime}\\
= & \mathbb{E}\left[\sum_{k=1}^{K}\sum_{l=1}^{K}\sum_{i=1}^{n_{k}}\sum_{j=1}^{n_{l}}W_{k,i}W_{l,j}U_{k,i}U_{l,j}{\rm vec}\left(e_{k}X_{0}^{\prime}\right){\rm vec}\left(e_{l}X_{0}^{\prime}\right)^{\prime}n_{l}^{-\frac{1}{2}}n_{k}^{-\frac{1}{2}}\right]-{\rm vec}(I_{K}){\rm vec}(I_{K})^{\prime}\\
= & \mathbb{E}\left[\sum_{k=1}^{K}\sum_{i=1}^{n_{k}}W_{k,i}^{2}U_{k,i}^{2}{\rm vec}\left(e_{k}X_{0}^{\prime}\right){\rm vec}\left(e_{k}X_{0}^{\prime}\right)^{\prime}n_{k}^{-1}\right]-{\rm vec}(I_{K}){\rm vec}(I_{K})^{\prime}\\
 & +\mathbb{E}\left[\sum_{k=1}^{K}\sum_{l\neq k}^{K}\sum_{i=1}^{n_{k}}\sum_{j=1}^{n_{l}}W_{k,i}W_{l,j}U_{k,i}U_{l,j}{\rm vec}\left(e_{k}X_{0}^{\prime}\right){\rm vec}\left(e_{l}X_{0}^{\prime}\right)^{\prime}n_{l}^{-\frac{1}{2}}n_{k}^{-\frac{1}{2}}\right]\\
 & +\mathbb{E}\left[\sum_{k=1}^{K}\sum_{i=1}^{n_{k}}\sum_{j\neq i}^{n_{k}}W_{k,i}W_{k,j}U_{k,i}U_{k,j}{\rm vec}\left(e_{k}X_{0}^{\prime}\right){\rm vec}\left(e_{k}X_{0}^{\prime}\right)^{\prime}n_{k}^{-\frac{1}{2}}n_{k}^{-\frac{1}{2}}\right],
\end{align*}
and for later use, we have $X_{0}=\sum_{k}X_{0,k}e_{k}=\sum_{p}U_{p}^{\prime}v_{n_{p}}e_{p}$.

\subsubsection{The First Term}

We have
\begin{align*}
 & \mathbb{E}\left[W_{k,i}^{2}U_{k,i}^{2}{\rm vec}\left(e_{k}X_{0}^{\prime}\right){\rm vec}\left(e_{k}X_{0}^{\prime}\right)^{\prime}n_{k}^{-1}\right]\\
= & \mathbb{E}\left[\sum_{p=1}^{K}\sum_{q=1}^{K}W_{k,i}^{2}U_{k,i}^{2}{\rm vec}\left(e_{k}U_{p}^{\prime}v_{n_{p}}e_{p}^{\prime}\right){\rm vec}\left(e_{k}U_{q}^{\prime}v_{n_{q}}e_{q}^{\prime}\right)^{\prime}n_{k}^{-1}\right]\\
= & \mathbb{E}\left[W_{k,i}^{2}U_{k,i}^{2}{\rm vec}\left(e_{k}U_{k}^{\prime}v_{n_{k}}e_{k}^{\prime}\right){\rm vec}\left(e_{k}U_{k}^{\prime}v_{n_{k}}e_{k}^{\prime}\right)^{\prime}n_{k}^{-1}\right]\\
 & +\mathbb{E}\left[\sum_{p\neq k}^{K}W_{k,i}^{2}U_{k,i}^{2}{\rm vec}\left(e_{k}U_{p}^{\prime}v_{n_{p}}e_{p}^{\prime}\right){\rm vec}\left(e_{k}U_{p}^{\prime}v_{n_{p}}e_{p}^{\prime}\right)^{\prime}n_{k}^{-1}\right],
\end{align*}
and with $p=k=q$, we have
\begin{align*}
 & \mathbb{E}\left[W_{k,i}^{2}U_{k,i}^{2}{\rm vec}\left(e_{k}U_{k}^{\prime}v_{n_{k}}e_{k}^{\prime}\right){\rm vec}\left(e_{k}U_{k}^{\prime}v_{n_{k}}e_{k}^{\prime}\right)^{\prime}n_{k}^{-1}\right]\\
= & \mathbb{E}\left[\sum_{p=1}^{n_{k}}\sum_{q=1}^{n_{k}}W_{k,i}^{2}U_{k,i}^{2}U_{k,p}U_{k,q}{\rm vec}\left(e_{k}e_{k}^{\prime}\right){\rm vec}\left(e_{k}e_{k}^{\prime}\right)^{\prime}n_{k}^{-2}\right]\\
= & \mathbb{E}\left[\sum_{p=1}^{n_{k}}\sum_{q=1}^{n_{k}}W_{k,i}^{2}U_{k,i}^{2}U_{k,p}U_{k,q}J_{k\otimes}^{e}n_{k}^{-2}\right]\\
= & \mathbb{E}\left[W_{k,i}^{2}U_{k,i}^{4}\right]J_{k\otimes}^{e}n_{k}^{-2}+\sum_{p\neq i}^{n_{k}}\mathbb{E}\left[W_{k,i}^{2}U_{k,i}^{2}U_{k,p}^{2}\right]J_{k\otimes}^{e}n_{k}^{-2}\\
= & \left[3\phi_{k,i}+\psi_{k,i}\left(n_{k}-1\right)\right]J_{k\otimes}^{e}n_{k}^{-2},
\end{align*}
where we use that ${\rm vec}\left(e_{k}e_{k}^{\prime}\right)=e_{k}\otimes e_{k}$
because $e_{k}$ is a $K\times1$ vector. Next, when $p\neq k$, we
have
\begin{align*}
 & \sum_{p\neq k}^{K}\mathbb{E}\left[W_{k,i}^{2}U_{k,i}^{2}{\rm vec}\left(e_{k}U_{p}^{\prime}v_{n_{p}}e_{p}^{\prime}\right){\rm vec}\left(e_{k}U_{p}^{\prime}v_{n_{p}}e_{p}^{\prime}\right)^{\prime}n_{k}^{-1}\right]\\
= & \sum_{p\neq k}^{K}\mathbb{E}\left[\sum_{r=1}^{n_{p}}\sum_{m=1}^{n_{p}}W_{k,i}^{2}U_{k,i}^{2}U_{p,r}U_{p,m}{\rm vec}\left(e_{k}e_{p}^{\prime}\right){\rm vec}\left(e_{k}e_{p}^{\prime}\right)^{\prime}n_{k}^{-1}n_{p}^{-1}\right]\\
= & \sum_{p\neq k}^{K}\mathbb{E}\left[\sum_{r=1}^{n_{p}}W_{k,i}^{2}U_{k,i}^{2}U_{p,r}^{2}\left(e_{p}\otimes e_{k}\right)\left(e_{p}^{\prime}\otimes e_{k}^{\prime}\right)n_{k}^{-1}n_{p}^{-1}\right]\\
= & \sum_{p\neq k}^{K}\psi_{k,i}n_{p}\left(e_{p}\otimes e_{k}\right)\left(e_{p}^{\prime}\otimes e_{k}^{\prime}\right)n_{k}^{-1}n_{p}^{-1}\\
= & \sum_{p\neq k}^{K}\psi_{k,i}\left(J_{p}^{e}\otimes J_{k}^{e}\right)n_{k}^{-1}\\
= & \psi_{k,i}\left[\left(I_{K}-J_{k}^{e}\right)\otimes J_{k}\right]n_{k}^{-1}.
\end{align*}
So, we have
\begin{align*}
 & \sum_{k=1}^{K}\sum_{i=1}^{n_{k}}\mathbb{E}\left[W_{k,i}^{2}U_{k,i}^{2}{\rm vec}\left(e_{k}X_{0}^{\prime}\right){\rm vec}\left(e_{k}X_{0}^{\prime}\right)^{\prime}n_{k}^{-1}\right]\\
= & \sum_{k=1}^{K}\sum_{i=1}^{n_{k}}\left[\left[3\phi_{k,i}+\psi_{k,i}\left(n_{k}-1\right)\right]n_{k}^{-2}J_{k\otimes}+\psi_{k,i}n_{k}^{-1}\left[\left(I_{K}-J_{k}\right)\otimes J_{k}\right]\right]\\
= & \sum_{k=1}^{K}\left[n_{k}^{-1}\left(3\bar{\phi}_{k}-\bar{\psi}_{k}\right)J_{k\otimes}^{e}+\bar{\psi}_{k}\left(I_{K}\otimes J_{k}^{e}\right)\right],
\end{align*}
where $\bar{\phi}_{k}$ and $\bar{\psi}_{k}$, $k=1,\ldots,K$ are
defined as
\[
\bar{\phi}_{k}=\frac{1}{n_{k}}\sum_{k=1}^{n_{k}}\phi_{k,i},\qquad\text{and}\qquad\bar{\psi}_{k}=\frac{1}{n_{k}}\sum_{k=1}^{n_{k}}\psi_{k,i},
\]
respectively.

\subsubsection{The Second Term}

For $l\neq k$, we have 
\begin{align*}
 & \mathbb{E}\left[W_{k,i}W_{l,j}U_{k,i}U_{l,j}{\rm vec}\left(e_{k}X_{0}^{\prime}\right){\rm vec}\left(e_{l}X_{0}^{\prime}\right)^{\prime}n_{l}^{-\frac{1}{2}}n_{k}^{-\frac{1}{2}}\right]\\
= & \mathbb{E}\left[\sum_{p=1}^{n_{k}}\sum_{q=1}^{n_{k}}W_{k,i}W_{l,j}U_{k,i}U_{l,j}{\rm vec}\left(e_{k}U_{p}^{\prime}v_{n_{p}}e_{p}^{\prime}\right){\rm vec}\left(e_{l}U_{q}^{\prime}v_{n_{q}}e_{q}^{\prime}\right)^{\prime}n_{l}^{-\frac{1}{2}}n_{k}^{-\frac{1}{2}}\right]\\
= & \mathbb{E}\left[W_{k,i}W_{l,j}U_{k,i}U_{l,j}{\rm vec}\left(e_{k}U_{k}^{\prime}v_{n_{k}}e_{k}^{\prime}\right){\rm vec}\left(e_{l}U_{l}^{\prime}v_{n_{l}}e_{l}^{\prime}\right)^{\prime}n_{l}^{-\frac{1}{2}}n_{k}^{-\frac{1}{2}}\right]\\
 & +\mathbb{E}\left[W_{k,i}W_{l,j}U_{k,i}U_{l,j}{\rm vec}\left(e_{k}U_{l}^{\prime}v_{n_{l}}e_{l}^{\prime}\right){\rm vec}\left(e_{l}U_{k}^{\prime}v_{n_{k}}e_{k}^{\prime}\right)^{\prime}n_{l}^{-\frac{1}{2}}n_{k}^{-\frac{1}{2}}\right].
\end{align*}
The first term is given by
\begin{align*}
 & \mathbb{E}\left[W_{k,i}W_{l,j}U_{k,i}U_{l,j}{\rm vec}\left(e_{k}U_{k}^{\prime}v_{n_{k}}e_{k}^{\prime}\right){\rm vec}\left(e_{l}U_{l}^{\prime}v_{n_{l}}e_{l}^{\prime}\right)^{\prime}n_{l}^{-\frac{1}{2}}n_{k}^{-\frac{1}{2}}\right]\\
= & \mathbb{E}\left[\sum_{r}^{n_{k}}\sum_{m}^{n_{l}}W_{k,i}W_{l,j}U_{k,i}U_{l,j}U_{k,r}U_{l,m}{\rm vec}\left(e_{k}e_{k}^{\prime}\right){\rm vec}\left(e_{l}e_{l}^{\prime}\right)^{\prime}n_{l}^{-1}n_{k}^{-1}\right]\\
= & \mathbb{E}\left[\sum_{r}^{n_{k}}\sum_{m}^{n_{l}}W_{k,i}W_{l,j}U_{k,i}U_{l,j}U_{k,r}U_{l,m}{\rm vec}\left(J_{k}^{e}\right){\rm vec}\left(J_{l}^{e}\right)^{\prime}n_{l}^{-1}n_{k}^{-1}\right]\\
= & \mathbb{E}\left[W_{k,i}W_{l,j}U_{k,i}^{2}U_{l,j}^{2}{\rm vec}\left(J_{k}^{e}\right){\rm vec}\left(J_{l}^{e}\right)^{\prime}n_{l}^{-1}n_{k}^{-1}\right]\\
= & {\rm vec}\left(J_{k}^{e}\right){\rm vec}\left(J_{l}^{e}\right)^{\prime}n_{l}^{-1}n_{k}^{-1}.
\end{align*}
The second term is given by
\begin{align*}
 & \mathbb{E}\left[W_{k,i}W_{l,j}U_{k,i}U_{l,j}{\rm vec}\left(e_{k}U_{l}^{\prime}v_{n_{l}}e_{l}^{\prime}\right){\rm vec}\left(e_{l}U_{k}^{\prime}v_{n_{k}}e_{k}^{\prime}\right)^{\prime}n_{l}^{-\frac{1}{2}}n_{k}^{-\frac{1}{2}}\right]\\
= & \mathbb{E}\left[\sum_{r=1}^{n_{l}}\sum_{m=1}^{n_{k}}W_{k,i}W_{l,j}U_{k,i}U_{l,j}U_{l,r}U_{k,m}{\rm vec}\left(e_{k}e_{l}^{\prime}\right){\rm vec}\left(e_{l}e_{k}^{\prime}\right)^{\prime}n_{l}^{-1}n_{k}^{-1}\right]\\
= & \mathbb{E}\left[W_{k,i}W_{l,j}U_{k,i}^{2}U_{l,j}^{2}{\rm vec}\left(e_{k}e_{l}^{\prime}\right){\rm vec}\left(e_{l}e_{k}^{\prime}\right)^{\prime}n_{l}^{-1}n_{k}^{-1}\right]\\
= & {\rm vec}\left(e_{k}e_{l}^{\prime}\right){\rm vec}\left(e_{l}e_{k}^{\prime}\right)^{\prime}n_{l}^{-1}n_{k}^{-1}\\
= & \left(J_{l}^{e}\otimes J_{k}^{e}\right)K_{K}n_{l}^{-1}n_{k}^{-1}.
\end{align*}
So, we have
\begin{align*}
 & \sum_{k=1}^{K}\sum_{l\neq k}^{K}\sum_{i=1}^{n_{k}}\sum_{j=1}^{n_{l}}\mathbb{E}\left[W_{k,i}W_{l,j}U_{k,i}U_{l,j}{\rm vec}\left(e_{k}X_{0}^{\prime}\right){\rm vec}\left(e_{l}X_{0}^{\prime}\right)^{\prime}n_{l}^{-\frac{1}{2}}n_{k}^{-\frac{1}{2}}\right]\\
= & \sum_{k=1}^{K}\sum_{l\neq k}^{K}\sum_{i=1}^{n_{k}}\sum_{j=1}^{n_{l}}\left[{\rm vec}\left(J_{k}^{e}\right){\rm vec}\left(J_{l}^{e}\right)^{\prime}n_{l}^{-1}n_{k}^{-1}+\left(J_{l}^{e}\otimes J_{k}^{e}\right)K_{K}n_{l}^{-1}n_{k}^{-1}\right]\\
= & \sum_{k=1}^{K}\sum_{l\neq k}^{K}\left[{\rm vec}\left(J_{k}^{e}\right){\rm vec}\left(J_{l}^{e}\right)^{\prime}+\left(J_{l}^{e}\otimes J_{k}^{e}\right)K_{K}\right]\\
= & \sum_{k=1}^{K}\left[{\rm vec}\left(J_{k}^{e}\right){\rm vec}\left(I_{K}\right)^{\prime}+\left(I_{K}\otimes J_{k}^{e}\right)K_{K}-J_{k\otimes}^{e}\left(K_{K}+I_{K^{2}}\right)\right]\\
= & {\rm vec}\left(I_{K}\right){\rm vec}\left(I_{K}\right)^{\prime}+K_{K}-2\sum_{k=1}^{K}J_{k\otimes}^{e},
\end{align*}
the last equality use the fact that $J_{k\otimes}^{e}={\rm vec}\left(J_{k}^{e}\right){\rm vec}\left(J_{k}^{e}\right)^{\prime}$
as $J_{k}^{e}=e_{k}e_{k}^{\prime}$, and $J_{k\otimes}^{e}K_{K}=J_{k\otimes}^{e}$.

\subsubsection{The Third Term}

We have $i\neq j$, then
\begin{align*}
 & \mathbb{E}\left[W_{k,i}W_{k,j}U_{k,i}U_{k,j}{\rm vec}\left(e_{k}X_{0}^{\prime}\right){\rm vec}\left(e_{k}X_{0}^{\prime}\right)^{\prime}n_{k}^{-\frac{1}{2}}n_{k}^{-\frac{1}{2}}\right]\\
= & \mathbb{E}\left[\sum_{p}^{n_{k}}\sum_{q}^{n_{k}}W_{k,i}W_{k,j}U_{k,i}U_{k,j}{\rm vec}\left(e_{k}U_{p}^{\prime}v_{n_{p}}e_{p}^{\prime}\right){\rm vec}\left(e_{k}U_{q}^{\prime}v_{n_{q}}e_{q}^{\prime}\right)^{\prime}n_{k}^{-1}\right]\\
= & \mathbb{E}\left[W_{k,i}W_{k,j}U_{k,i}U_{k,j}{\rm vec}\left(e_{k}U_{k}^{\prime}v_{n_{k}}e_{k}^{\prime}\right){\rm vec}\left(e_{k}U_{k}^{\prime}v_{n_{k}}e_{k}^{\prime}\right)^{\prime}n_{k}^{-1}\right]\\
= & \sum_{r}^{n_{k}}\sum_{m}^{n_{k}}\mathbb{E}\left[W_{k,i}W_{k,j}U_{k,i}U_{k,j}U_{k,r}U_{k,m}{\rm vec}\left(e_{k}e_{k}^{\prime}\right){\rm vec}\left(e_{k}e_{k}^{\prime}\right)^{\prime}n_{k}^{-2}\right]\\
= & \mathbb{E}\left[W_{k,i}W_{k,j}U_{k,i}U_{k,j}U_{k,i}U_{k,j}+W_{k,i}W_{k,j}U_{k,i}U_{k,j}U_{k,j}U_{k,i}\right]J_{k\otimes}^{e}n_{k}^{-2}\\
= & 2\mathbb{E}\left[W_{k,i}W_{k,j}U_{k,i}^{2}U_{k,j}^{2}\right]J_{k\otimes}^{e}n_{k}^{-2}\\
= & 2J_{k\otimes}^{e}n_{k}^{-2}.
\end{align*}
So we have
\begin{align*}
 & \mathbb{E}\left[\sum_{k=1}^{K}\sum_{i=1}^{n_{k}}\sum_{j\neq i}^{n_{k}}W_{k,i}W_{k,j}U_{k,i}U_{k,j}{\rm vec}\left(e_{k}X_{0}^{\prime}\right){\rm vec}\left(e_{k}X_{0}^{\prime}\right)^{\prime}n_{k}^{-\frac{1}{2}}n_{k}^{-\frac{1}{2}}\right]\\
= & \mathbb{E}\left[\sum_{k=1}^{K}\sum_{i=1}^{n_{k}}\sum_{j\neq i}^{n_{k}}2J_{k\otimes}^{e}n_{k}^{-2}\right]=\sum_{k=1}^{K}2J_{k\otimes}^{e}\left(1-n_{k}^{-1}\right).
\end{align*}

\subsubsection{Combine}

Now we have
\begin{align*}
 & \mathbb{E}\left[\sum_{k=1}^{K}\sum_{i=1}^{n_{k}}W_{k,i}U_{k,i}{\rm vec}\left(e_{k}X_{0}^{\prime}\right)^{\prime}n_{k}^{-\frac{1}{2}}\right]\left[\sum_{l=1}^{K}\sum_{j=1}^{n_{k}}W_{l,j}U_{l,j}{\rm vec}\left(e_{l}X_{0}^{\prime}\right)^{\prime}n_{l}^{-\frac{1}{2}}\right]-{\rm vec}\left(I_{K}\right){\rm vec}\left(I_{K}\right)^{\prime}\\
= & K_{K}+\sum_{k=1}^{K}\left[n_{k}^{-1}\left(3\bar{\phi}_{k}-\bar{\psi}_{k}\right)J_{k\otimes}^{e}+\bar{\psi}_{k}\left(I_{K}\otimes J_{k}^{e}\right)-2J_{k\otimes}^{e}+2J_{k\otimes}^{e}\left(1-n_{k}^{-1}\right)\right]\\
= & K_{K}+\Upsilon_{K}^{e},
\end{align*}
where $\Upsilon_{K}^{e}=\sum_{k=1}^{K}\Psi_{k}^{e}$ with
\begin{align*}
\Psi_{k}^{e} & =n_{k}^{-1}\left(3\bar{\phi}_{k}-\bar{\psi}_{k}\right)J_{k\otimes}^{e}+\bar{\psi}_{k}\left(I_{K}\otimes J_{k}^{e}\right)-2J_{k\otimes}^{e}+2J_{k\otimes}^{e}\left(1-n_{k}^{-1}\right)\\
 & =n_{k}^{-1}\left(3\bar{\phi}_{k}-2-\bar{\psi}_{k}\right)J_{k\otimes}^{e}+\bar{\psi}_{k}\left(I_{K}\otimes J_{k}^{e}\right).
\end{align*}

\subsection{The Form of Matrix $\mathcal{I}_{A}^{(2)}$}

We first define the $K\times1$ vector $\bar{S}$ with elements
\[
\bar{S}_{k}=\sum_{i=1}^{n_{k}}W_{k,i}U_{k,i}F{}_{k,i}U_{k}-\left(n_{k}-1\right),
\]
and obviously we have $\mathbb{E}\left(S_{k}S_{l}\right)=0$ for $k\neq l$.
And we need to compute
\begin{align*}
\mathbb{E}\left(\bar{S}_{k}^{2}\right) & =\mathbb{E}\left(\sum_{i=1}^{n_{k}}\sum_{j=1}^{n_{k}}W_{k,i}W_{k,j}U_{k,i}U_{k,j}F_{k,i}U_{k}U_{k}^{\prime}F_{k,j}^{\prime}\right)-\left(n_{k}-1\right)^{2}\\
 & =\mathbb{E}\left(\sum_{i=1}^{n_{k}}\sum_{j=1}^{n_{k}}W_{k,i}W_{k,j}\left(U_{k,i}U_{k,j}\right)F_{k,i}\left(U_{k}U_{k}^{\prime}\right)F_{k,j}^{\prime}\right)-\left(n_{k}-1\right)^{2}\\
 & =\mathbb{E}\left(\sum_{i=1}^{n_{k}}W_{k,i}^{2}U_{k,i}^{2}F_{k,i}\left(U_{k}U_{k}^{\prime}\right)F_{k,i}^{\prime}\right)-\left(n_{k}-1\right)^{2}\\
 & \quad+\mathbb{E}\left(\sum_{i=1}^{n_{k}}\sum_{j\neq i}^{n_{k}}W_{k,i}W_{k,j}\left(U_{k,i}U_{k,j}\right)F_{k,i}\left(U_{k}U_{k}^{\prime}\right)F_{k,j}^{\prime}\right).
\end{align*}
Based on the following results on $F_{k,i}=\tilde{e}_{i}^{\prime}v_{n_{k}}^{\perp}v_{n_{k}}^{\perp\ \prime}$,
for $i\neq j$ we have
\begin{align*}
F_{k,i}F_{k,i}^{\prime} & =\tilde{e}_{i}^{\prime}\left(v_{n_{k}}^{\perp}v_{n_{k}}^{\perp\ \prime}\right)\tilde{e}_{i}=\tilde{e}_{i}^{\prime}\left(I_{n_{k}}-v_{n_{k}}v_{n_{k}}^{\prime}\right)\tilde{e}_{i}=1-n_{k}^{-1}\\
F_{k,i}\tilde{e}_{i}\tilde{e}_{i}^{\prime}F_{k,i}^{\prime} & =\left(\tilde{e}_{i}^{\prime}v_{n_{k}}^{\perp}v_{n_{k}}^{\perp\ \prime}\tilde{e}_{i}\right)\left(\tilde{e}_{i}^{\prime}v_{n_{k}}^{\perp}v_{n_{k}}^{\perp\ \prime}\tilde{e}_{i}\right)=\left(1-n_{k}^{-1}\right)^{2}\\
F_{k,i}\tilde{e}_{j}\tilde{e}_{i}^{\prime}F_{k,j}^{\prime} & =\left(\tilde{e}_{i}^{\prime}v_{n_{k}}^{\perp}v_{n_{k}}^{\perp\ \prime}\tilde{e}_{j}\right)\left(\tilde{e}_{i}^{\prime}v_{n_{k}}^{\perp}v_{n_{k}}^{\perp\ \prime}\tilde{e}_{j}\right)=\left(-n_{k}^{-1}\right)^{2}=n_{k}^{-2},
\end{align*}
we have
\begin{align*}
 & \mathbb{E}\left[W_{k,i}^{2}U_{k,i}^{2}F_{k,i}\left(U_{k}U_{k}^{\prime}\right)F_{k,i}^{\prime}\right]\\
= & \mathbb{E}\left[\sum_{p}^{n_{k}}\sum_{q}^{n_{k}}W_{k,i}^{2}U_{k,i}^{2}\left(U_{k,p}U_{k,q}\right)F_{k,i}\left(\tilde{e}_{p}\tilde{e}_{q}^{\prime}\right)F_{k,i}^{\prime}\right]\\
= & \mathbb{E}\left[W_{k,i}^{2}U_{k,i}^{4}F_{k,i}\left(\tilde{e}_{i}\tilde{e}_{i}^{\prime}\right)F_{k,i}^{\prime}\right]+\sum_{p\neq i}^{n_{k}}\mathbb{E}\left[W_{k,i}^{2}U_{k,i}^{2}U_{k,p}^{2}F_{k,i}\left(\tilde{e}_{p}\tilde{e}_{p}^{\prime}\right)F_{k,i}^{\prime}\right]\\
= & 3\phi_{k,i}F_{k,i}\left(\tilde{e}_{i}\tilde{e}_{i}^{\prime}\right)F_{k,i}^{\prime}+\sum_{p\neq i}^{n_{k}}\psi_{k,i}F_{k,i}\left(\tilde{e}_{p}\tilde{e}_{p}^{\prime}\right)F_{k,i}^{\prime}\\
= & 3\phi_{k,i}F_{k,i}\left(\tilde{e}_{i}\tilde{e}_{i}^{\prime}\right)F_{k,i}^{\prime}+\psi_{k,i}n_{k}^{-2}\left(n_{k}-1\right)\\
= & 3\phi_{k,i}\left(1-n_{k}^{-1}\right)^{2}+\psi_{k,i}n_{k}^{-2}\left(n_{k}-1\right).
\end{align*}
So
\[
\sum_{i=1}^{n_{k}}\mathbb{E}\left[W_{k,i}^{2}U_{k,i}^{2}F_{k,i}\left(U_{k}U_{k}^{\prime}\right)F_{k,i}^{\prime}\right]=3n_{k}\bar{\phi}_{k}\left(1-n_{k}^{-1}\right)^{2}+\bar{\psi}_{k}\left(1-n_{k}^{-1}\right),
\]
and for $i\neq j$, we have
\begin{align*}
 & \mathbb{E}\left[W_{k,i}W_{k,j}\left(U_{k,i}U_{k,j}\right)F_{k,i}\left(U_{k}U_{k}^{\prime}\right)F_{k,j}^{\prime}\right]\\
= & \mathbb{E}\left[\sum_{p}^{n_{k}}\sum_{q}^{n_{k}}W_{k,i}W_{k,j}\left(U_{k,i}U_{k,j}\right)\left(U_{k,p}U_{k,q}\right)F_{k,i}\left(\tilde{e}_{p}\tilde{e}_{q}^{\prime}\right)F_{k,j}^{\prime}\right]\\
= & \mathbb{E}\left[W_{k,i}W_{k,j}\left(U_{k,i}^{2}U_{k,j}^{2}\right)F_{k,i}\left(\tilde{e}_{i}\tilde{e}_{j}^{\prime}\right)F_{k,j}^{\prime}\right]+\mathbb{E}\left[W_{k,i}W_{k,j}\left(U_{k,i}^{2}U_{k,j}^{2}\right)F_{k,i}\left(\tilde{e}_{j}\tilde{e}_{i}^{\prime}\right)F_{k,j}^{\prime}\right]\\
= & F_{k,i}\left(\tilde{e}_{i}\tilde{e}_{j}^{\prime}+\tilde{e}_{j}\tilde{e}_{i}^{\prime}\right)F_{k,j}^{\prime}=\left(1-n_{k}^{-1}\right)^{2}+n_{k}^{-2}.
\end{align*}
So,
\[
\mathbb{E}\left[\sum_{j\neq i}^{n_{k}}W_{k,i}W_{k,j}\left(U_{k,i}U_{k,j}\right)F_{k,i}\left(U_{k}U_{k}^{\prime}\right)F_{k,j}^{\prime}\right]=\left(n_{k}-1\right)\left[\left(1-n_{k}^{-1}\right)^{2}+n_{k}^{-2}\right],
\]
and this leads to 
\begin{align*}
\mathbb{E}\left(\bar{S}_{k}^{2}\right) & =3n_{k}\bar{\phi}_{k}\left(1-n_{k}^{-1}\right)^{2}+\bar{\psi}_{k}\left(1-n_{k}^{-1}\right)+n_{k}\left(n_{k}-1\right)\left[\left(1-n_{k}^{-1}\right)^{2}+n_{k}^{-2}\right]-\left(n_{k}-1\right)^{2},\\
\mathbb{E}\left(S_{k}^{2}\right) & =\lambda_{k}^{-2}\left(n_{k}-1\right)^{-2}\mathbb{E}\left(\bar{S}_{k}^{2}\right)=\lambda_{k}^{-2}n_{k}^{-1}\left[3\bar{\phi}_{k}-1+\left(\bar{\psi}_{k}+1\right)\left(n_{k}-1\right)^{-1}\right],
\end{align*}
and define the matrix $\Xi$ as $\Xi_{kk}=\mathbb{E}\left(S_{k}^{2}\right)$
and $\Xi_{kl}=\mathbb{E}\left(S_{k}S_{j}\right)=0$ for $k\neq l$,
we have $\mathcal{I}_{A}^{(2)}=\frac{1}{4}E_{d}^{\prime}\Xi E_{d}$.

\subsection{The Form of Matrix $\mathcal{I}_{A}^{(12)}$}

We first need to compute
\begin{align*}
 & \mathbb{E}\left[\sum_{i=1}^{n_{k}}W_{k,i}U_{k,i}F_{k,i}U_{k}-\left(n_{k}-1\right)\right]\left[\sum_{l=1}^{K}\sum_{j=1}^{n_{l}}W_{l,j}U_{l,j}{\rm vec}\left(e_{l}X_{0}^{\prime}\right)^{\prime}n_{l}^{-\frac{1}{2}}-{\rm vec}(I_{K})^{\prime}\right]\\
= & \mathbb{E}\left[\sum_{l=1}^{K}\sum_{i=1}^{n_{k}}\sum_{j=1}^{n_{l}}W_{k,i}W_{l,j}U_{k,i}U_{l,j}\left(F_{k,i}U_{k}\right){\rm vec}\left(e_{l}X_{0}^{\prime}\right)^{\prime}n_{l}^{-\frac{1}{2}}\right]-\left(n_{k}-1\right){\rm vec}(I_{K})^{\prime}\\
= & \mathbb{E}\left[\sum_{i=1}^{n_{k}}W_{k,i}^{2}U_{k,i}^{2}\left(F_{k,i}U_{k}\right){\rm vec}\left(e_{k}X_{0}^{\prime}\right)^{\prime}n_{k}^{-\frac{1}{2}}\right]-\left(n_{k}-1\right){\rm vec}(I_{K})^{\prime}\\
 & +\mathbb{E}\left[\sum_{i=1}^{n_{k}}\sum_{j\neq i}^{n_{k}}W_{k,i}W_{k,j}U_{k,i}U_{k,j}\left(F_{k,i}U_{k}\right){\rm vec}\left(e_{k}X_{0}^{\prime}\right)^{\prime}n_{k}^{-\frac{1}{2}}\right]\\
 & +\mathbb{E}\left[\sum_{l\ne k}^{K}\sum_{i=1}^{n_{k}}\sum_{j=1}^{n_{l}}W_{k,i}W_{l,j}U_{k,i}U_{l,j}\left(F_{k,i}U_{k}\right){\rm vec}\left(e_{l}X_{0}^{\prime}\right)^{\prime}n_{l}^{-\frac{1}{2}}\right].
\end{align*}
For the first term, we have
\begin{align*}
 & \mathbb{E}\left[W_{k,i}^{2}U_{k,i}^{2}\left(F_{k,i}U_{k}\right){\rm vec}\left(e_{k}X_{0}^{\prime}\right)^{\prime}n_{k}^{-\frac{1}{2}}\right]\\
= & \mathbb{E}\left[\sum_{q}^{K}\sum_{p}^{n_{k}}W_{k,i}^{2}U_{k,i}^{2}\left(F_{k,i}U_{k,p}\tilde{e}_{p}\right){\rm vec}\left(e_{k}U_{q}^{\prime}v_{n_{q}}e_{q}^{\prime}\right)^{\prime}n_{k}^{-\frac{1}{2}}\right]\\
= & \mathbb{E}\left[\sum_{p}^{n_{k}}W_{k,i}^{2}U_{k,i}^{2}U_{k,p}\left(F_{k,i}\tilde{e}_{p}\right){\rm vec}\left(e_{k}U_{k}^{\prime}v_{n_{k}}e_{k}^{\prime}\right)^{\prime}n_{k}^{-\frac{1}{2}}\right]\\
= & \mathbb{E}\left[\sum_{p}^{n_{k}}\sum_{r}^{n_{k}}W_{k,i}^{2}U_{k,i}^{2}U_{k,p}U_{k,r}\left(F_{k,i}\tilde{e}_{r}\right){\rm vec}\left(e_{k}e_{k}^{\prime}\right)^{\prime}n_{k}^{-1}\right]\\
= & \mathbb{E}\left[W_{k,i}^{2}U_{k,i}^{4}\left(F_{k,i}\tilde{e}_{i}\right){\rm vec}\left(e_{k}e_{k}^{\prime}\right)^{\prime}n_{k}^{-1}\right]+\mathbb{E}\left[\sum_{p\neq i}^{n_{k}}W_{k,i}^{2}U_{k,i}^{2}U_{k,p}^{2}\left(F_{k,i}\tilde{e}_{p}\right){\rm vec}\left(e_{k}e_{k}^{\prime}\right)^{\prime}n_{k}^{-1}\right]\\
= & 3\phi_{k,i}\left[\left(F_{k,i}\tilde{e}_{i}\right){\rm vec}\left(J_{k}^{e}\right)^{\prime}n_{k}^{-1}\right]+\sum_{p\neq i}^{n_{k}}\psi_{k,i}\left(F_{k,i}\tilde{e}_{p}\right){\rm vec}\left(J_{k}^{e}\right)^{\prime}n_{k}^{-1}\\
= & 3\phi_{k,i}\left(1-n_{k}^{-1}\right){\rm vec}\left(J_{k}^{e}\right)^{\prime}n_{k}^{-1}-\psi_{k,i}\left(n_{k}-1\right)n_{k}^{-1}{\rm vec}\left(J_{k}^{e}\right)^{\prime}n_{k}^{-1}\\
= & \left(3\phi_{k,i}-\psi_{k,i}\right)\left(1-n_{k}^{-1}\right){\rm vec}\left(J_{k}^{e}\right)^{\prime}n_{k}^{-1}.
\end{align*}
So
\begin{align*}
\mathbb{E}\left[\sum_{i}^{n_{k}}W_{k,i}^{2}U_{k,i}^{2}\left(F_{k,i}U_{k}\right){\rm vec}\left(e_{k}X_{0}^{\prime}\right)^{\prime}n_{k}^{-\frac{1}{2}}\right] & =\left(3\bar{\phi}_{k}-\bar{\psi}_{k}\right)\left(1-n_{k}^{-1}\right){\rm vec}\left(J_{k}\right)^{\prime},
\end{align*}
and for the second term, we have $i\neq j$ and
\begin{align*}
 & \mathbb{E}\left[W_{k,i}W_{k,j}U_{k,i}U_{k,j}\left(F_{k,i}U_{k}\right){\rm vec}\left(e_{k}X_{0}^{\prime}\right)^{\prime}n_{k}^{-\frac{1}{2}}\right]\\
= & \mathbb{E}\left[\sum_{q=1}^{K}\sum_{p=1}^{n_{k}}W_{k,i}W_{k,j}U_{k,i}U_{k,j}U_{k,p}\left(F_{k,i}\tilde{e}_{p}\right){\rm vec}\left(e_{k}U_{q}^{\prime}v_{n_{q}}e_{q}^{\prime}\right)^{\prime}n_{k}^{-\frac{1}{2}}\right]\\
= & \mathbb{E}\left[\sum_{p=1}^{n_{k}}W_{k,i}W_{k,j}U_{k,i}U_{k,j}U_{k,p}\left(F_{k,i}\tilde{e}_{p}\right){\rm vec}\left(e_{k}U_{k}^{\prime}v_{n_{k}}e_{k}^{\prime}\right)^{\prime}n_{k}^{-\frac{1}{2}}\right]\\
= & \mathbb{E}\left[\sum_{p=1}^{n_{k}}\sum_{r=1}^{n_{k}}W_{k,i}W_{k,j}U_{k,i}U_{k,j}U_{k,p}U_{k,r}\left(F_{k,i}\tilde{e}_{p}\right){\rm vec}\left(e_{k}e_{k}^{\prime}\right)^{\prime}n_{k}^{-1}\right]\\
= & \mathbb{E}\left[W_{k,i}W_{k,j}U_{k,i}^{2}U_{k,j}^{2}\left(F_{k,i}\tilde{e}_{i}\right){\rm vec}\left(J_{k}\right)^{\prime}n_{k}^{-1}\right]+\mathbb{E}\left[W_{k,i}W_{k,j}U_{k,i}^{2}U_{k,j}^{2}\left(F_{k,i}\tilde{e}_{j}\right){\rm vec}\left(J_{k}\right)^{\prime}n_{k}^{-1}\right]\\
= & \left(F_{k,i}\tilde{e}_{i}\right){\rm vec}\left(J_{k}^{e}\right)^{\prime}n_{k}^{-1}+\left(F_{k,i}\tilde{e}_{j}\right){\rm vec}\left(J_{k}^{e}\right)^{\prime}n_{k}^{-1}\\
= & \left(1-n_{k}^{-1}\right){\rm vec}\left(J_{k}^{e}\right)^{\prime}n_{k}^{-1}-n_{k}^{-1}{\rm vec}\left(J_{k}^{e}\right)^{\prime}n_{k}^{-1}\\
= & \left(1-2n_{k}^{-1}\right){\rm vec}\left(J_{k}^{e}\right)^{\prime}n_{k}^{-1}.
\end{align*}
So
\[
\mathbb{E}\left[\sum_{i=1}^{n_{k}}\sum_{j\neq i}^{n_{k}}W_{k,i}W_{k,j}U_{k,i}U_{k,j}\left(F_{k,i}U_{k}\right){\rm vec}\left(e_{k}X_{0}^{\prime}\right)^{\prime}n_{k}^{-\frac{1}{2}}\right]=\left(n_{k}-1\right)\left(1-2n_{k}^{-1}\right){\rm vec}\left(J_{k}^{e}\right)^{\prime},
\]
as for the third term, we have $k\neq l$, and
\begin{align*}
 & \mathbb{E}\left[W_{k,i}W_{l,j}U_{k,i}U_{l,j}\left(F_{k,i}U_{k}\right){\rm vec}\left(e_{l}X_{0}^{\prime}\right)^{\prime}n_{l}^{-\frac{1}{2}}\right]\\
= & \mathbb{E}\left[\sum_{q=1}^{K}\sum_{p=1}^{n_{k}}W_{k,i}W_{l,j}U_{k,i}U_{l,j}U_{k,p}\left(F_{k,i}\tilde{e}_{p}\right){\rm vec}\left(e_{l}U_{q}^{\prime}v_{n_{q}}e_{q}^{\prime}\right)^{\prime}n_{l}^{-\frac{1}{2}}\right]\\
= & \mathbb{E}\left[\sum_{p=1}^{n_{k}}W_{k,i}W_{l,j}U_{k,i}U_{l,j}U_{k,p}\left(F_{k,i}\tilde{e}_{p}\right){\rm vec}\left(e_{l}U_{l}^{\prime}v_{n_{l}}e_{l}^{\prime}\right)^{\prime}n_{l}^{-\frac{1}{2}}\right]\\
= & \mathbb{E}\left[\sum_{p=1}^{n_{k}}\sum_{r=1}^{n_{l}}W_{k,i}W_{l,j}U_{k,i}U_{k,p}U_{l,j}U_{l,r}\left(F_{k,i}\tilde{e}_{p}\right){\rm vec}\left(e_{l}e_{l}^{\prime}\right)^{\prime}n_{l}^{-1}\right]\\
= & \mathbb{E}\left[W_{k,i}W_{l,j}U_{k,i}^{2}U_{l,j}^{2}\left(F_{k,i}\tilde{e}_{i}\right){\rm vec}\left(e_{l}e_{l}^{\prime}\right)^{\prime}n_{l}^{-1}\right]\\
= & \left(F_{k,i}\tilde{e}_{i}\right){\rm vec}\left(J_{l}^{e}\right)^{\prime}n_{l}^{-1}\\
= & \left(1-n_{k}^{-1}\right){\rm vec}\left(J_{l}^{e}\right)^{\prime}n_{l}^{-1}.
\end{align*}
So, we have
\[
\mathbb{E}\left[\sum_{l\ne k}^{K}\sum_{i=1}^{n_{k}}\sum_{j=1}^{n_{l}}W_{k,i}W_{l,j}U_{k,i}U_{l,j}\left(F_{k,i}U_{k}\right){\rm vec}\left(e_{l}X_{0}^{\prime}\right)^{\prime}n_{l}^{-\frac{1}{2}}\right]=\left(n_{k}-1\right){\rm vec}\left(I_{K}-J_{k}^{e}\right)^{\prime},
\]
and
\begin{align*}
 & \mathbb{E}\left[\sum_{i=1}^{n_{k}}W_{k,i}U_{k,i}F_{k,i}U_{k}-\left(n_{k}-1\right)\right]\left[\sum_{l=1}^{K}\sum_{j=1}^{n_{k}}W_{k,j}U_{k,j}{\rm vec}\left(e_{k}X_{0}^{\prime}\right)^{\prime}n_{k}^{-\frac{1}{2}}-{\rm vec}(I_{K})^{\prime}\right]\\
= & \left(3\bar{\phi}_{k}-\bar{\psi}_{k}\right)\left(1-n_{k}^{-1}\right){\rm vec}\left(J_{k}^{e}\right)^{\prime}+\left(n_{k}-1\right)\left(1-2n_{k}^{-1}\right){\rm vec}\left(J_{k}^{e}\right)^{\prime}\\
 & +\left(n_{k}-1\right){\rm vec}\left(I_{K}-J_{k}^{e}\right)^{\prime}-\left(n_{k}-1\right){\rm vec}(I_{K})^{\prime}\\
= & \left[\left(3\bar{\phi}_{k}-\bar{\psi}_{k}-2\right)\left(1-n_{k}^{-1}\right)\right]{\rm vec}\left(J_{k}^{e}\right)^{\prime}.
\end{align*}
We finally arrive at the following expression for $\Theta$, with
$\mathcal{I}_{A}^{(21)}=\frac{1}{2}E_{d}^{\prime}\Theta E_{d}$, 
\[
\Theta=\mathbb{E}\left[S\left(\sum_{k=1}^{K}\sum_{j=1}^{n_{k}}W_{k,j}U_{k,j}{\rm vec}\left(e_{k}X_{0}^{\prime}\right)^{\prime}n_{k}^{-\frac{1}{2}}-{\rm vec}(I_{K})^{\prime}\right)\right]=\sum_{k=1}^{K}e_{k}\left(e_{k}^{\prime}\Theta\right),
\]
where its $k$-th row $e_{k}^{\prime}\Theta$ is given by
\begin{align*}
e_{k}^{\prime}\Theta & =-\left(n_{k}-1\right)^{-1}\lambda_{k}^{-1}\left[\left(3\bar{\phi}_{k}-\bar{\psi}_{k}-2\right)\left(1-n_{k}^{-1}\right)\right]{\rm vec}\left(J_{k}^{e}\right)^{\prime}\\
 & =-\lambda_{k}^{-1}n_{k}^{-1}\left(3\bar{\phi}_{k}-\bar{\psi}_{k}-2\right){\rm vec}\left(J_{k}^{e}\right)^{\prime}.
\end{align*}

\end{document}